SPECTRALLY AND POWER EFFICIENT
OPTICAL COMMUNICATION SYSTEMS

A DISSERTATION
SUBMITTED TO THE DEPARTMENT OF ELECTRICAL ENGINEERING
AND THE COMMITTEE ON GRADUATE STUDIES
OF STANFORD UNIVERSITY
IN PARTIAL FULFILLMENT OF THE REQUIREMENTS
FOR THE DEGREE OF
DOCTOR OF PHILOSOPHY

Jose Krause Perin
June 2018

I certify that I have read this dissertation and that, in my opinion, it is fully adequate in scope and quality as a dissertation for the degree of Doctor of Philosophy.

\_\_\_\_\_\_\_\_\_\_\_\_\_\_\_\_\_\_\_\_\_\_\_\_\_\_\_\_\_\_\_\_\_\_\_\_\_\_

(Joseph M. Kahn)    Principal Adviser

I certify that I have read this dissertation and that, in my opinion, it is fully adequate in scope and quality as a dissertation for the degree of Doctor of Philosophy.

\_\_\_\_\_\_\_\_\_\_\_\_\_\_\_\_\_\_\_\_\_\_\_\_\_\_\_\_\_\_\_\_\_\_\_\_\_\_

(Olav Solgaard)

I certify that I have read this dissertation and that, in my opinion, it is fully adequate in scope and quality as a dissertation for the degree of Doctor of Philosophy.

\_\_\_\_\_\_\_\_\_\_\_\_\_\_\_\_\_\_\_\_\_\_\_\_\_\_\_\_\_\_\_\_\_\_\_\_\_\_

(Boris Murmann)

Approved for the Stanford University Committee on Graduate Studies

\_\_\_\_\_\_\_\_\_\_\_\_\_\_\_\_\_\_\_\_\_\_\_\_\_\_\_\_\_\_\_\_\_\_\_\_\_\_

# Abstract

Increased traffic demands globally and in particular in short-reach links in data centers will require optical communication systems to continue scaling at an accelerated pace. Nevertheless, energy constraints start to limit the bit rate that can be practically transmitted over optical systems both at the shortest distances in data centers and at the longest distances in ultra-long submarine links. Short-reach links in data centers face strict constraints on power consumption, size, and cost, which will demand low-power solutions that scale to bit rates beyond 100 Gbit/s per wavelength, while accommodating increased losses due to longer fiber plant, multiplexing of more wavelengths, and possibly optical switching. At the longest distances, submarine optical cables longer than about 5,000 km face energy constraints due to power feed limits at the shores, which restricts the electrical power available to the undersea optical amplifiers, ultimately limiting the optical power and throughput per fiber.

This dissertation addresses fundamental challenges towards designing spectrally and power efficient optical communication systems.

The first part of this dissertation focuses on short-reach optical systems for intra- and inter-data center applications.

Chapter 2 evaluates higher-order modulation formats compatible with direct detection (DD) that are best suited to replaced on/off keying (OOK) in next-generation data center links that support 100 Gbit/s per wavelength. We show that four-level pulse-amplitude modulation (4-PAM) outperforms orthogonal frequency-division multiplexing (OFDM) due to its relatively low complexity and higher tolerance to noise and distortion. And in fact, 4-PAM was later adopted by the IEEE 802.3bs task force to enable 400 Gbit/s using $8\times50$ Gbit/s and $4\times100$ Gbit/s transceivers. The work in Chapter 2 was done in collaboration with Dr. Milad Sharif, who has conducted the research and analyses for single-carrier modulation formats.

Chapter 3 focuses on how to improve the limited receiver sensitivity of 4-PAM systems proposed in Chapter 2 by using avalanche photodiodes (APDs) or semiconductor optical amplifiers (SOAs). We showed that APDs and SOAs improve the receiver sensitivity by 4 to 6 dB, which will extend the lifetime of 4-PAM and other DD-compatible modulation formats. The work in Chapter 3 was also done in collaboration with Dr. Milad Sharif, who studied the benefits and drawbacks of using

SOAs.

Chapter 4 focuses on the design of DSP-free coherent receiver architectures for low-power short-reach systems. As demonstrated in Chapters 2 and 3, DD-compatible formats face significant challenges to scale beyond 100 Gbit/s per wavelength. Moreover, these systems already face tight practical constraints even when counting on amplification, either by using APDs or SOAs. The underlying reason behind these challenges is that DD-compatible systems only leverage one degree of freedom of the optical channel, namely its intensity. Coherent receivers allow four degrees of freedom, two quadratures in two polarizations. But coherent receivers have been traditionally realized using high-speed analog-to-digital converters (ADCs) and digital signal processing (DSP), which are prohibitively power hungry for data center applications. We proposed low-power coherent receivers architectures that completely preclude the need of high-speed DSP and ADCs, while achieving similar performance to their DSP-based counterparts. The work in Chapter 4 was done in collaboration with Dr. Anujit Shastri, who designed and simulated the polarization recovery system based on cascaded phase shifters and maker tone detection.

The second part of this dissertation focuses on ultra-long submarine optical links, where energy constraints due to limited power feed voltage at the shores ultimately limits the amount of information that can be practically transmitted per fiber.

Chapter 5 focuses on the channel power optimization of long-haul submarine systems limited by energy constraints. The throughput of submarine transport cables is approaching fundamental limits imposed by amplifier noise and Kerr nonlinearity. Energy constraints in ultra-long submarine links exacerbate this problem, as the throughput per fiber is further limited by the electrical power available to the undersea optical amplifiers. Recent works have studied how employing more spatial dimensions can mitigate these limitations. This chapter addresses the fundamental question of how to optimally use each spatial dimension. Specifically, we discuss how to optimize the channel power allocation in order to maximize the information-theoretic capacity under an electrical power constraint. Our formulation accounts for amplifier physics, Kerr nonlinearity, and power feed constraints. We show that the optimized channel power allocation increases the capacity of submarine links by about 70% compared to the theoretical capacity of a recently proposed high-capacity system. Our solutions also provide new insights on the optimal number of spatial dimensions, amplifier operation, and nonlinear regime operation.

Chapter 6 presents the concluding remarks of this dissertation and recommendations for future work.

# Acknowledgments

I am very grateful for the opportunity of pursing my PhD at Stanford. I have learned a lot, grown a lot, and had the pleasure of working with many truly brilliant people. Although a great part of the work as a graduate student is done alone, many people have contributed along the way, and I would like to express my sincere gratitude to them.

I would like to first thank my principal adviser Prof. Joseph M. Kahn for his continuous support, guidance, and encouragement over the past five years. I have learned a lot from Prof. Kahn's approach to scientific research, from his commitment to teaching, and from his unwavering pursuit of excellence. I could have not asked for a more insightful, generous, and caring research adviser.

I would also like to thank members of my oral defense committee: Prof. Olav Solgaard, Prof. Boris Murmann, Prof. Sanjay Lall, and Prof. Bernard Widrow, who generously agreed to be part of my committee. I greatly appreciate their time and I am honored to have them in my oral defense committee. I want to thank Prof. Solgaard and Prof. Murmann for also serving as members of my dissertation reading committee and taking time out of their busy schedules to read this dissertation.

I would also like to thank the Coordenação de Aperfeiçoamento de Pessoal de Nível Superior (CAPES) – a Brazilian federal government agency – for awarding me a fellowship for three years of my graduate studies.

I would also like to thank the collaboration and funding from our industry partners: Maxim Integrated, Google, and Corning Inc.

I am also very grateful for the continuous support and guidance that I received from my former professors at Universidade Federal do Espírito Santo (UFES), in Brazil. In particular, Prof. Moisés Ribeiro, who over the years has always demonstrated great interest in my personal and academic success.

I am also thankful for of all past and current members of Prof. Kahn's Optical Communications Group: Milad Sharif, Anujit Shastri, Sercan Arik, Daulet Askarov, Ian Roberts, Ruo Yu Gu, Karthik Choutagunta, Michael Taylor, Brandon Buscaino, Elaine Chou, and Hrishikesh Srinivas. In particular, I would like to thank Milad Sharif and Anujit Shastri, who were my collaborators and greatly contributed to part of the work in this dissertation.

I am also very grateful for the many friends I have made during my time at Stanford. Their friendship and support made my life at Stanford much more enjoyable.

Last but certainly not least, I am also very grateful for my parents, Luiz Fernando and Elzina, and my brother Luiz Carlos. Their many selfless sacrifices allowed me to be where I am today. And despite the distance, they never stopped supporting and encouraging me. There are no words to describe my love and gratitude for them.

# Contents

# List of Tables

# List of Figures

# Chapter 1

# Introduction

One of the pillars supporting the information age is the ability to transmit ever-larger amounts of information across countries and continents. Optical communication systems have been remarkably successful in fulfilling that task. Over the past three decades, pivotal technologies such as erbium-doped fiber amplifiers (EDFAs), wavelength-division multiplexing (WDM), and coherent detection employing digital compensation of fiber impairments have enabled data transmission of tens of terabits per second across transoceanic distances.

Over the next few years, traffic demands are expected to continue growing at an accelerated pace. According to the Cisco forecast shown in Fig. 1.1, global Internet protocol (IP) traffic has approached 1 zettabyte ($10^{21}$ bytes) per year in 2016, and it is expected to grow at a compound annual rate of 24%, resulting in roughly a three-fold increase in traffic over fiver years.

Meeting this projected traffic growth will be particularly challenging at the shortest distances, as shifting computing paradigms have transformed how information is distributed, processed, and stored. For instance, web-based applications, content streaming, and cloud computing have turned personal computers and mobile devices into "mere" client interfaces, while most of the computing heavy lifting is realized remotely in large computing facilities known as data centers. As a result, overall traffic and growth rate is even larger in short-reach optical links within data centers. As shown in Fig. 1.2a, IP traffic is already above 10 zettabytes (greater than Fig. 1.1), and it is expected to grow at a compound annual rate of 25%, resulting in a three-fold increase over five years.

Fig. 1.2b illustrates the projected traffic distribution by destination in 2021. About 71% of the global data center IP traffic is expected to reside within data centers, while user-destined traffic will account for only 14%. The remaining 15% will be between data centers. This trend will be accentuated by machine learning applications, whereby the end user makes simple queries that, nevertheless, require significant computing power.

As traffic demands continue to soar globally and in particular in data centers, optical communication systems must continue to scale at an accelerated pace. However, energy constraints start to limit the bit rate that can be practically transmitted over those systems both at the shortest distances and at the longest distances [1]. At the shortest distances, optical systems for data centers face strict constraints on power consumption, size, and cost, factors that were usually secondary in designing high-performance optical systems. At the longest distances, submarine optical cables longer than about 5,000 km face energy constraints due to power feed limits at the shores, which restricts the electrical power available to the undersea optical amplifiers, ultimately limiting the optical power and throughput per fiber.

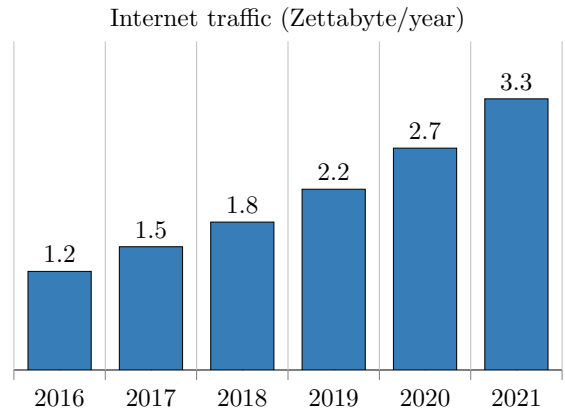

Figure 1.1: Global IP traffic forecast. A zettabyte equals $10^{21}$ bytes. Source: Cisco Global IP Traffic Growth, 2016–2021.

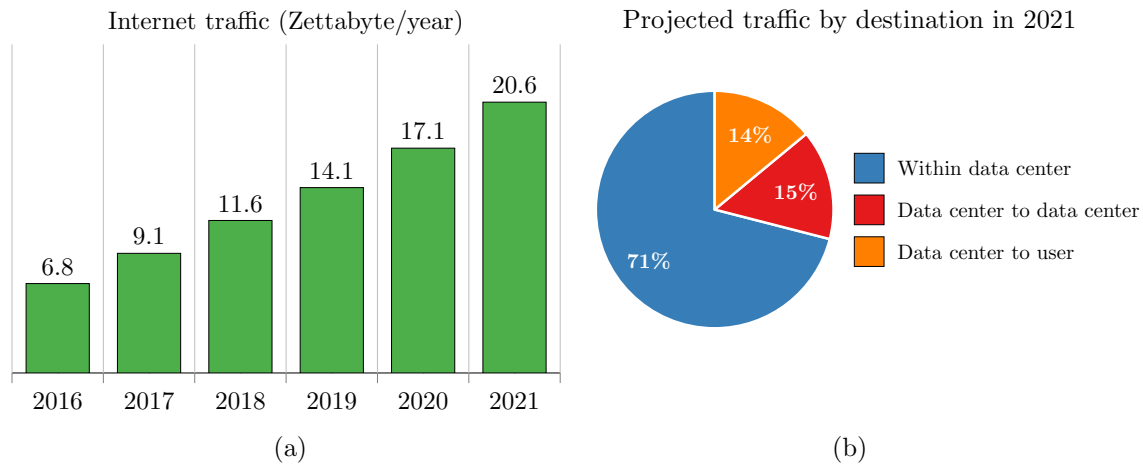

Figure 1.2: (a) Global IP traffic growth in data centers. (b) Projected traffic by destination in 2021. About 71% of all traffic is expected to reside within data centers. Source: Cisco Global Cloud Index, 2016–2021.

In this dissertation, we propose spectrally and power efficient optical systems for short-reach links in data centers and ultra-long links in submarine systems. Given the evident differences in those systems, different strategies are warranted. In data center applications, we propose low-power coherent detection systems that completely avoid high-speed analog-to-digital converters (ADC) and digital signal processors (DSP). At the longest distances, we optimize the channel power allocation to maximize the information-theoretic capacity per fiber under an electrical power constraint.

The subsequent subsections detail the problems faced in data centers and submarine systems and review part of the terminology and technicalities of each problem.

## 1.1 Data center links

Fig. 1.3a shows an exemplary data center, and Fig. 1.3b displays the interior of a large data center containing numerous rows of computer clusters. Hyper-scale data centers today can accommodate over 100,000 servers. These systems are typically interconnected following a two-tier topology [2], as illustrated in Fig. 1.3c. In this configuration all servers in a rack connect to top-of-the-rack switches that are connected to leaf switches, which in turn connect to every spine switch. In some cases, neighboring data centers may be interconnected by connecting their leaf switches.

The short links of a few hundred meters, shown in black in Fig. 1.3c, typically use vertical-cavity surface-emitting lasers (VCSEL) with multi-mode fiber (MMF) due to low manufacturing costs, low power consumption, and ease of coupling light into the fiber, while other links in the data center use single-mode fibers (SMF), which allow transmission over longer distances.

Throughout this dissertation, we will refer to intra-data center links as the SMF links reaching up to 10 km that connect different switches in a data center, shown in green in Fig. 1.3c. Inter-data center links reaching up to 100 km connect switches of neighboring data centers and are shown in blue in Fig. 1.3c.

In today's data centers these links are realized by multiplexing several wavelengths carrying conventional on/off keying (OOK) modulated signals. For instance, 100 Gbit/s links are achieved by multiplexing four wavelengths, each carrying 25 Gbit/s OOK signals. The IP traffic forecast shown in Fig. 1.1 suggests that data center links will need to scale to higher bit rates per wavelength. In fact, one of the industry goals was to develop transceivers capable of transmitting 100 Gbit/s per wavelength. However, in addition to higher throughput per wavelength, next-generation transceivers will likely need to tolerate higher fiber losses due to longer fiber plant, reduced power per channel in order to accommodate more wavelengths while complying with eye safety regulations, and possibly optical switches that will complement power-hungry electronic switches. Therefore, the challenge of next-generation optical systems for data center is to support higher bit rates per wavelength, while offering reasonable power margin.

To satisfy strict constraints in power consumption and cost, research focused initially on modulation formats compatible with direct detection, i.e., detection of information encoded in the optical intensity. However, Chapters 2 and 3 show that these direct-detected (DD) systems are extremely constrained beyond 100 Gbit/s, and in the long term, they likely cannot offer satisfactory power margin even when leveraging optical amplification or avalanche photodiodes. The underlying reason behind these challenges is that DD-compatible systems only leverage one degree of freedom of the optical channel, namely its intensity. Coherent detection enables four degrees of freedom of SMF, namely two quadratures in two polarizations, and improves noise tolerance by up to 20 dB by mixing a weak signal with a strong local oscillator. Nevertheless, commercial coherent transceivers today require high-speed ADCs and DSP, making them prohibitively power-hungry and costly for data centers applications. To address these challenges, in Chapter 4, we detail low-power DSP-free coherent and differentially coherent architectures that allow high-spectral efficiency and performance comparable to their DSP-based counterparts, while consuming much less power.

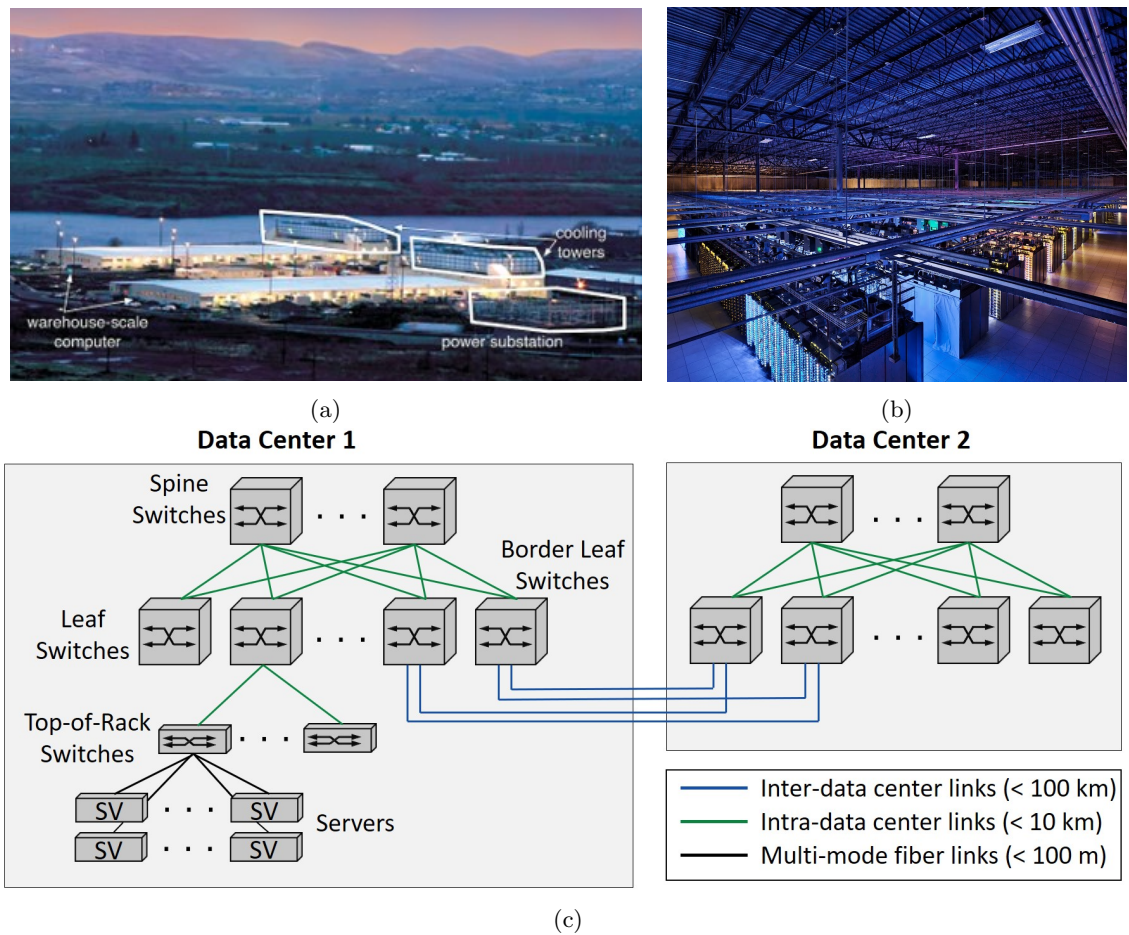

Figure 1.3: (a) hyperscale data center, (b) server racks inside a data center, and (c) typical two-tier topology allowing connectivity between neighboring data centers.

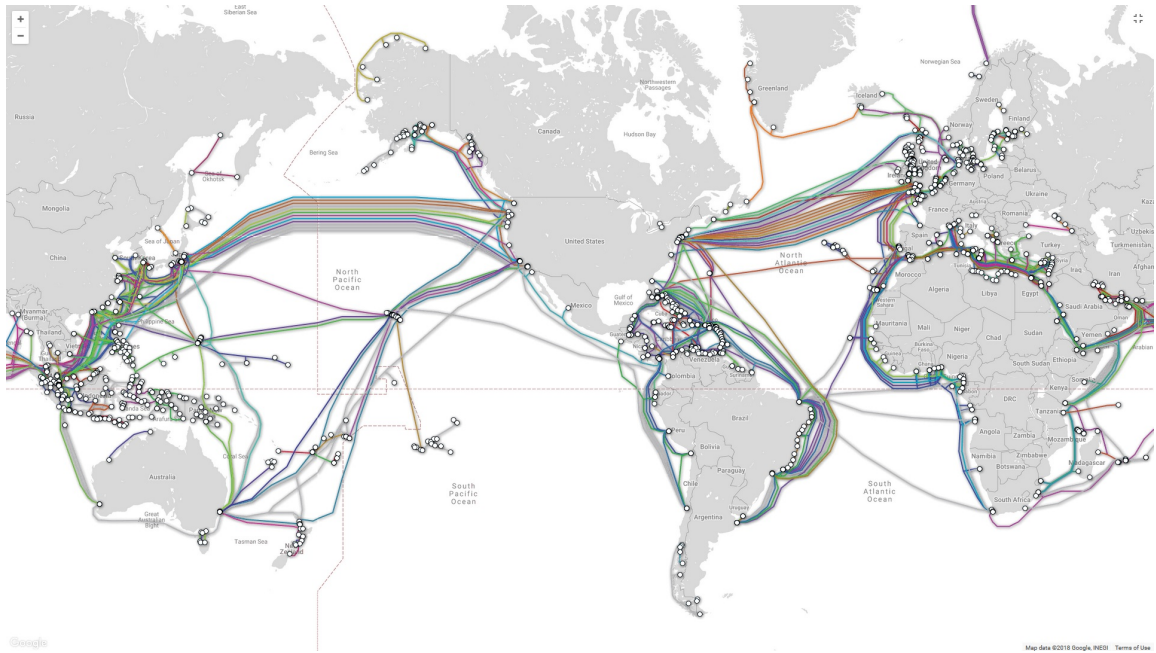

Figure 1.4: Map of deployed submarine cables. White nodes represent landing points, and cable color is to ease visualization. Source: www.submarinecablemap.com/.

## 1.2 Long-haul submarine links

As evidenced by a recent surge in deployment, submarine systems are of great and increasing importance to society and information technology. On October 7, 2017, The Economist reported that 100,000 km of submarine cable was laid in 2016, up from 16,000 km in 2015. This is consistent with the \$ 9.2-billion investment on submarine links between 2016 and 2018, five times as much as in the previous three years.

Fig. 1.4 shows a world map of deployed undersea optical communication cables. The white nodes represent cable landing points. Trans-atlantic cables connecting the United States (US) to Europe reach over 6,000 km, while trans-pacific cables connecting the US to Asia reach over 11,000 km, which is roughly the same length of cables connecting the US to the coast of Brazil.

The submerged cables and other equipment are reinforced to support the water pressure at the sea bed, and they are designed to operate uninterruptedly for 25 years. Fig. 1.5a shows a cross-section of a modern submarine cable. To compensate for the optical fiber attenuation of roughly 0.16 dB/km for state-of-the-art SMFs, optical repeaters are periodically positioned every $\sim 50$ km. A submarine-grade repeater is show in Fig. 1.5b. These repeaters are powered from the shores, where the dielectric constant of cables limits the feed voltages to 12–15 kV. Hence, the power available to each repeater is limited.

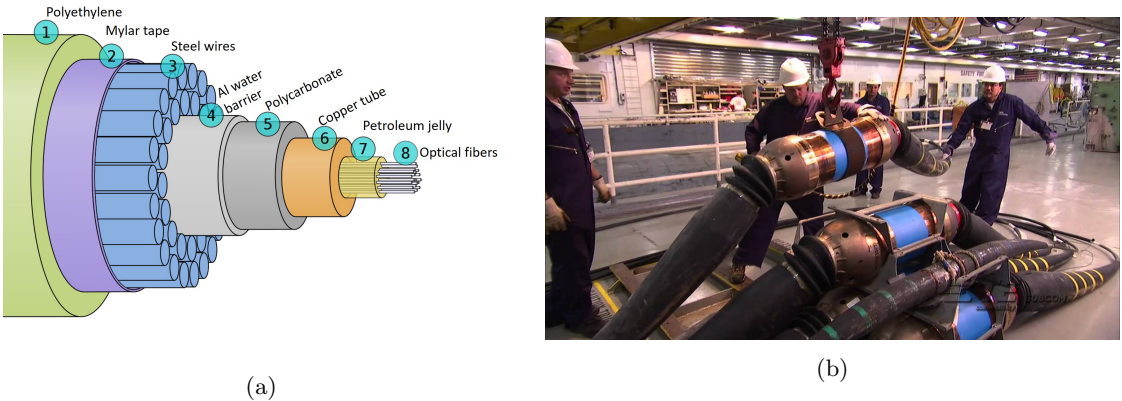

Figure 1.5: (a) Cross-section of a modern submarine cable, as described in US Patent No. 4,278,835 (Source: Wikimedia), and (b) TE SubCom submarine repeater, i.e., an EDFA for submarine optical links.

To illustrate the effect of this limitation, let us consider the example of a 10,000-km-long cable, which is typical of cables connecting the US to Asia or to Brazil. Assuming that amplifiers are positioned every 50 km results in a total of 200 amplifiers. All these amplifiers are powered from the shores where the feed voltage is 12 kV. For maximum electrical power transfer, the source resistance (cables) must be equal to the load resistance (repeaters). Therefore, the energy dissipated in the cables must be equal to the energy available to the repeaters. Assuming cable resistance of $\sim 1\Omega/\text{km}$, leads to 3,600 W of power for all repeaters, or equivalently 18 W for each of the 200 repeaters. Typically, 10% of the repeater power is spent in operations that do no contribute directly to optical amplification such as cooling and monitoring [3]. Therefore, 16.2 W is available for amplification. Assuming that the cable contains eight fiber pairs. The EDFA for each fiber would have roughly 1 W of electrical power. Unfortunately, EDFAs are not remarkably power efficient; typical electric-to-optical power conversion efficiency ranges from 1.5% to 5% [3, 4]. Thus, 5% efficiency results in approximately 17 dBm of available optical power at the output of each amplifier. As a result, from all the electrical power fed to the cable, only roughly 2% becomes useful optical power. This limit in optical power naturally poses a strict limit in the throughput per fiber.

To mitigate this problem, recent works have turned to an insight from Shannon's capacity that establishes that in energy-constrained systems, we can maximize capacity by employing more dimensions while transmitting less data (power) in each. In fact, recent works have studied how employing more spatial dimensions (modes, cores, or fibers) through spatial-division multiplexing (SDM) improves capacity and power efficiency of submarine systems [3, 5–7]. In complement to that work, we address the fundamental question of how to optimally use each spatial dimension under an energy constraint. Specifically, in Chapter 5, we demonstrate how to optimize the optical power of each WDM channel in order to maximize the information-theoretic capacity per spatial dimension given a constraint in the total electrical power. Our formulation accounts for amplifier physics, Kerr nonlinearity, and power feed constraints. Modeling amplifier physics is critical for translating energy constraints into parameters that govern the channel capacity such as amplification bandwidth, noise, and optical power. We show that the optimized channel power allocation almost doubles the capacity of submarine links compared to recently published works leveraging SDM. Our solutions also provide new insights on the optimal number of spatial dimensions, amplifier operation, and nonlinear regime operation.

In the second part of this dissertation, Chapter 5 formulates the problem of optimizing the power allocation under an energy constraint. Chapter 5 also details the models used for amplifier physics, Kerr nonlinearity, as well as the optimization algorithms used to solve the resulting non-convex problem.

# Part I

# Data Center Optical Systems

# Chapter 2

# Data Center Links Beyond On/Off Keying

Scaling the capacity of data center links has long relied on using multiple wavelengths or multiple fibers to carry conventional on-off keying (OOK) signals. Current 100 Gbit/s transceivers, for instance, use either ten multi-mode fibers (MMFs) each carrying 10 Gbit/s OOK signals, or four wavelengths of 25 Gbit/s OOK in one single-mode fiber (SMF), which is the case of the module shown in Fig. 2.1. This strategy cannot scale much further, however, as 400 Gbit/s links, for instance, would require 16 lanes of 25 Gbit/s, resulting in prohibitively high cost, size, and power consumption. Recent research has focused on spectrally efficient modulation formats compatible with direct detection (DD) [8–11] to enable 100 Gbit/s per wavelength. These "single-laser 100 G links" are intended to minimize optical component count, power consumption and size [12], and may facilitate optical switching in future data center networks.

Several modulation formats have been proposed to realize single-laser 100 G links, including pulse-amplitude modulation (PAM) [13, 14], carrierless amplitude-and-phase (CAP) [14, 15], quadrature amplitude modulation (QAM) [16], orthogonal multi-pulse modulation (OMM) [17], and orthogonal frequency-division multiplexing (OFDM), often referred to as discrete multi-tone (DMT) [14, 18]. All attempt to provide higher spectral efficiency than OOK, while offering similar complexity and power consumption.

In this chapter, we review and compare the most promising modulation formats to enable single-laser 100G links. As a complement to prior simulation-based studies, we derive analytical models to evaluate performance and complexity of different modulation formats, since analysis is more generally applicable and fosters insight into design optimization and the relative merits of the various schemes. In Section 2.1, we start by reviewing the main impairments of the optical fiber in short-reach links. In Section 2.2, we review important characteristics of intra- and inter-data center links. In Section 2.3, we discuss modulation formats compatible with direct detection focusing primarily on multicarrier formats based on OFDM. Single-carrier formats were studied in collaboration with Dr. Sharif in [8], and four-level PAM (4-PAM) was shown to outperform other single-carrier formats. Hence, we briefly review 4-PAM in Section 2.3.1. In Section 2.4, we compare these different modulation formats in terms of receiver sensitivity and required optical signal-to-noise ratio (OSNR) to achieve a target bit error rate (BER). In Section 2.5, we compare these different modulation formats in terms of system complexity and power consumption. Section 2.6 summarizes the main conclusions of this chapter.

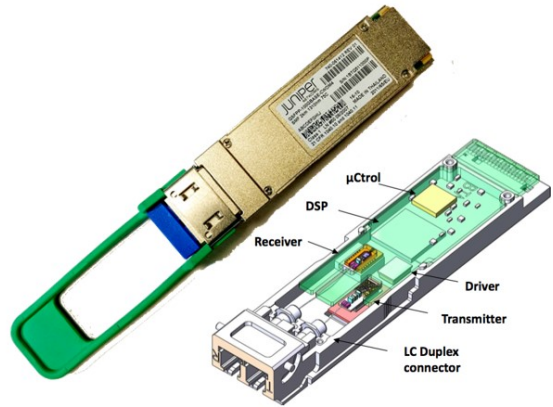

Figure 2.1: Example of 100 Gbit/s transceiver based on $4 \times 25$ Gbit/s for intra-data center links up to 2 km of SMF. The module size is 18.4 mm $\times$ 50 mm $\times$ 8.5 mm with power consumption of roughly 4 W. Images courtesy of Juniper Networks and Oclaro.

## 2.1　Optical fiber impairments

In short-reach links, the two primary impairments introduced by propagation over SMF is loss and chromatic dispersion (CD). Other phenomena such as polarization mode dispersion (PMD) and Kerr nonlinearity are generally negligible due to the short link length, and, in particular for Kerr nonlinearity, due to the relatively small optical power levels. Data center transceivers are designed to be eye safe and consequently the maximum power per fiber cannot exceed 14 dBm near 1310 nm or 17 dBm near 1550 nm [19].

Fig. 2.2 shows attenuation (top) and CD (bottom) coefficients in the two wavelength windows of interest: near 1310 nm, known as O-band, and near 1550 nm, known as C-band. Intra-data center links typically operate near 1310 (O-band) to minimize the amount of dispersion. Inter-data center links and long-haul communications generally operate near 1550 nm (C-band), since in that band standard SMF exhibits the smallest attenuation and that is the band of operation of erbium-doped fiber amplifiers (EDFAs).

The loss introduced by fiber attenuation only affects the total power margin of the system, and naturally must be accounted in the system power budget.

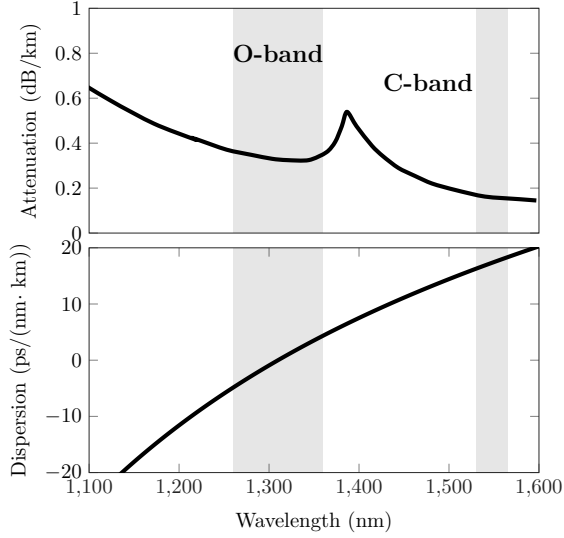

Figure 2.2: Attenuation (top) and dispersion (bottom) coefficients of standard SMF (SMF28).

CD, on the other hand, leads to power fading in intensity-modulated direct-detected (IM-DD) links, which ultimately limits the reach and the bit rate that can be practically transmitted over the fiber.

CD arises as signals at different frequencies propagate through the optical fiber with different velocities. Thus, CD can be modeled as a phase shift in the electric field:

$$\frac{E(f; z = L)}{E(f; z = 0)} = e^{-j\theta}, \qquad \theta = -0.5\beta_2(2\pi f)^2 z \tag{2.1}$$

where $E(f; z)$ is the Fourier transform of the electric field at distance $z$ along the fiber, and $\beta_2 = -(\lambda^2/2\pi c)D(\lambda)$, where $D(\lambda)$ is the dispersion parameter shown in the bottom plot of Fig. 2.2.

However, in DD systems the information is encoded in the optical signal intensity (instantaneous power) and not on the electric field. CD is not a linear operation in the intensity, and thus a simple transfer function between input and output optical power cannot be derived. In the small-signal (or small-dispersion) regime, we can derive an approximated transfer function given by [20]

$$H_{\text{IM-DD}}(f; z) = \frac{P(f; z = L)}{P(f; z = 0)} \approx \cos\theta - \alpha\sin\theta, \tag{2.2}$$

where $P(f; z)$ is the Fourier transform of the optical power signal at distance $z$, and $\alpha$ is the transient chirp parameter, which is not a property of the optical fiber. Chirp refers to the phenomenon of instantaneous variation of the optical carrier frequency upon intensity modulation. In optical communication systems, chirp is usually introduced by the optical modulator. In high-speed directly modulated lasers (DMLs) and electro-absorption modulators (EAMs), transient chirp is dominant [21] and arises due to the intimate relationship between real and imaginary refractive indexes dictated by causality and described by the Kramers-Kronig relations [22]. As a result, an intensity modulation of $P(t)$ is accompanied by a phase shift $\Delta\phi(t) = \frac{\alpha}{2} \ln P(t)$. In DMLs, the parameter $\alpha$ is always positive. In EAMs, the magnitude of $\alpha$ is typically smaller than in DMLs, but $\alpha$ can also be negative.

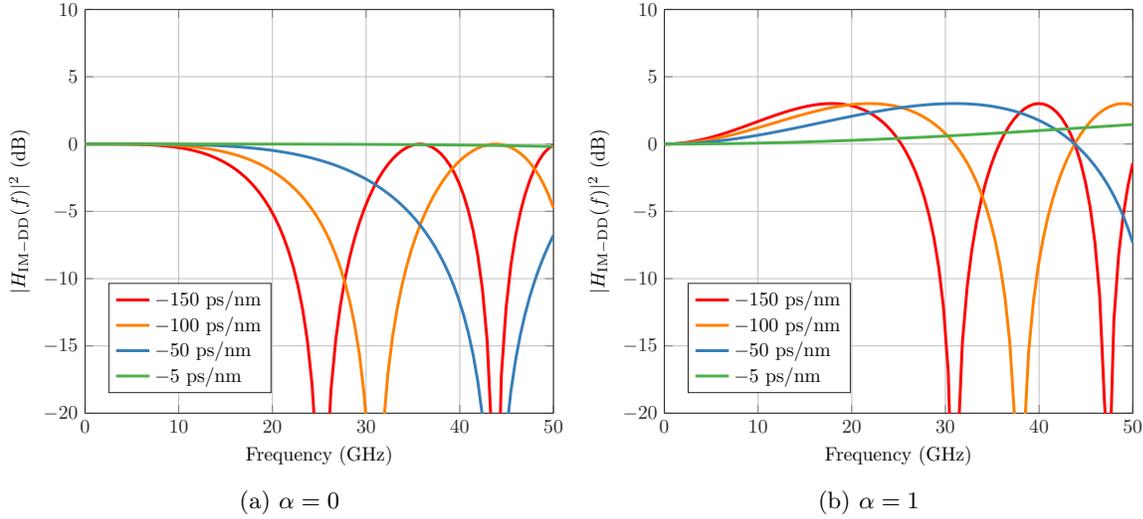

Figure 2.3: Small-signal fiber frequency response for (a) $\alpha = 0$ and (b) $\alpha = 1$.

Fig 2.3 plots $H_{\text{IM-DD}}(f; z)$ for several values of dispersion and for (a) $\alpha = 0$ and (b) $\alpha = 1$. Note that for $\theta$ small, if $D\alpha > 0$, the second term in (2.2) is positive and hence reduces the magnitude of the fiber frequency response at low frequencies. Conversely, if $D\alpha < 0$, the second term becomes negative, which causes the magnitude of the fiber frequency response at low frequencies to increase, i.e., dispersion provides some gain. Naturally, the second case is preferable, as the fiber frequency response compensates for the modulator bandwidth limitations and consequently reduces the power penalty. For this reason, if, $\alpha > 0$ we should use wavelengths shorter than the zero-dispersion wavelength so that $D < 0$. Hence, the combined effect of chirp and CD can have a positive effect on the signal by boosting the frequencies that are typically attenuated by bandwidth limitations of the optical modulator and transmitter electronics.

Nonetheless, the combined effect of CD and modulator chirp leads to power fading. Due to the periodicity of $H_{\text{IM-DD}}(f; z)$, the small-signal frequency response of the fiber is characterized by several notches. As dispersion increases the frequency of the first notch becomes smaller. Fig. 2.4 shows the frequency of the first notch of the IM-DD channel frequency response for several values of transient chirp parameter $\alpha$. To allow receiver-side linear equalization of single-carrier formats, the first notch cannot fall below half of the symbol rate; otherwise, the noise enhancement penalty becomes exceedingly high. Hence, for 56 Gbaud 4-PAM, the first notch cannot fall below 28 GHz.

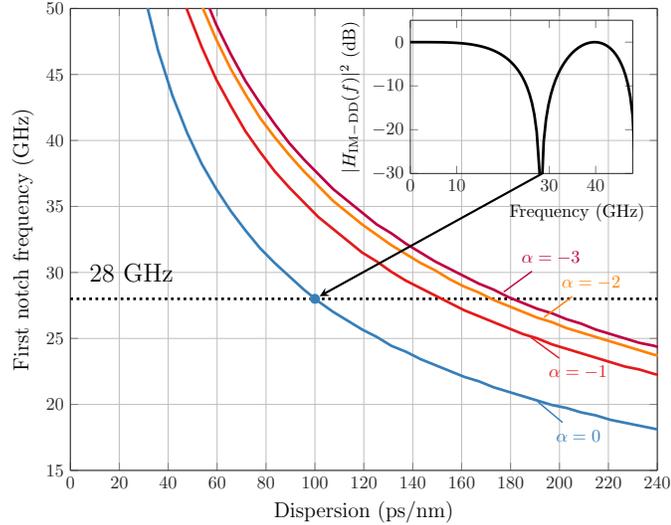

Figure 2.4: Frequency of first notch of IM-DD channel frequency response for several values of chirp parameter.

From Fig. 2.4, we can see that linear equalization is only effective up to about 100 ps/nm. Chirp increases the first notch frequency, but the maximum dispersion is still below 200 ps/nm. Some line coding techniques such as duobinary 4-PAM [23] and Tomlinson-Harashima [24] precoding can narrow the transmitted signal bandwidth, but even if the bandwidth is halved, the maximum tolerable dispersion is only on the order of 300 ps/nm.

Therefore, CD limits the tolerable dispersion to hundreds of ps/nm. In standard SMF, this corresponds to transmission distances of roughly 17 km near 1250 nm, and only 6 km near 1550 nm.

This strict limitation and the unique requirements of data center links may motivate reevaluation of optical fiber CD characteristics. When power consumption is the primary concern, fibers with small CD or optical CD compensation should be preferred, since electronic compensation will inevitably be more power hungry. For instance, dispersion shifted fibers (DSFs) with zero-dispersion wavelength near 1550 nm would allow small-dispersion systems that can leverage EDFAs. Note that nonlinear fiber effects, which can be exacerbated by DSF, are negligible in intra-data center links, since they are short (up to a few km) and operate with relatively small power levels due to eye safety constraints. The DSF CD slope near 1550 nm should be small in order to maximize the number of WDM channels supported. Dispersion-flattened optical fibers with zero-dispersion wavelengths near both 1310 nm and 1550 nm bands would allow operability of intra-data center links in both bands.

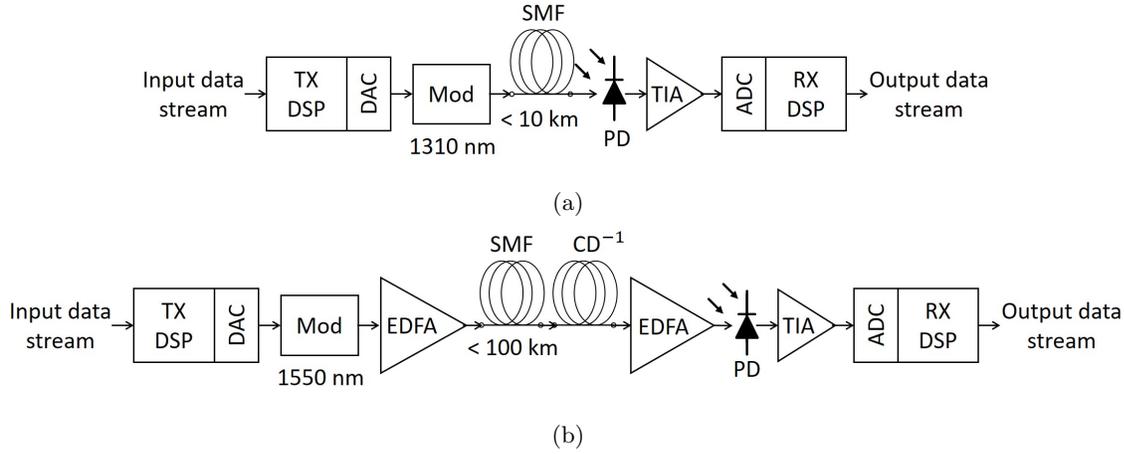

Figure 2.5: System-level diagrams of (a) intra-data center links and (b) inter-data center links.

## 2.2 Modeling intra- and inter-data center links

Fig. 2.5a shows the block diagram of a generic DD intra-data center link. The transmitter encodes the incoming bits into symbols and may perform some additional digital signal processing (DSP), which depends on the particular modulation format as discussed in Section 2.3. The analog signal generated by the digital-to-analog converter (DAC) drives an optical modulator, which in present intra-data center transceivers is typically a DML or an EAM. Future intra-data center transceivers will likely shift to Mach-Zenhder modulators (MZMs), which are already use in inter-data center transceivers due to negligible chirp, high bandwidth, and the ability to modulate both quadratures of the electric field. A thorough review of DMLs, EAMs, and MZMs are given in [25], [26], and [27], respectively.

Intra-data center links reach up to 10 km and typically operate near 1310 nm to minimize CD. Intra-data center links are typically unamplified, resulting in low power margin. In these unamplified links the dominant noise is thermal noise from the receiver electronics, in particular the trans-impedance amplifier (TIA). Typical high-speed TIAs have 3-dB bandwidth of 20–70 GHz and input-referred noise ($\bar{I}_n$) of 20–50 pA/$\sqrt{\text{Hz}}$ [28, Table 2], where $\bar{I}_n^2 = N_0$ is the one-sided power spectrum density (PSD) of thermal noise. Avalanche photodiodes (APDs) and semiconductor optical amplifiers (SOAs) may be used to improve the receiver sensitivity, and they are studied in detail in Chapter 3.

After analog-to-digital conversion (ADC), the receiver performs equalization to mitigate the intersymbol interference (ISI) introduced by bandwidth limitations of the components along the link. As discussed in Section 2.1, in short-reach links CD is accurately modeled by a linear filter, and thus receiver-side electronic equalization is effective to compensate for CD-induced distortion, as shown in the performance curves of Section 2.4.

Fig. 2.5b shows an example system model for an inter-data center link. Inter-data center links reach up to 100 km and operate near 1550 nm to leverage EDFAs. CD is significant and consequently simple receiver-side linear electronic equalization is not effective. Transmitter-side predistortion or self-coherence with an unmodulated carrier such as single-sideband modulation (Section 2.3.3) allow effective electronic CD compensation. Alternatively, CD may be compensated optically by dispersion-shifted fibers (DCFs) or tunable fiber Bragg gratings (FBGs) [29], depicted in Fig. 2.5b by the block $\text{CD}^{-1}$. Though optical CD compensation is less flexible than electronic equalization, it is more power-efficient, since in the optical domain CD compensation by DCFs of FBGs is a passive operation.

The optical amplifier introduces amplified spontaneous emission (ASE) noise whose one-sided PSD per real dimension is given by [30]

$$S_{\text{ASE}} = 1/2 \text{NF}(G_{\text{AMP}} - 1)h\nu, \tag{2.3}$$

where $G_{\text{AMP}}$ is the amplifier gain, $h\nu$ is the photon energy, and NF is the equivalent amplifier noise figure and depends on the number of amplifiers in the link as well as their individual noise figures. In the case of $N_A$ identical amplifiers each with noise figure $\text{NF}_1$, the equivalent noise figure is $\text{NF} = N_A \text{NF}_1$.

Direct detection causes mixing between signal and ASE, resulting in the signal-spontaneous beat noise, which is the dominant noise at the receiver. The signal-spontaneous beat noise one-sided PSD is given by

$$S_{\text{sig-spont}} = 4G_{\text{AMP}} R \bar{P}_{rx} S_{\text{ASE}}, \tag{2.4}$$

where $R$ is the receiver photodiode responsivity and $\bar{P}$ is the received average optical power.

## 2.3 Modulation formats compatible with direct detection

### 2.3.1 Pulse-amplitude modulation

PAM and other single-carrier techniques were studied by Sharif et al. in [8]. PAM was shown to outperform other single-carrier formats due to its relatively low complexity and high tolerance to modulator nonlinearities. This section reviews PAM and extends the analysis in [8] to inter-data center links where amplifier noise is dominant.

In $M$-PAM, the information is encoded in $M$ intensity levels. At the transmitter, the intensity modulator driving signal is generated by a $\log_2 M$-bit DAC. The transmitter may also realize other operations, such as pulse shaping and pre-equalization or preemphasis, but there are important considerations. Firstly, these operations require higher-resolution DACs, which at high sampling rates (> 50 GS/s) are power-hungry and have narrow bandwidths on the order of 10–15 GHz. Secondly, preemphasis increases the signal peak-to-average power ratio (PAPR), resulting in signals with high excursion, which requires components with high dynamic range in order to avoid distortion. Lastly, after pulse shaping and preemphasis filtering, a relatively large DC bias must be added to make the $M$-PAM signal non-negative, and thus compatible with intensity modulation. This DC bias directly affects the receiver sensitivity and it was shown to cause a 3-dB power penalty in 100 Gbit/s 4-PAM systems for intra-data center links [8].

At the receiver, the optical signal is direct detected, filtered, and converted to the digital domain where adaptive equalization is performed. The equalizer may be a simple feedforward equalizer (FFE) or a decision-feedback equalizer (DFE). Alternatively, the receiver may perform maximum likelihood sequence detection (MLSD). Provided that CD is small, the IM-DD channel is accurately modeled as a linear channel. In this regime, an FFE exhibited only a 1-dB penalty with respect to the optimal and more complex MLSD [8]. For large CD, the fiber IM-DD channel is no longer approximately linear, and FFE or DFE are less effective.

#### 2.3.1.1  Performance evaluation

The performance of an $M$-PAM system is determined by the noise variance at each intensity level. There are three scenarios of interest. The first consists of unamplified links in which the receiver uses a positive-intrinsic-negative (PIN) photodiode and thermal noise is dominant. In the next scenario, the receiver uses an avalanche photodiode (APD), which offers higher sensitivity, but shot noise becomes significant and will affect the noise variance at each level differently. APD-based receivers are discussed in detail in Chapter 3. Lastly, in amplified systems with either SOAs or EDFAs, the signal-amplified spontaneous emission (ASE) beat noise is dominant, resulting in different noise variances at the different intensity levels. Although the signal-ASE beat noise is not Gaussian, it can be approximated as Gaussian, as systems with forward error correction (FEC) operate at relatively high error rates. For each of these scenarios, we can compute the total noise variance at the $k$th intensity level:

$$\sigma_k^2 \approx \begin{cases} N_0 \Delta f, & \text{PIN photodiode} \\ 4 G_{\text{AMP}} R P_k S_{\text{ASE}} \Delta f, & \text{optically amplified} \end{cases} \tag{2.5}$$

where $\Delta f = |H_{rx}(0) H_{eq}(0)|^{-2} \int_0^\infty |H_{rx}(f) H_{eq}(f)|^2 df$ is the receiver one-sided noise bandwidth, where $H_{rx}(f)$ is receiver equivalent frequency response and $H_{eq}(f)$ is the equalizer's equivalent continuous-time frequency response. $N_0$ is the one-sided thermal noise PSD at the receiver, $R$ is the photodiode responsivity, $P_k$ is the optical power of the $k$th intensity level at the input of the PIN or the optical amplifier, and $G_{\text{AMP}}$ is the amplifier gain.

Assuming that all the noises involved are Gaussian distributed and uncorrelated, the BER is given by

$$\begin{aligned}\text{BER} \approx \frac{1}{M\log_2 M}\bigg[&Q\bigg(\frac{G_{eff}(d_1-P_0)}{\sigma_0}\bigg)\\&+\sum_{k=1}^{M-2}\bigg(Q\bigg(\frac{G_{eff}(P_k-d_k)}{\sigma_k}\bigg)+Q\bigg(\frac{G_{eff}(d_{k+1}-P_k)}{\sigma_k}\bigg)\bigg)\\&+Q\bigg(\frac{G_{eff}(P_{M-1}-d_{M-1})}{\sigma_{M-1}}\bigg)\bigg]\end{aligned} \quad (2.6)$$

where $Q(\cdot)$ is the well-known $Q$-function and $G_{eff}$ is the effective gain of the receiver; i.e., $G_{eff}=R$ for PIN-based receivers and $G_{eff}=RG_{\text{AMP}}$ for amplified systems. Equation (2.6) assumes that ISI is negligible or was compensated by FFE or DFE. In compensating for ISI, the equalizer causes the well-known phenomenon of noise enhancement, incurring a performance penalty. The effect of noise enhancement is accounted by the receiver noise bandwidth $\Delta f$ in (2.5), which would otherwise be $\Delta f=R_s/2$, where $R_s$ is the symbol rate.

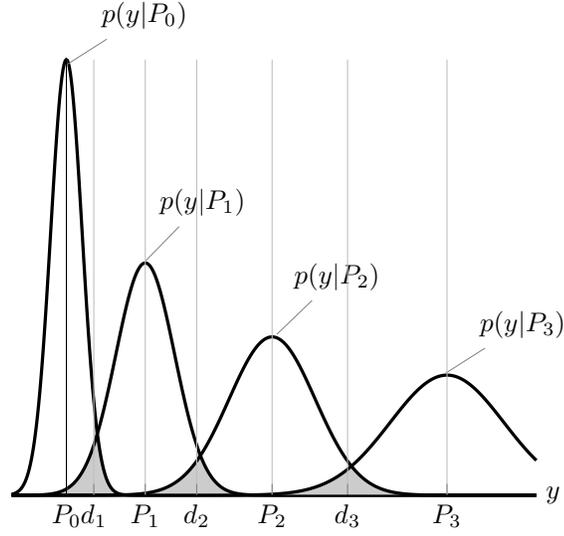

Figure 2.6: Example of optimized levels and their corresponding noise conditional probability density functions.

The intensity levels $\{P_0,\ldots,P_{M-1}\}$ and the decision thresholds $\{d_1,\ldots,d_{M-1}\}$ are typically equally spaced, but they can be appropriately optimized to minimize the BER. While the exact optimization is intractable, nearly optimal performance is achieved by setting the intensity levels sequentially according to the following heuristics [31]:

$$P_k = P_{k-1} + \frac{Q^{-1}(P_e)}{G_{eff}}(\sigma_k+\sigma_{k-1}) \quad (2.7)$$

where $\sigma_k^2$ is given by (2.5). Given $P_{k-1}$, we can determine $\sigma_{k-1}^2$ and solve for $P_k$ using (2.7). Following this procedure, all error events will have equal probability $P_e = \frac{\text{BER} \log_2 M}{2(M-1)}$.

This procedure may be realized in an iterative fashion to account for the modulator non-ideal extinction ratio $r_{ex}$. That is, ideally modulators would have minimum power $P_{min} = 0$. However, in practice the minimum power outputted by the modulator is limited by its extinction ratio such that $P_{min} = r_{ex}P_{max}$, where $P_{max}$ is the maximum power outputted by the modulator. Practical high-speed modulators exhibit $r_{ex}$ on the order of $-10$ to $-20$ dB. Returning to the level optimization procedure, at the first iteration, $P_0^{(0)} = 0$, and all other levels are calculated according to (2.7). At the $i$th iteration, $P_0^{(i)} = r_{ex} P_{M-1}^{(i-1)}$ [31]. We repeat this process until the required extinction ratio is achieved with reasonable accuracy. Fig. 2.6 shows optimized intensity levels with their respective conditional probability density functions (PDFs) of the noise. Each error event shown by the shaded areas has equal probability $P_e$. The decision thresholds are set at the midpoint of the intensity levels. Alternatively, the receiver could sweep the decision thresholds until the BER is minimized. This is equivalent to the point where the conditional PDF of neighboring levels intersect, which corresponds to the maximum likelihood decision. Even when the noise is not Gaussian, a similar level spacing optimization procedure based on the saddle point approximation can be applied to calculate the optimal intensity levels and decision thresholds [31].

For the unamplified systems, we characterize the performance in terms of the receiver sensitivity, defined as the average optical power $\bar{P}_{rx} = 1/M \sum_{k=1}^{M} P_k$ required to achieve a target BER, defined by the FEC code threshold. In amplified systems, it is more convenient to characterize the performance in terms of the required OSNR: $\text{OSNR}_{\text{req}} = \frac{G_{\text{AMP}}\bar{P}}{2S_{eq}B_{ref}}$, where $B_{ref}$ is the reference bandwidth for measuring the OSNR. $B_{ref}$ is typically 0.1 nm, corresponding to $B_{ref} \approx 12.5$ GHz near 1550 nm.

### 2.3.2 Orthogonal frequency-division multiplexing or discrete multitone

In OFDM, the information is encoded on narrowband and orthogonal subcarriers. In data center literature, OFDM is commonly referred to as discrete multitone (DMT), which is terminology borrowed from wireline communications literature, where DMT is often used to describe an OFDM signal transmitted at baseband.

OFDM, in principle, offers higher spectral efficiency than 4-PAM, since the individual subcarriers can be modulated using higher-order QAM. Two variants of OFDM were originally proposed for intensity-modulated data center links: DC-biased OFDM (DC-OFDM) and asymmetrically clipped optical (ACO)-OFDM. These OFDM variants differ in how they meet the non-negativity constraint of the intensity-modulated optical channel, and they achieve different tradeoffs between power efficiency and spectral efficiency. In DC-OFDM, a relatively high DC bias is added to minimize clipping distortion. By contrast, in ACO-OFDM, the entire negative excursion of the signal is clipped, and clipping distortion is avoided by encoding information only on the odd subcarriers [32].

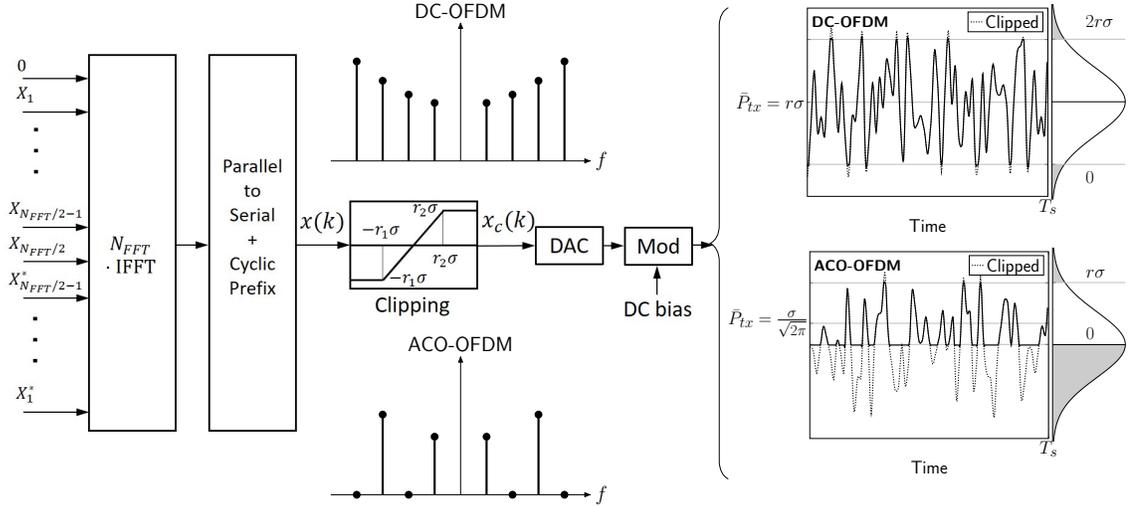

Figure 2.7: Block diagram of the OFDM transmitter for DC- and ACO-OFDM. Example time-domain waveforms are shown on the right.

Fig. 2.7 shows a general block diagram of an OFDM transmitter. A discrete-time OFDM symbol is generated by performing an $N_{\text{FFT}} \cdot \text{IFFT}(\cdot)$ operation, where the symbol transmitted on the $n$th subcarrier, $X_n$ is uniformly chosen from a $M_n$-QAM constellation with average power $P_n = \mathbb{E}(|X_n|^2)$. The constellation size $M_n$ and power $P_n$ are determined from a bit loading and power allocation algorithm, as discussed in Section 2.3.2.2.

To obtain a real-valued time-domain signal $x[k]$, $X_n$ must satisfy the Hermitian symmetry condition: $X_n = X^*_{N-n}$. For ACO-OFDM, we have the additional constraint $X_n = 0$, for $n$ even. That is, the even subcarriers are not modulated, as illustrated in Fig. 2.7. This condition ensures that clipping distortion does not fall on the data-bearing odd subcarriers.

By the central limit theorem, for an IFFT length $N_{\text{FFT}}$ sufficiently large, the OFDM signal is approximately Gaussian-distributed with zero mean and variance

$$\sigma^2 = \mathbb{E}(|x[k]|^2) = 2 \sum_{n=1}^{N_{\text{FFT}}/2-1} P_n. \tag{2.8}$$

After parallel-to-serial conversion and cyclic prefix insertion, the discrete-time OFDM signal $x[k]$ is clipped at levels $-r_1\sigma$ and $r_2\sigma$ to reduce the required dynamic range of the DAC and subsequent components:

$$x_c[k] = \begin{cases} -r_1\sigma, & x[k] \leq -r_1\sigma \\ x[k], & -r_1\sigma < x[k] < r_2\sigma \\ r_2\sigma, & x[k] \geq r_2\sigma \end{cases}, \quad (2.9)$$

where $r_1 = r_2 = r$ for DC-OFDM; $r_1 = 0$, and $r_2 = r$ for ACO-OFDM. The parameters $r_1$ and $r_2$ are referred to as clipping ratios. This definition allows us to easily calculate the clipping probability: $P_c = Q(r_1) + Q(r_2)$, where $Q(\cdot)$ is the Q-function for the tail probability of a Gaussian distribution. Note that a clipping event does not necessarily result in a bit error event.

In DC-OFDM, the clipping ratio $r_1 = r_2 = r$ determines the tradeoff between clipping distortion and quantization noise, as discussed in Section 2.3.2.3. In ACO-OFDM, $r_1 = 0$ and $r_2 = r$. The distortion caused by clipping the entire negative excursion only falls onto the even subcarriers, which purposely do not carry data [32].

The clipped OFDM signal $x_c[k]$ is converted to the analog domain by the DAC and an appropriate DC bias is added to make the signal non-negative. Fig. 2.7 shows example time-domain waveforms of DC-OFDM and ACO-OFDM indicating the different clipping strategies. The average optical power $\bar{P}_{tx}$ for each OFDM variant is given by

$$\bar{P}_{tx} = \begin{cases} r\sigma, & \text{DC-OFDM} \\ \frac{\sigma}{\sqrt{2\pi}}, & \text{ACO-OFDM} \end{cases}, \quad (2.10)$$

where for ACO-OFDM, $\bar{P}_{tx}$ follows directly from calculating the mean value of the clipped Gaussian distribution and assuming $r$ large [32]. Equation (2.10) clearly indicates the average-power advantage of ACO-OFDM over DC-OFDM, as generally $r > \sqrt{2\pi}$.

### 2.3.2.1 Performance evaluation

The performance of the OFDM signal depends on the received SNR of each data-bearing subcarrier. Assuming that the noises involved are white and consequently equal in all subcarriers, we can write the noise variance at the $n$th subcarrier for the same noise scenarios as in Section 2.3.1:

$$\sigma_n^2 = \begin{cases} f_s \frac{N_0}{2}, & \text{PIN photodiode} \\ f_s(2G_{\text{AMP}}R\bar{P}_{rx}S_{ASE}), & \text{optically amplified} \end{cases}, \quad (2.11)$$

where $\bar{P}_{rx}$ is the average optical power at the receiver input; i.e., the input of the PIN photodiode, or the optical amplifier. Moreover,

$$f_s = \frac{2pR_b}{\log_2 M} \frac{N_{\text{FFT}} + N_{\text{CP}}}{N_{\text{FFT}}} r_{os} \quad (2.12)$$

is the sampling rate of the OFDM signal, where $p = 1$ or 2 for DC-OFDM or ACO-OFDM, respectively, accounts for the loss in spectral efficiency by not modulating the even subcarriers. Here, $M$ is the nominal constellation size, $R_b$ is the bit rate, $N_{\text{CP}}$ is the cyclic prefix length and should be larger than the channel memory length, $r_{os} = N_{\text{FFT}}/(pN_u)$ is the oversampling ratio of the OFDM signal, where $N_u$ is the number of subcarriers used to transmit data.

After DD, the SNR at the $n$th subcarrier is given by

$$\text{SNR}_n = \frac{N_{\text{FFT}} G_{eff} P_{n,rx}}{\sigma_n^2 + \sigma_Q^2} \tag{2.13}$$

where $P_{n,rx}$ is the power of the $n$th subcarrier referred to the receiver input; i.e., to the input of the PIN photodiode, APD, or optical amplifier. Note that (2.13) could be easily modified to include any receiver-side bandwidth limitation by accounting for how signal and noise power are attenuated by the receiver frequency response at each subcarrier. As OFDM usually requires high-resolution DAC and ADC, quantization noise must be included. Computation of quantization noise variance $\sigma_Q^2$ is detailed in Section 2.3.2.3.

The BER is given by the average of the bit error probability in each subcarrier weighted by the number of bits in each subcarrier:

$$\text{BER} = \frac{\sum_{n=1}^{N_{\text{FFT}}/2-1} \log_2(M_n) \cdot P_{QAM}(\text{SNR}_n; M_n)}{\sum_{n=1}^{N_{\text{FFT}}/2-1} \log_2(M_n)} \tag{2.14}$$

where $P_{QAM}(\text{SNR}_n; M_n)$ gives the bit error probability for an uncoded $M$-QAM constellation in an additive white Gaussian noise channel with a given SNR. There are analytical expressions for $P_{QAM}(\text{SNR}_n; M_n)$ for square and non-square QAM constellations [33].

### 2.3.2.2 Power allocation and bit loading

The non-flat frequency response of the channel causes some subcarriers to be attenuated more than others. Thus, to use all subcarriers effectively, we must perform power allocation, bit loading, or a combination of the two. We consider two alternatives: (i) constant bit loading and preemphasis (channel inversion), and (ii) optimized bit loading and power allocation.

In the preemphasis or channel inversion approach all subcarriers have the same constellation size $M$, but their power is inversely proportional to the channel gain at their corresponding frequencies: $P_n \propto |G_{ch}(f_n)|^{-2}$, where $G_{ch}(f_n)$ is simply the frequency response of the channel at the $n$th subcarrier. As a result, at the receiver, all subcarriers have the same power and SNR, provided the noise PSD is constant over the signal band.

In the optimized bit loading and power allocation method, the constellation size of each subcarrier is determined by solving the margin maximization problem [34]. In this optimization problem, we minimize the total power subject to a bit rate constraint. Formally,

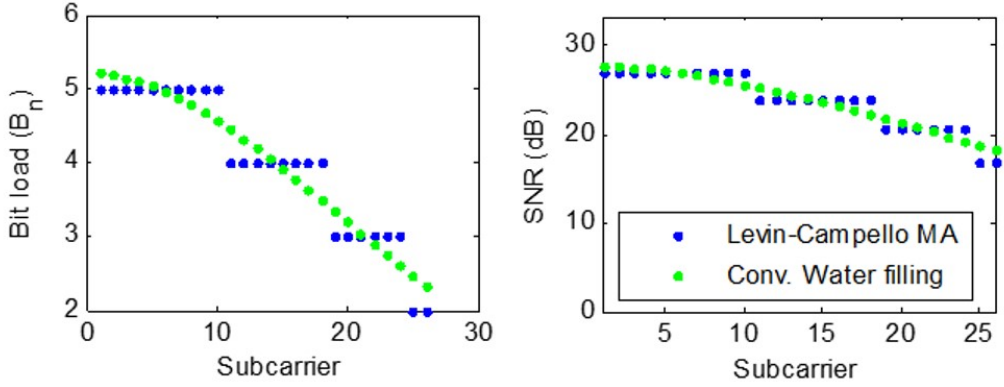

Figure 2.8: Comparison between bit loading (left) and power allocation (right) done by the Levin-Campello and conventional water filling algorithms. $B_n$ refers to the number of bits in each subcarrier. Hence, the constellation size is $2^{B_n}$.

$$\min_{P_n} \sigma^2 = 2 \sum_{n=1}^{N_{\text{FFT}}/2-1} P_n$$
$$\text{subject to } b = \sum_{n=1}^{N_{\text{FFT}}/2-1} \log_2\left(1 + \Gamma P_n \text{GNR}_n\right). \tag{2.15}$$

Here, $0 < \Gamma \leq 1$ is a coding gap, which represents the SNR penalty for using a suboptimal and practical coding scheme instead of a capacity-achieving coding scheme. $\text{GNR}_n$ is defined as the channel gain-to-noise ratio at the $n$th subcarrier. Note that $\text{GNR}_n$ is related to the SNR at the $n$th subcarrier by $\text{SNR}_n = P_n \text{GNR}_n$. The solution to the optimization problem in (2.15) minimizes the average optical power, since $\bar{P}_{tx} \propto \sigma = \sqrt{P_t}$, as in (2.10).

The optimization problem (2.15) can be solved via Lagrange multipliers, resulting in the conventional water-filling solution. However, in practice, we employ the Levin-Campello (LC) algorithm [35] to obtain constellations with integer numbers of bits. Fig. 2.8 shows a comparison between LC and conventional water filling algorithms. Roughly speaking, the LC algorithm transfers bits from bad (more attenuated) subcarriers to good subcarriers, so that bad subcarriers can achieve the target BER at smaller SNRs, and thus requiring less power than in the preemphasis method. Implementation of the LC algorithm is described in [34]. This algorithm has two stages. In the first stage, an arbitrary bit distribution is made efficient. Efficiency in this context means that there is no movement of a bit from one subcarrier to another that can reduce the signal power. The next stage is the so-called B-tightening stage, where the number of bits in appropriate subcarriers is increased

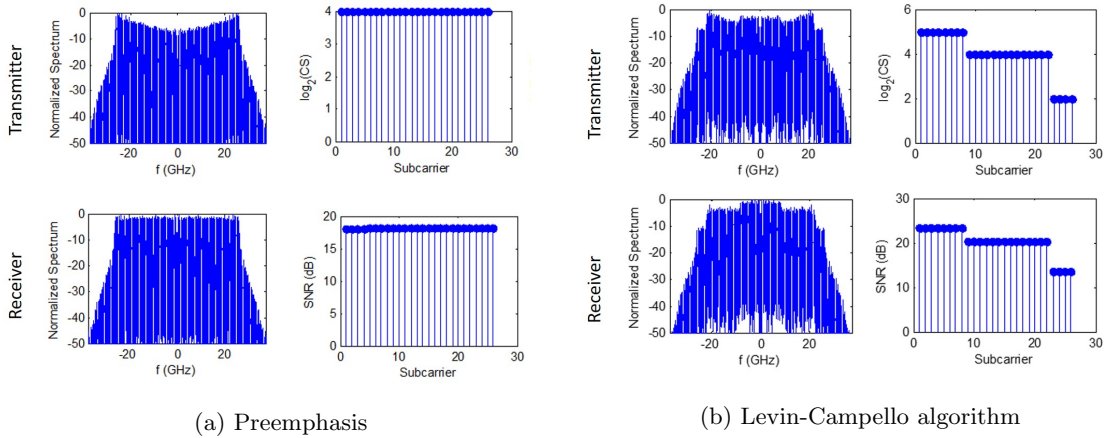

(a) Preemphasis

(b) Levin-Campello algorithm

Figure 2.9: Comparison between (a) preempahsis and (b) Levin-Campello algorithm for power allocation and bit loading. Figures show power spectrum (left) and bit loading (right) at the transmitter (top) and receiver (bottom). The acronym CS stands for the QAM constellation size.

or reduced to ensure that the constraint in the bit rate is met.

Fig. 2.9 shows a comparison between preemphasis and LC algorithm. Note that preemphasis uses the same bit loading for all subcarriers and consequently the power of outer subcarriers must be increased to compensate for the channel attenuation. On the other hand, the LC algorithm allocates fewer bits on the more-attenuated subcarriers, which allows them to achieve target BER using less power.

#### 2.3.2.3 Clipping versus quantization trade-off

OFDM is characterized by high peak-to-average power ratio (PAPR) and noise-like time-domain waveforms. As a result DACs and ADCs for OFDM systems must have high dynamic range in order to minimize clipping, and they must have high effective resolution in order to minimize quantization noise. These conflicting requirements lead to a trade-off between clipping distortion and quantization noise. As the effective resolution of DACs/ADCs is limited at roughly 6 bits for sampling rates higher than 30 GS/s, it is necessary to properly optimize clipping and quantization. Studying clipping and quantization allows us to derive the optimal clipping ratio, required effective resolution, and effect of quantization on SNR.

**Clipping distortion**

Clipping is necessary to reduce the required dynamic range of DAC/ADC and other components. Here, we extend the theory derived in [32] for ACO-OFDM to encompass both DC- and ACO-OFDM with two clipping levels. Assuming $x[k] \sim \mathcal{N}(0, \sigma^2)$, we can apply Bussgang's theorem [36], and (2.9) can be written as

$$x_c[k] = Kx[k] + d[k], \tag{2.16}$$

where $d[k]$ is a random process that is uncorrelated with $x[k]$, i.e., $\mathbb{E}(x[k]d[k]) = 0$. Here, $K$ is a constant that depends only on the nonlinear amplitude distortion [36], which is clipping in this case. It can be shown that

$$K = 1 - Q(r_1) - Q(r_2). \tag{2.17}$$

Note that for $r_1 = 0$ and $r_2 \to \infty$ (i.e., ACO-OFDM with clipping only at the zero level), $K = 1/2$, as previously shown in [32]. For ACO-OFDM, it can be further shown that $d[k]$ only has frequency components on the even subcarriers, which intentionally do not carry data [32].

For DC-OFDM, $d[k]$ does cause distortion on the data-bearing subcarriers. The variance of $d[k]$ is given by

$$\mathrm{Var}(d[k]) = \mathrm{Var}(x_c[k]) - K^2\sigma^2 \tag{2.18}$$

where $\mathrm{Var}(\cdot)$ is a function of $r_1$, $r_2$, and $\sigma^2$, which can be obtained from the distribution of $x_c[k]$, i.e., a Gaussian distribution clipped at $-r_1\sigma$ and $r_2\sigma$.

**Quantization**

Quantization noise is typically modeled as an additive, uniformly distributed white noise, whose variance is given by

$$\sigma_Q = (1 - P_c)\frac{\Delta X}{12 \cdot 2^{2\mathrm{ENOB}}}, \tag{2.19}$$

where $\Delta X$ denotes the dynamic range of the quantizer, and ENOB is the effective number of bits of the quantizer. Practical quantizers introduce noise and distortion, which effectively lowers their resolution. ENOB specifies the resolution of an ideal quantizer that obtains the same resolution of a practical quantizer subject to noise and distortion. Note that the clipping probability reduces the quantization noise variance, since at the clipped levels there is no error due to quantization, provided they are also quantization levels.

The dynamic range of the quantizer depends on the input signal statistics. At the transmitter, the input signal is the clipped OFDM signal. Therefore, the quantization noise variance at the transmitter is given by

$$\sigma_{Q,tx} = \begin{cases} (1 - P_c)\frac{r_{tx}^2\sigma^2}{3 \cdot 2^{2\mathrm{ENOB}}}, & \text{DC-OFDM} \\ (1 - P_c)\frac{r_{tx}^2\sigma^2}{12 \cdot 2^{2\mathrm{ENOB}}}, & \text{ACO-OFDM} \end{cases}. \tag{2.20}$$

For a given transmitter clipping ratio, the signal excursion of DC-OFDM is twice the signal excursion of ACO-OFDM. As a result, quantization noise variance for DC-OFDM is four times greater. Moreover, assuming negligible clipping distortion at data-bearing subcarriers, we have $P_c \approx 0$ for DC-OFDM, and $P_c \approx 1/2$ for ACO-OFDM, which further reduces the quantization noise in ACO-OFDM relative to DC-OFDM.

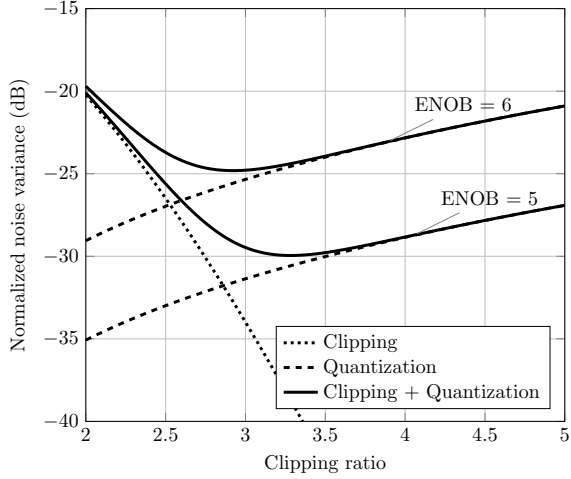

Figure 2.10: Clipping and quantization noise variance normalized by the signal power $\sigma^2$ as a function of clipping ratio for DC-OFDM.

At the receiver, the signal has undergone linear filtering by the channel with overall frequency response $G_{ch}(f)$. A DC-OFDM signal can still be considered Gaussian-distributed with variance

$$\sigma_{rx}^2 = 2 \sum_{n=1}^{N_{\text{FFT}/2-1}} P_n |G_{ch}(f_n)|^2. \tag{2.21}$$

Thus the dynamic range of the quantization for DC-OFDM at the receiver is given by $\Delta X_{rx} = 2r_{rx}\sigma_{rx}$, where $r_{rx}$ is the clipping ratio at the receiver.

ACO-OFDM, on the other hand, is highly asymmetric. As an approximation, we can consider the received ACO-OFDM signal as non-negative with mean $\sigma/\sqrt{2\pi}$ (assuming all filters have unit DC gain), with the positive tail approximated by a Gaussian of variance $\sigma_{rx}^2$. For ACO-OFDM the sum in (2.21) is over the odd subcarriers only. Thus the dynamic range of the quantizer for ACO-OFDM at the receiver is given by $\Delta X_{rx} = \sigma/\sqrt{2\pi} + r_{rx}\sigma_{rx}$. This approximation is not ultimately important, as we optimize the clipping ratio both at the transmitter and at the receiver to minimize the power penalty. It is just a convenient way to express the clipping and quantization levels in terms of the signal power. This facilitates the analysis of clipping and quantization noises, as well as the required ENOB.

Hence, the quantization noise variance at the receiver is given by

$$\sigma_{Q,rx} = \begin{cases} \frac{r_{rx}^2 \sigma_{rx}^2}{3 \cdot 2^{2\text{ENOB}}}, & \text{DC-OFDM} \\ \frac{\left(\sigma/\sqrt{2\pi} + r_{rx}\sigma_{rx}\right)^2}{12 \cdot 2^{2\text{ENOB}}}, & \text{ACO-OFDM} \end{cases} \quad (2.22)$$

**Optimal clipping ratio**

Note that the clipping noise variance (2.18) and the quantization noise variance (2.20), (2.22) depend on the clipping ratio $r$. Clipping noise decreases as $r$ increases, and quantization noise does the opposite.

Fig. 2.10 shows clipping and quantization noise variances normalized by the signal power $\sigma^2$ as a function of the clipping ratio for DC-OFDM. We focus on DC-OFDM, since the clipping ratio directly affects the required DC bias and consequently the overall power penalty.

There is a clear tradeoff between clipping and quantization noises. Although the minimum total noise is achieved around $r = 2.8$ for ENOB = 5, and $r = 3.8$ for ENOB = 6, we must choose the clipping ratio so as to make clipping noise negligible compared to quantization noise. This is because clipping noise has several undesired characteristics, such as non-white power spectrum, whereas quantization noise can be accurately modeled as a bounded uniform white noise. Indeed, minimum optical power is achieved for clipping ratios in the range of 3.7 to 4.5, where clipping noise becomes negligible, as can be seen in Fig. 2.10.

**Required DAC/ADC resolution**

Assuming that all subcarriers have the same power and bit loading, and considering the limit when quantization noise becomes dominant, equation (2.13) reduces to

$$\text{SNR}_n = \frac{K^2 N P_u}{\sigma_{Q,tx}^2 + \sigma_{Q,rx}^2} \quad (2.23)$$

where $\sigma_{Q,tx}^2$ and $\sigma_{Q,rx}^2$ are given by (2.20) and (2.22), respectively. Note that although quantization noise is uniformly distributed, after the FFT operation at the OFDM receiver, the noise is approximately Gaussian distributed by to the central limit theorem. Note also that $\sigma_{Q,tx}^2$ and $\sigma_{Q,rx}^2$ are proportional to the signal power, and that in the case of equal bit loading and power allocation we have $\sigma^2 = N_u P_n$. Thus, $\text{SNR}_n$ has a ceiling in the quantization-noise limited regime. This can be verified by plotting the SNR as a function fo the received power 16- and 64-QAM DC-OFDM, as shown in Fig. 2.11. For infinite DAC/ADC resolution, in the thermal-noise limited regime, the SNR increases linearly with the received power. After a certain threshold the SNR increases sub-linearly with, until it reaches a ceiling due the laser intensity noise. When ADC noise is included, the SNR ceiling is smaller and is reached at lower power than in the intensity-noise limited regime. This indicates that, at high SNR, quantization noise is the limiting noise for the performance of OFDM signals. Thus, neglecting intensity noise and shot noise, as done in (2.13), should not cause significant error.

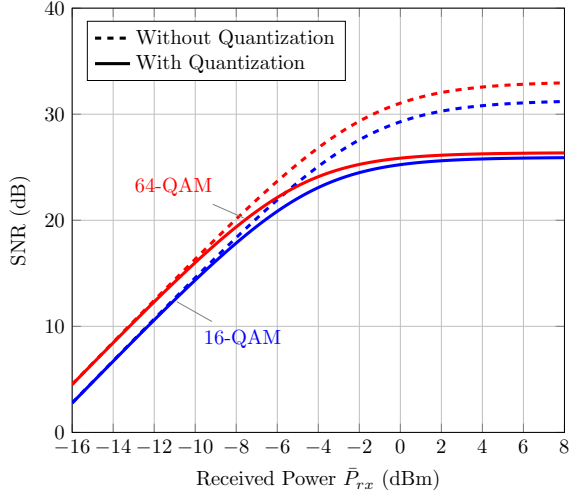

Figure 2.11: SNR as a function of the received power including and disregarding quantization noise. These curves were obtained for ENOB = 6, and other parameters as given in Table 2.1.

We can solve (2.23) for the ENOB as a function of $\text{SNR}_{req}$ that leads to the target BER:

$$\text{ENOB}_{req} = \begin{cases} \frac{1}{2} \log_2 \left( \frac{2r^2}{3r_{os}} \text{SNR}_{req} \right), & \text{DC-OFDM} \\ \frac{1}{2} \log_2 \left( \frac{r^2 + 2(1/\sqrt{2\pi}+r)^2}{12 r_{os}} \text{SNR}_{req} \right), & \text{ACO-OFDM} \end{cases} \quad (2.24)$$

This value of ENOB is actually a lower bound, as we have neglected thermal noise and filtering; however, it is useful to provide a first estimate of the required resolution for DC- and ACO-OFDM, allowing $\text{SNR}_{req}$ to be calculated based on the target BER and the nominal constellation size of the OFDM signal.

Fig. 2.12 shows the required ENOB for DC- and ACO-OFDM with 16-QAM and 64-QAM constellation as a function of the clipping ratio. ACO-OFDM requires fewer bits since the signal excursion is half of the DC-OFDM. However, the difference does not go up to 1 bit as one might expect because with ACO-OFDM, clipping reduces the signal power by 1/4, since $K \approx 1/2$. This result shows that ENOB must be at least 5 for 16-QAM, and at least 6 for 64-QAM. This result agrees well with the rule of thumb $\text{ENOB}_{req} \approx \log_2(\sqrt{M}) + 3$ for the resolution required of ADC to detect filtered single-carrier QAM signals [37].

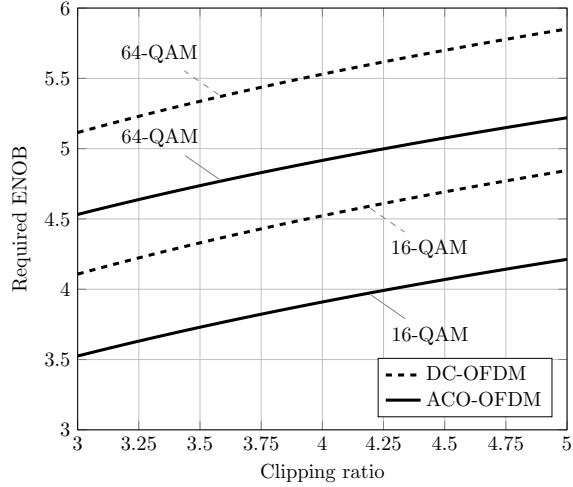

Figure 2.12: Required ENOB to achieve target BER of $1.8 \times 10^{-4}$ for DC-OFDM (dashed lines) and ACO-OFDM (solid lines) with 16- and 64-QAM nominal constellation sizes.

### 2.3.3 Single-sideband orthogonal frequency-division multiplexing

In SSB-OFDM, the subcarriers corresponding to the negative-frequency sideband are not modulated. The SSB-OFDM signal can still be directly detected, provided that a sufficiently strong unmodulated optical carrier is also transmitted. After DD, the mixing of the unmodulated carrier and the SSB-OFDM signal yields a real-valued double-sideband (DSB)-OFDM signal carrying the same information as the original SSB-OFDM signal. This DSB-OFDM signal does not experience the power fading characteristic of the IM-DD channel shown in Fig. 2.3. In fact, the DSB-OFDM signal only experiences phase distortion, which can be effectively compensated by electronic equalization.

The negative sideband of an intensity-modulated OFDM signal can be suppressed electronically, as indicated in the diagram of Fig. 2.13, or using an optical bandpass filter, resulting in a format known as vestigial-sideband (VSB) OFDM. The transmitter laser and the optical filter must have fine wavelength stabilization in order to ensure filtering of the correct signal band. SSB modulation has generally better performance than VSB modulation [38], hence we restrict our attention to SSB-OFDM.

Fig. 2.13 shows the block diagram of a SSB-OFDM transmitter. The negative sideband subcarriers are set to zero, and the resulting complex time-domain signal $x[k]$ may be written in terms of a real-valued DSB-OFDM signal $s[k]$:

$$x[k] = x[k] + j\mathcal{H}\{s[k]\}, \tag{2.25}$$

where $\mathcal{H}\{\cdot\}$ denotes the Hilbert transform.

After clipping, and digital-to-analog conversion, the resulting signals drive an dual-quadrature

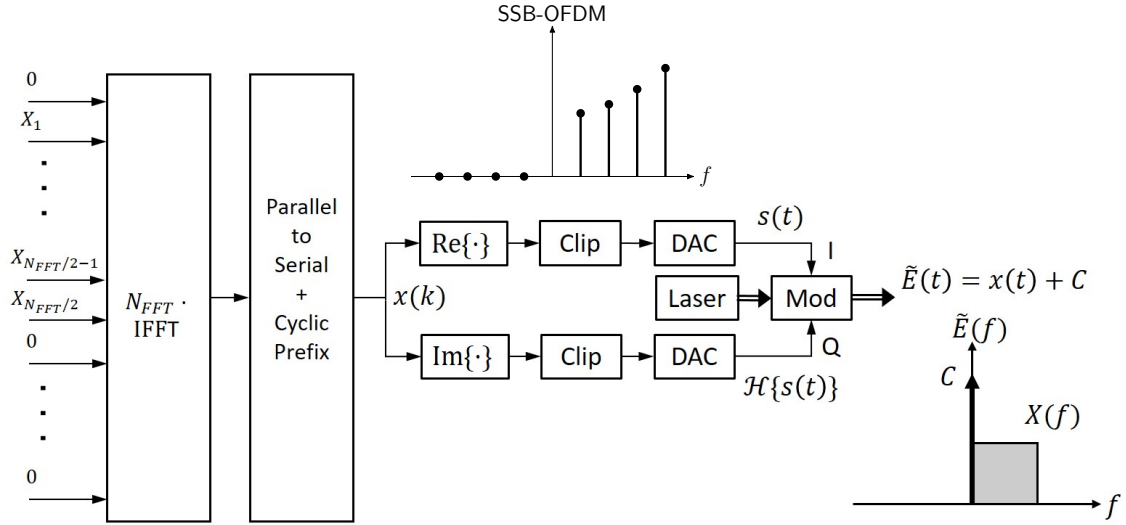

Figure 2.13: Block diagram of SSB-OFDM transmitter. Output electric field consists of a SSB-OFDM signal plus a strong unmodulated carrier.

(I&Q) modulator. The output electric field contains the SSB-OFDM signal $x(t)$ and a carrier component $C$. The carrier-to-signal power ratio (CSPR), defined as $\text{CSPR} = P_s/P_c = \frac{1}{|C|^2} \sum_{n=1}^{N_{\text{FFT}}/2-1} P_n$, affects the system performance. The signal propagates through the fiber, whose complex impulse response due to CD is $h_{CD}(t)$. The received signal $y(t)$, after DD is given by

$$y(t) \approx 2RG_{\text{AMP}}\sqrt{P_C}s(t) * g(t) + 2R\sqrt{G_{\text{AMP}}(P_c + P_s)}n(t) + RG_{\text{AMP}}|x(t) * h_{CD}(t)|^2. \quad (2.26)$$

The constant terms and the ASE-ASE beat noise were neglected. $n(t)$ is a white Gaussian noise whose one-sided PSD is $S_{ASE}$ (2.3), and $g(t)$ is a real-valued impulse response whose Fourier transform is given by [39]

$$G(f) = \begin{cases} H_{CD}(f)e^{-j\varphi_C}, & f > 0 \\ 2\cos\varphi_C, & f = 0 \\ H_{CD}^*(-f)e^{j\varphi_C}, & f < 0 \end{cases} \quad (2.27)$$

where $\varphi_C = \arg C$, $H_{CD}(f) = \exp(-0.5j\beta_2(2\pi f)^2 L)$. Note that $G(f)$ only causes phase distortion and therefore the desired signal $s(t)$ does not experience power fading. The second term in (2.26) is the noise component corresponding to the carrier-ASE beat noise and signal-ASE beat noise. The last term in (2.26) accounts for the signal-signal beating interference (SSBI), which is minimized by increasing the CSPR. Nonetheless, the SSB-OFDM receiver must employ some form of SSBI cancellation.

The SNR at the $n$th subcarrier is given by:

$$\text{SNR}_n = \frac{N_{\text{FFT}} P_{n,rx} \cdot \text{CSPR}}{(1+\text{CSPR})\frac{F_n \lambda}{hc} f_s + 2/3 r^2 P_s \cdot \text{CSPR} \cdot 2^{-2\text{ENOB}} + \gamma(\text{CSPR})} \quad (2.28)$$

where $0 \leq \gamma(\text{CSPR}) << 1$ accounts for imperfect SSI cancellation. $\gamma(\text{CSPR})$ may be interpreted as the remaining power of the SSBI term after SSBI cancellation. This approximation is possible since, by the central limit theorem, any noise after the FFT operation is approximately Gaussian distributed. $P_s = \sum_{n=1}^{N_{\text{FFT}}/2-1} P_{n,rx}$ is the signal power at the optical amplifier input, where $P_{n,rx}$ is the power of the $n$th subcarrier referred to the input of the optical amplifier. The three terms in the denominator of $\text{SNR}_n$ in (2.28) account for, respectively, signal-ASE beat noise, quantization noise, and imperfect SSBI cancellation. Knowing the SNR at each subcarrier, we can compute the BER according to (2.14).

The OSNR required is given by $\text{OSNR}_{\text{req}} \approx \frac{G_{\text{AMP}} P_C}{2 S_{sp} B_{ref}}$. In contrast to the DC-OFDM discussed in Section 2.3.2, the OSNR required no longer depends on the clipping ratio at the transmitter, but it now depends on the carrier power $P_c = |C|^2$.

Several SSBI cancellation techniques have been proposed with different efficacies and complexities. In [39], SSBI cancellation is performed by using the received signal $y[k]$ to estimate the SSBI term by computing $|y[k] + j\mathcal{H}\{y[k]\}|^2$ and subtracting it from the received signal. A similar procedure is proposed in [40], where the interference estimate is computed by linearization of the receiver. Due to noise, these techniques are most effective at high OSNR. Moreover, calculating the SSBI estimate in the frequency domain simplifies the Hilbert transform calculation, but it requires frequency-domain convolution to implement the squaring operation. Another technique is based on non-linear equalization based on truncated Volterra series [41]. The number of taps $N_{taps}$ in a Volterra non-linear equalizer grows rapidly as the memory length increases, and a simple time-domain implementation has complexity $\mathcal{O}(N_{taps}^2)$. In [41], the Volterra nonlinear equalizer had 28 taps.

Another SSBI cancellation technique proposed in [40] is based on the so-called Kramers-Kronig (KK) receiver [42, 43]. In contrast to previous techniques, the KK receiver reconstructs the phase of the electrical field from the detected intensity waveform. This reconstruction is only possible if the electric field signal is minimum phase. As discussed in [42], the minimum-phase condition is guaranteed by transmitting a sufficiently strong carrier. For minimum-phase signals, the phase $\hat{\phi}[k]$ can be estimated from the detected intensity $P[k]$:

$$\hat{\phi}[k] = \mathcal{F}^{-1}\left\{\mathcal{H}\{\ln\sqrt{P[k]}\}\right\} = \mathcal{F}^{-1}\left\{j\text{sgn}(\omega)\mathcal{F}\{\ln\sqrt{P[k]}\}\right\}, \quad (2.29)$$

where $\mathcal{F}\{\cdot\}$ and $\mathcal{F}^{-1}\{\cdot\}$ denote direct and inverse discrete-time Fourier transform, respectively. $\text{sgn}(\omega)$ is the sign function and it equals 1, for $\omega > 0$; $-1$, for $\omega < 0$; and 0, for $\omega = 0$. The electric field $\hat{E}[k]$ can then be reconstructed:

$$\hat{E}[k] = \sqrt{P[k]}e^{j\phi[k]} \tag{2.30}$$

The reconstructed electric field in (2.30) corresponds to the SSB-OFDM signal at the receiver, which can be detected as a conventional OFDM signal by removing cyclic prefix, computing the FFT, performing one-tap frequency-domain equalization, and finally performing symbol detection.

The KK phase retrieval technique outlined in equations (2.29) is not restricted to SSB-OFDM signals. In fact, the KK phase retrieval technique was utilized to reconstruct a SSB 4-PAM signal in [43], and to reconstruct a $M$-QAM signal in [42]. Note that for QAM, the information on the negative-frequency sideband is not redundant. Hence, the transmitted signal must be frequency-shifted by $R_s/2$ with respect to the carrier, where $R_s$ is the signal rate. Consequently, the spectral efficiency of KK $M$-QAM is halved: $0.5 \log_2 M$, which is the same spectrum efficiency achieved by $\sqrt{M}$-PAM modulation. Moreover, this is the same spectral efficiency achieved by carrierless amplitude and phase (CAP) modulation [8] without the SSB requirement and additional complexity of KK phase retrieval. However, CAP does not allow electronic CD compensation. For these reasons, the so-called KK receiver does not improve spectral efficiency or receiver sensitivity.

The KK phase retrieval does permit electronic CD compensation, but at arguably higher DSP complexity than the techniques described previously. The logarithm and square root computations require high-precision arithmetic as well as upsampling by a large factor in order to correctly represent $\ln\sqrt{P[k]}$ in the frequency domain. In [42], an upsampling factor of three was recommended.

## 2.4 Performance comparison

In this section we compare the different modulation formats discussed in Section 2.3 for the intra- and inter-data center links of Section 2.2. In intra-data center links, the system performance is quantified by computing the receiver sensitivity, which is the received power $\bar{P}_{rx}$ necessary to achieve a certain target BER, usually determined by the FEC code threshold. For all scenarios studied in this chapter, we consider a weak FEC code such as RS(255, 239), which has a net coding gain of 5.6 dB at BER $= 10^{-12}$, an input BER threshold of $1.8 \times 10^{-4}$ to achieve $10^{-12}$ BER, and overhead of $\sim 7\%$. Note that the FEC choice is not critical for the performance comparison, since all schemes would benefit from a stronger FEC code. While stronger codes can provide higher gains, their complexity is prohibitive for low-cost and low-power applications such as data center links. On the other hand, without coding, single-laser 100 G links presumably cannot achieve the required $10^{-12}$ BER, and are difficult to analyze, since it is difficult to formulate system models that are accurate to such low BERs.

As inter-data center links are optically amplified, their performance is more conveniently quantified in terms of the OSNR necessary to achieve the target BER.

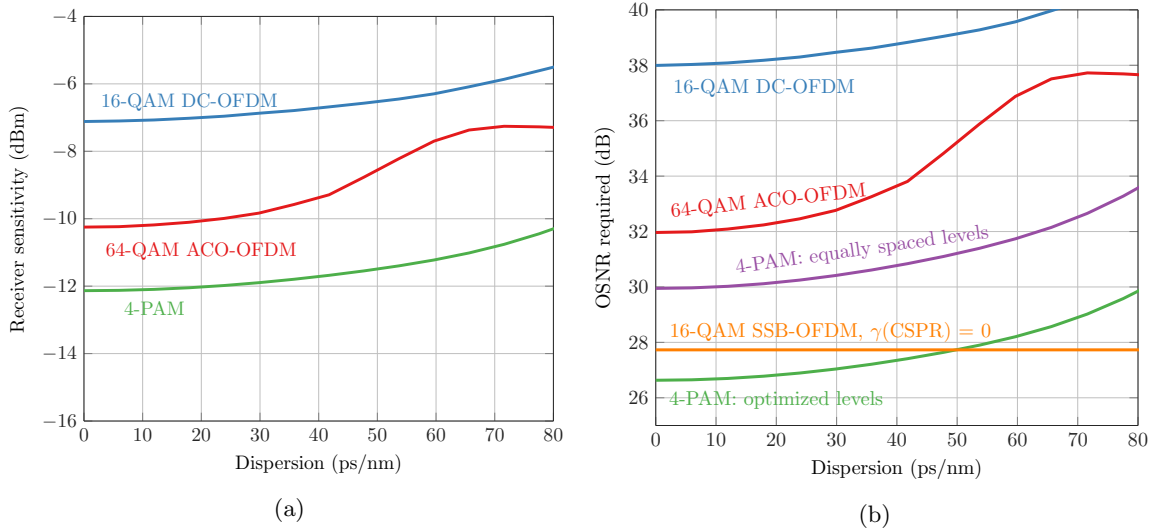

Figure 2.14: Performance comparison of DD-compatible modulation schemes vs chromatic dispersion at 112 Gbit/s. Unamplified systems are characterized in terms receiver sensitivity 2.14a, while amplified systems are characterized in terms of OSNR required 2.14b. The $x$-axis may be interpreted as total dispersion in intra-data center links or residual dispersion after optical CD compensation in inter-data center links.

Fig. 2.14a shows the receiver sensitivity achieve with each modulation format vs. dispersion for unamplified systems based on PIN photodiode. Fig. 2.14b shows the required OSNR in amplified systems. The dispersion axis may be interpreted as total CD in intra-data center links, or residual CD after optical CD compensation in inter-data center links. The results obtained with the simplified equations presented in this chapter are typically within 2 dB of the Monte Carlo simulations including other phenomena such as intensity noise and modulator imperfections. The simulation parameters are summarized in Table 2.1.

4-PAM outperforms all other candidates in all considered scenarios. Level spacing optimization improves OSNR required by roughly 3 dB (Fig. 2.14b). After roughly 50 ps/nm of dispersion, the penalty due to CD increases steeply. This penalty poses a limit in the reach of intra-data center links and restricts the maximum residual dispersion after optical CD compensation in inter-data center links.

DC-OFDM has a significant penalty due to the relatively high DC bias required to meet the non-negativity constraint of the intensity-modulated optical channel. Although ACO-OFDM has better performance, it requires prohibitively high DAC/ADC sampling rates (equation (2.12)), as the even subcarriers cannot be used to transmit data. In fact, the ACO-OFDM performance curves are not monotonic because, as dispersion increases, subcarriers near the first notch of the IM-DD channel frequency response achieve poor SNR and are not used.

Similarly to 4-PAM, CD mitigation through linear equalization is only effective when CD is

Table 2.1: Parameters used in Monte Carlo simulations for determining receiver sensitivity and OSNR required of DD-compatible modulation schemes.

| | | |
|---|---|---|
| | Bit rate ($R_b$) | 112 Gbit/s |
| | Target BER | $1.8 \times 10^{-4}$ |
| | Laser linewidth | 200 kHz |
| Tx | Relative intensity noise | $-150$ dB/Hz |
| | Modulator bandwidth | 30 GHz |
| | Chirp parameter ($\alpha$) | 0 |
| | Extinction ratio ($r_{ex}$) | $-15$ dB |
| | Responsivity ($R$) | 1 A/W |
| PIN & TIA | Bandwidth | 30 GHz |
| | TIA input-referred noise ($\sqrt{N_0}$) | 30 pA/$\sqrt{\text{Hz}}$ |
| Optical | Gain ($G_{\text{AMP}}$) | 20 dB |
| Amplifier | Noise figure ($F_n$) | 5 dB |
| | Number of amplifiers ($N_A$) | 1 |
| $M$-PAM | ADC ENOB | 5 bits |
| Rx | Oversampling rate ($r_{os}$) | 5/4 |
| | FFE number of taps ($N_{taps}$) | 9 |
| | ADC ENOB | 5 bits* |
| OFDM | FFT length ($N_{\text{FFT}}$) | 256 |
| Rx | Oversampling rate ($r_{os}$) | 1.23 |
| | Cyclic prefix length ($N_{\text{CP}}$) | 10 |

*6 bits for ACO-OFDM

small. Bit loading and power allocation would allow OFDM variants to better exploit the power-faded optical channel resulting from considerable CD, but such systems are unlikely to be practical, since DD also leads to intermodulation products that fall in the signal band.

Fig. 2.14b shows the required OSNR for a SSB-OFDM with $\gamma(\text{CSPR}) = 0$ (i.e., perfect SSBI cancellation). The required OSNR does not vary with dispersion because, as mentioned above, the detected DSB-OFDM does not experience power fading. The CSPR has been optimized for all cases. The $\sim 28$-dB OSNR required for $\gamma(\text{CSPR}) = 0$ is similar to the OSNR required using Kramers-Kronig technique in [40].

## 2.5 Complexity comparison

The previous section compared the performance of the various modulation formats and detection techniques in terms of receiver sensitivity and OSNR required. This section focuses on the overall complexity and power consumption of these schemes.

Table 2.2 summarizes the main complexity differences between the various schemes discussed in this paper. This comparison covers spectral efficiency, modulator type, complexity of the optical receiver, number of ADCs and their sampling rate and ENOB, capability to electronically compensate for CD, and DSP operations required at the receiver.

Table 2.2: Complexity comparison of DD-compatible modulation formats.

| Scheme | SE (b/s/Hz) | Mod. type | Optical receiver | ADC (GS/bit) | # ADCs / ENOB | Digital CD comp. | DSP operations |
|---|---|---|---|---|---|---|---|
| 4-PAM | 2 | IM | 1 PD | $0.5r_{os}$ | 1 / 4 | Very low | TD-EQ |
| 16-QAM DC-OFDM | 4 | IM | 1 PD | $0.5r_{os}r_{\text{CP}}$ | 1 / 5 | Very low | IFFT/FFT, 1-tap FD-EQ |
| 16-QAM SSB-OFDM | 4 | I&Q | 1 PD | $0.5r_{os}r_{\text{CP}}$ | 1 / 5 | Moderate | FD-EQ, SSBI cancellation |
| KK 4-PAM | 2 | I&Q | 1 PD | $0.5r_{os}$ | 1 / 5 | Moderate | SSB filtering, KK-PE, and TD-EQ |

$r_{os}$ denotes oversampling ratio, and $r_{\text{CP}} = (N_{\text{FFT}} + N_{\text{CP}})/N_{\text{FFT}}$ is the oversampling ratio due to cyclic prefix in OFDM.
Acronyms: spectral efficiency (SE), photodiode (PD), time-domain equalizer (TD-EQ), frequency-domain equalizer (FD-EQ), phase estimation (PE), single-input single output (SISO), carrier recovery (CR), and not applicable (NA).

Fig. 2.15 shows a coarse estimate of power consumption in 28-nm CMOS for various modulation schemes at 100 Gbit/s. The power consumption of DSP is estimated using the power consumption models presented in [44]. First, the number of real additions and real multiplications is counted for all DSP operations (summarized in Table 2.2). Then, the power consumption is obtained by computing how much energy a given operation consumes. For instance, a real addition in 28-nm CMOS with 6-bit precision consumes 0.28 pJ, while a real multiplication with 6-bit precision consumes 1.66 pJ [44]. The power consumption estimates for DACs and ADCs given in [44] assume that the power consumption scales linearly with resolution and sampling rate. The DAC figure of merit is 1.56 pJ/conv-step, while the ADC figure of merit is 2.5 pJ/conv-step [44]. The resolution of the DACs and ADCs, as well as the DSP arithmetic precision, is assumed equal to ENOB + 2, where ENOB is given in Table 2.2. Only OFDM formats are assumed to need high-resolution DAC, since 4-PAM may avoid it, if pulse shaping or preemphasis are not performed. For all cases, the oversampling ratio assumed is $r_{os} = 5/4$.

Fig. 4.13 compare the power consumption of DD-compatible schemes at 100 Gbit/s for (a) a CD-compensated link where the residual CD is at most 80 ps/nm, and for (b) an 80-km uncompensated CD link. As expected, 4-PAM is more power efficient than the other formats. Compared to OFDM schemes, 4-PAM benefits from requiring lower sampling frequency, lower resolution, and performing time-domain equalization, which is more power efficient than frequency-domain equalization for short filters. However, in the high-uncompensated-CD regime, SSB modulation is the only viable choice. The SSBI cancellation in SSB-OFDM is assumed to be a Volterra nonlinear equalizer with 14 taps in (a) and 28 taps in (b). The power consumption of KK 4-PAM is excessively high due to the phase estimation using 3-times upsampling for computation of the Hilbert transform, as discussed in Section 2.3.3. Although not shown in Fig. 4.13, the power consumption of 4-PAM with MLSD in an uncompensated link would also be excessively high, since the complexity of the MLSD receiver grows exponentially with the memory length of the Viterbi decoder. We do not include MLSD 4-PAM in the comparison of Fig. 4.13 due to the lack of models to translate branch metric computations into power consumption.

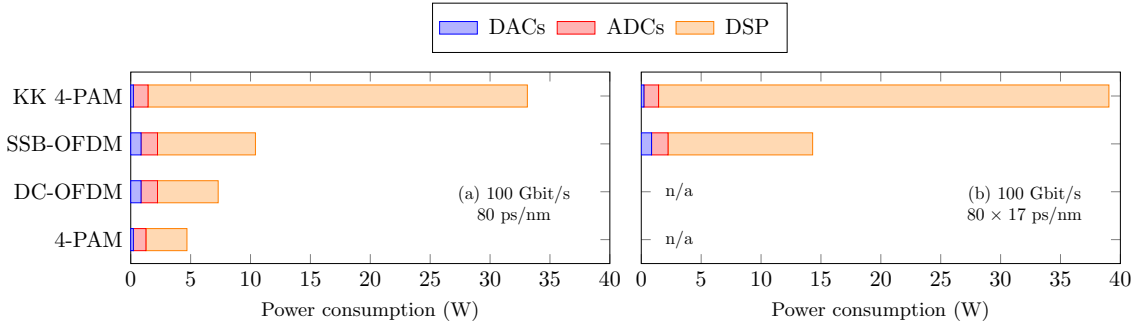

Figure 2.15: Coarse estimate of power consumption of high-speed DACs, ADCs, and DSP for various DD-compatible modulation schemes at 100 Gbit/s. DSP power consumption estimates are made for 28-nm CMOS using the models presented in [44]. In (a) we assume that CD is compensated optically and the residual CD is at most 80 ps/nm at 100 Gbit/s. In (b) we assume uncompensated transmission up to 80 km near 1550 nm.

## 2.6 Summary

We have evaluated the performance, complexity, and power consumption of 4-PAM and OFDM variants for intra- and inter-data center links. 4-PAM outperforms all the other modulation formats due to its relatively low complexity and high tolerance to noise and distortion.

In unamplifed intra-data center links operating near 1310 nm and reaching up to 10 km, dispersion is small enough that CD can be modeled as a linear filter that causes power fading. This power fading will limit the reach and ultimately the throughput that can be practically transmitted over a SMF. Another challenge is that systems designed to achieve 400 Gbit/s or 1 Tbit/s over a single fiber will have small power margin even when using single-laser 100 Gbit/s link with 4-PAM. Eye safe systems cannot exceed 14 dBm per fiber, which limits the power per channel. Accounting for all the losses, an eye-safe $4 \times 100$ Gbit/s system would only have about 5 dB of margin. Consequently, these systems would not support increased losses due to longer fiber plant, more wavelengths, and possibly optical switching. Chapter 3 studies the use of APDs to mitigate this problem.

In amplified inter-data center links operating near 1550 nm and reaching up to 100 km, CD is significant and must be compensated. As CD is a nonlinear operation in IM-DD systems, receiver-side linear equalization is not effective. Transmitter-side predistortion, for instance, could compensate for CD, but any electronic compensation technique will inevitably be more power-hungry than simple optical CD compensation, which can be realized by dispersion-matched DCFs or FBGs. Alternatively, future data center links may favor DSF with small dispersion in the C-band, so that CD compensation is avoided altogether. However, even in small CD regime the OSNR require for 100 Gbit/s 4-PAM systems is roughly 30 dB, which may also not support data center network evolution in the long term.

# Chapter 3

# Improving the Receiver Sensitivity of Intra-Data Center Links

Chapter 2 showed that although 4-PAM outperforms other competing techniques, it may not offer sufficient power margin to support data center network evolution in the long term. As shown in Fig 2.14a, 100 Gbit/s 4-PAM links have receiver sensitivity of roughly −10 dBm. Therefore, an eye-safe 400 Gbit/s link using four wavelength-division-multiplexed (WDM) channels is expected to have an optical power margin of under 5 dB after accounting for all the losses [8]. Practical systems, however, will require significantly higher margins to accommodate component aging and increased optical losses due to longer fiber plant, more wavelengths requiring lower per channel power to maintain eye safety, and possibly optical switches. Thus, it is desirable to improve receiver sensitivity while minimizing system power dissipation, cost and size. Simply using stronger forward error-correction (FEC) codes, for instance, would improve sensitivity, but would dramatically increase power consumption and latency.

Avalanche photodiodes (APD) and semiconductor amplifiers (SOA) are promising alternatives to improve receiver sensitivity, with reasonable additional cost and power consumption. This chapter studies the benefits and drawbacks of APDs in 100 Gbit/s-per-wavelength links for intra-data center applications. SOAs were studied in collaboration with Dr. Sharif in [31].

The remainder of this chapter is organized as follows. In Section 3.1, we review recent progress in the design of high-speed APDs. In Section 3.2, we present the system model used to evaluate the performance of APD-based data center links. In Section 3.3, we evaluate the performance of APD-based sytems in WDM systems. In Section 3.4, we consider practical considerations of APD-based receivers. Section 3.5 summarizes the main findings of this chapter.

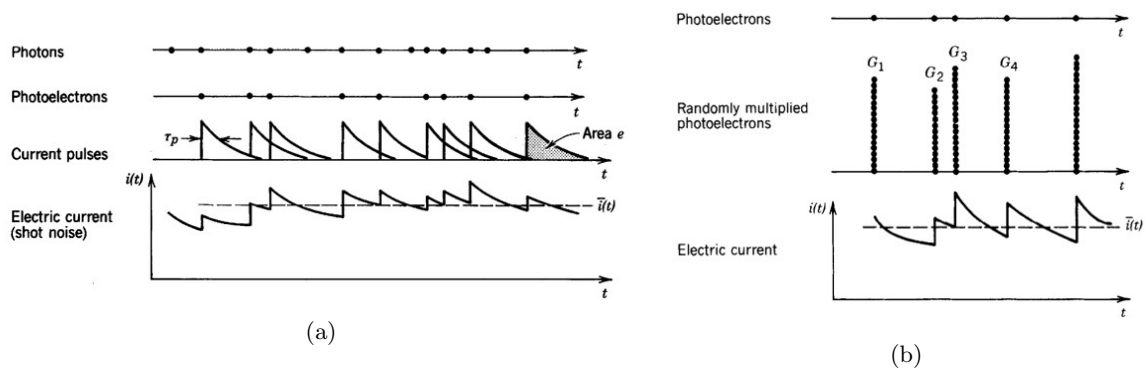

Figure 3.1: Photodetection in a conventional PIN photodetector and in an APD. Source: Bahaa Saleh et al. "Fundamentals of Photonics," 1991.

## 3.1  Avalanche photodiodes

APDs provide internal electrical gain through impact ionization. APDs are operated in high reverse bias resulting in a large junction electric field. As a result, carriers generated by the photoelectric effect gain enough energy to excite new carriers through the process of impact ionization. These new carriers, in turn, will excite other carriers unleashing an avalanche effect.

The gain provided by the APD comes at the expense of excess shot noise due to the inherently stochastic nature of the impact ionization process. Fig. 3.1 illustrates this problem by depicting the photocurrent generation in a positive-intrinsic-negative (PIN) photodiode and in an APD. In a PIN photodiode, shot noise arises from variations in the detect current due to the superposition of current pulses generated by each photon event. On the other hand, in APDs, each photoelectron is multiplied by a random gain resulting from impact ionization, which introduces another form of randomness in the detected current.

In addition to excess shot noise, the avalanche process increases the carrier transit time through the multiplication region, an effect known as avalanche buildup. As a result, the APD bandwidth decreases as the gain increases. This dependency is often expressed as a constraint on the gain-bandwidth product (GBP) of an APD.

APDs have been widely adopted in 10 Gbit/s links for metro and access networks [52], as they are more cost-effective than optical pre-amplification followed by a PIN photodetector. 100 Gbit/s systems pose a greater challenge, however, as they require APDs with both small impact ionization factor $k_A$ (i.e., small noise) and wide bandwidth. Recent advances in APD technology have improved these characteristics. Impact ionization factors have been reduced by using a multiplication layer of InAlAs ($k_A \sim 0.2$) [47] and Si ($k_A < 0.1$) [51]. Bandwidths have been increased by new designs that decouple bandwidth from responsivity, which is normally reduced in high-speed APDs as the absorption region is made thinner to reduce transit time. These new designs include resonant cavities

Table 3.1: Characteristics of published APDs.

| Ref. | Responsivity $R$ at 1310 nm (A/W) | Impact ionization $k_A$ | Dark current at G = 10 (nA) | Low-gain BW (GHz) / GBP (GHz) | Structure and materials |
|---|---|---|---|---|---|
| [45] | 0.74 | 0.18 | 40 | 24 / 290 | Resonant-cavity InGaAs–InAlAs |
| [46] | 0.65 | 0.2 | 50* | 40 / 115 | Waveguide InGaAs–InAlAs |
| [47] | 0.27 | 0.18–0.27 | 200 | 27 / 120 | Waveguide InGaAs–InAlAs |
| [48] | 0.17 | 0.1–0.2 | 60 | 28 / 320 | Waveguide InGaAs–InAlAs |
| [49] | 0.68 | 0.15–0.25 | 1000* | 37.5 / 140 | Waveguide evanescently coupled photodiode InGaAs–InAlAs |
| [50] | 0.42 | 0.2 | 65 | 27 / 220 | p-down inverted InGaAs–InAlAs |
| [51] | 0.55 | 0.08–0.18 | 1000 | 14 / 340 | Separate absorption, charge, and multiplication (SACM) Ge–Si |

* At 90% of breakdown voltage.

APDs [45], waveguide APDs [46, 47, 49, 53], and thin-multiplication-layer APDs in which both excess noise and avalanche buildup time are reduced by the dead zone effect [54]. Recent works have also investigated using bit-synchronous sinusoidal biasing to increase the APD gain-bandwidth product (GBP) [55, 56].

Table 3.1 shows typical values of responsivity, impact ionization factor $k_A$, dark current, low-gain bandwidth and GBP, and structures and materials for published APDs. Fig 3.2 shows a few examples of APD structures.

### 3.1.1 Shot noise

Shot-noise is significant in APD-based receivers. The received optical powers for optical communication systems in data centers are relatively high (e.g., above $\sim -13$ dBm from (3.3)). At such power levels, far from the quantum regime, shot noise can be described accurately as a white Gaussian noise whose one-sided PSD, ignoring the APD frequency response, is given by the well-known expression [30]

$$S_{sh} = 2qG_{\mathrm{APD}}^2 F_A(G_{\mathrm{APD}})(R\bar{P}_{rx} + I_d), \quad F_A(G_{\mathrm{APD}}) = k_A G_{\mathrm{APD}} + (1 - k_A)(2 - 1/G_{\mathrm{APD}}), \quad (3.1)$$

where $q$ is the electron charge, $G_{\mathrm{APD}}$ is the APD gain, $R$ is the APD responsivity, $\bar{P}_{rx}$ is the received optical power, $I_d$ is the APD dark current, and $F_A(G_{\mathrm{APD}})$ is the APD excess noise factor.

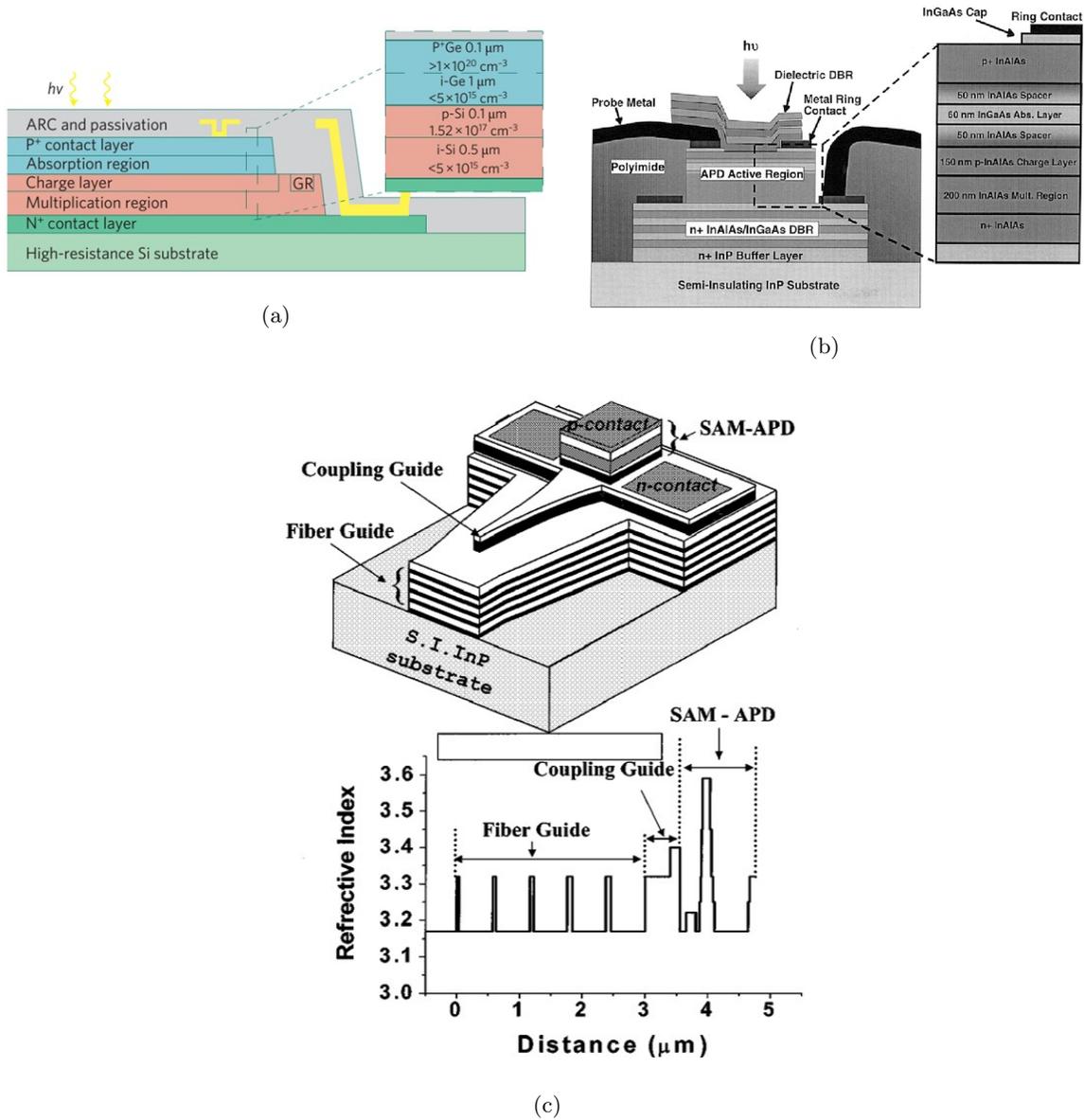

Figure 3.2: Examples of APD structures: (a) SiGe [51], (b) resonant-cavity APD [45], and (c) twin-waveguide APD [57].

The Gaussian approximation allows us to calculate the BER without relying on moment generating functions and the saddle point approximation, but it must be used with caution. As shown in [58], the Gaussian approximation can be inadequate for evaluating receiver performance when ISI is introduced by an APD in the avalanche buildup time-limited regime. In this work, however, the modulator, APD, and receiver electronics (and possibly fiber propagation) all cause considerable ISI. Moreover, as discussed in Section IV.B, there is only a small sensitivity penalty for operating the APD at relatively small gains, such that the deterministic transit time and RC time constant limit the bandwidth more than avalanche buildup. Hence, the Gaussian approximation is sufficiently accurate to predict the performance at the relatively high BERs ($\sim 10^{-4}$) at which coded systems can operate.

### 3.1.2 APD bandwidth and the gain-bandwidth product

Equation (3.1) does not account for the APD frequency response, which filters both signal and shot noise. Exact computation of the filtered shot noise variance at the APD output would require knowledge of the second-order statistics of the impulse response, which is generally not tractable analytically and requires computationally intensive simulations [50], [58]. To circumvent this problem, simpler models, such as parametric or deterministic impulse response functions, are customarily employed [59]. In this work, we model the impulse response of the APD by a deterministic exponential decay, which results in a frequency response

$$H_{\text{APD}}(f) = \left(1 + j\frac{f}{B(G_{\text{APD}})}\right)^{-1}, \tag{3.2}$$

where $B(G_{\text{APD}})$ is the 3-dB bandwidth of the APD at gain $G_{\text{APD}}$. Operating regimes for this model are illustrated in Fig. 3.3. At low gains, the bandwidth is generally independent of gain, since in this regime the major bandwidth limitation comes from carrier transit time and parasitic capacitance (RC time constant). As the gain increases, avalanche buildup time dominates, leading to a fixed GBP. In this model, an APD frequency response can be characterized by its low-gain bandwidth and by its GBP. Table 3.1 shows typical values of these parameters for state-of-the-art APDs.

The choice of a deterministic exponential decay for the impulse response is consistent with the transit-time/RC limited regime, but it does not capture the intrinsic correlation between an APD's gain and its impulse response. As shown in [59], however, this problem can be mitigated by using a shot noise equivalent bandwidth, as opposed to the APD's 3-dB bandwidth, in computing the shot noise variance. This definition of noise bandwidth approximately captures the fluctuations in the impulse response as well as in the gain, and may exceed an APD's 3-dB bandwidth by up to 30% [59]. This definition also captures the effect of dead space in APDs. Dead space is the distance a newly generated carrier must propagate before gaining sufficient energy to impact ionize other carriers. This effect is particularly important in thin APDs [54].

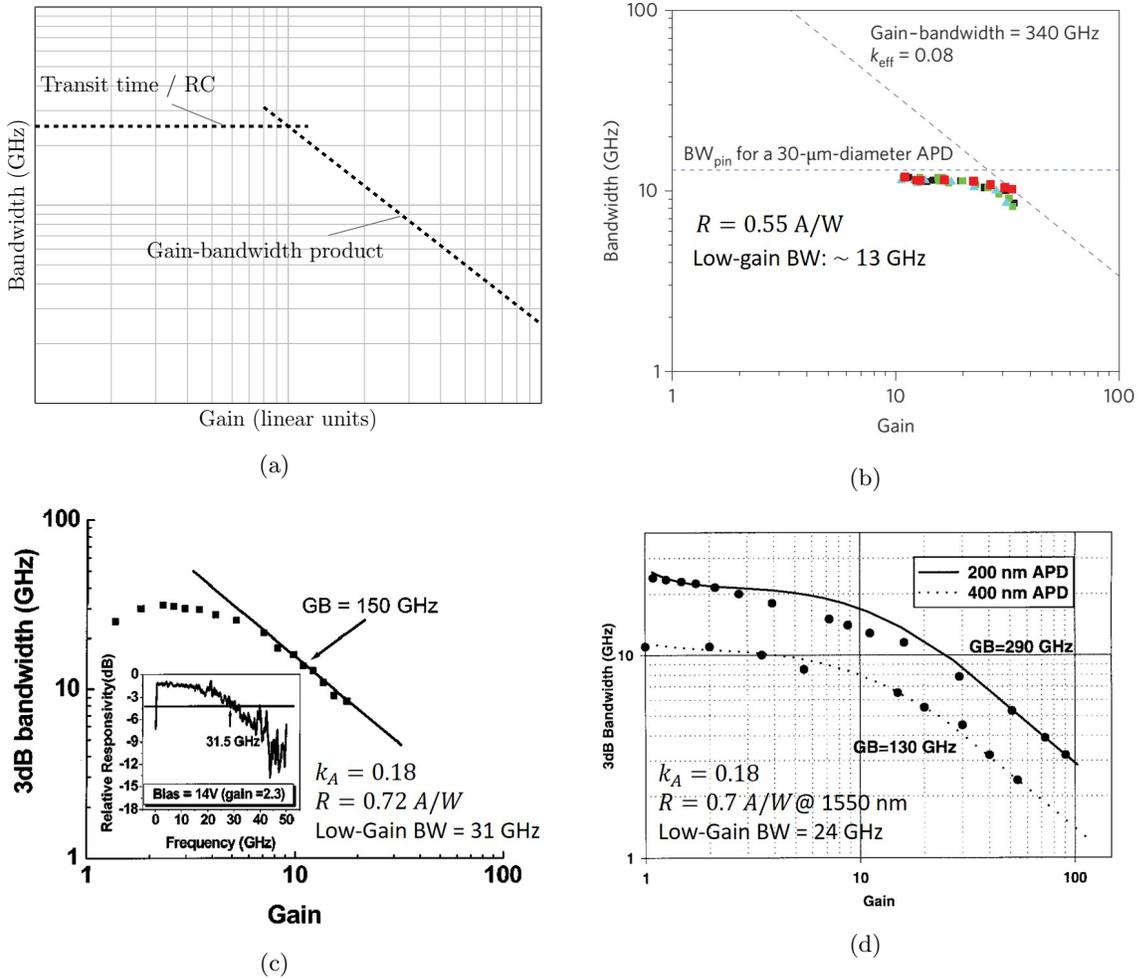

Figure 3.3: (a) Generic bandwidth-vs-gain curve for avalanche photodiodes illustrating the two regimes of operations: low-gain operation where bandwidth is limited by transit-time and RC time constants, and high-gain operation where gain is limited by avalanche buildup time given rise to a fixed gain-bandwidth product. Some examples of real devices are also shown: (b) SiGe APD [51], (c) resonant-cavity APD [45], and (d) twin-waveguide APD [57].

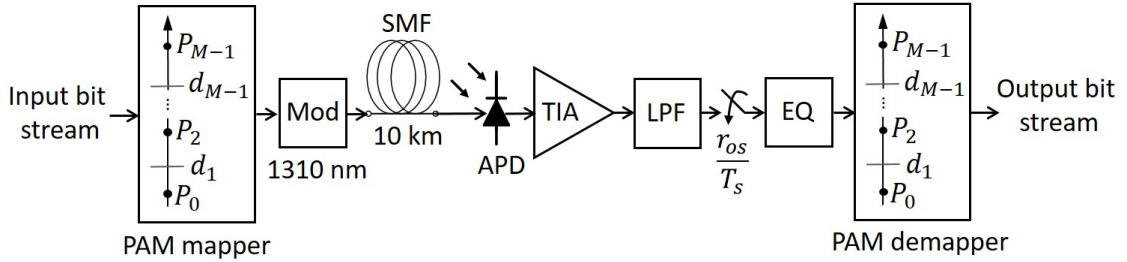

Figure 3.4: Block diagram for APD-based system.

## 3.2 System model for APD-based intra-data center links

A general block diagram for a system using multi-level intensity modulation and APD-based direct detection is shown in Fig. 3.4. This diagram is similar to the diagram for intra-data center link in Fig. 2.5a.

At the transmitter, a stream of input bits is mapped onto $M$-PAM symbols with non-negative intensity levels $\{P_0, \ldots, P_{M-1}\}$. Digital pulse shaping can reduce the signal bandwidth and pre-compensate for the modulator frequency response, but requires a high-speed digital-to-analog converter (DAC) and, more importantly, enforcing the non-negativity of intensity modulation requires an additional DC bias, which was shown to cause a 3-dB power penalty [8]. Hence, we assume a multi-level PAM encoder with a rectangular pulse shape. Programmable intensity levels can enable pre-compensation for modulator non-linearity and transmission of unequally spaced intensity levels to improve receiver sensitivity, as discussed in Section 2.3.1.

The encoder output drives an optical modulator, which could be an directly modulated lasers (DML), an electro-absorption modulators (EAM), or a Mach-Zenhder modulator (MZM). The intensity-modulated signal is launched into an SMF. The received signal after fiber propagation is detected using an APD-based receiver, as illustrated in Fig. 3.4. After photo-detection and transimpedance amplifier (TIA), the receiver realizes sampling, equalization, and detection. Typical TIAs have 3-dB bandwidth of 20–70 GHz and input-referred noise $\bar{I}_n$ of 20–50 pA/$\sqrt{\text{Hz}}$ [28, Table 2], where $\bar{I}_n^2 = N_0$ is the one-sided power spectrum density of thermal noise.

Due to the strong bandwidth limitations of the modulator, optical fiber, and other components such as the APD, equalization is necessary. To facilitate the analysis, we assume a symbol-rate LE with analog noise-whitening filter cascaded by an electrical filter matched to the received pulse shape. The fixed symbol-rate LE requires accurate knowledge of the channel response and precise timing recovery. In practice, a receiver employing a fixed anti-aliasing filter with an adaptive fractionally spaced LE can achieve performance approaching the ideal symbol-rate LE, while compensating for timing errors and not requiring prior knowledge of the channel.

After equalization, symbol-by-symbol detection is performed using decision thresholds $\{d_1, \ldots, d_{M-1}\}$,

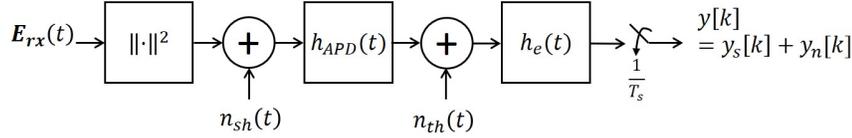

Figure 3.5: Equivalent baseband block diagram for APD-based receiver.

which may be optimized based on the statistics of the received noise, as described in Section 2.3.1.

As in Chapter 2, the system is assumed to use a simple Reed-Solomon code such as RS(239, 255) with BER threshold of $1.8 \times 10^{-4}$ and 7% overhead.

### 3.2.1 Performance evaluation

Fig. 3.5 illustrates the receiver equivalent baseband model for the APD-based receiver. The APD filters both signal and shot noise with impulse response $h_{\mathrm{APD}}(t)$. In Fig. 3.5, $h_e(t)$ denotes the baseband equivalent transfer function of the post-detection electrical filter. In an ideal APD-based system, $h_e(t)$ can be approximated by the cascade of the noise whitening filter, matched filter, and the continuous-time equivalent of the LE.

Throughout this chapter, we quantify the system performance relative to an ideal ISI-free 107-Gbit/s 4-PAM system with a thermal noise-limited PIN receiver. The receiver sensitivity of the reference system can be computed as a function of the target BER [8]:

$$\bar{P}_{req,ref} = \sqrt{\frac{R_b S_{th}(M-1)^2}{2R^2 \log_2 M}} Q^{-1}\left(\frac{M \log_2 M \cdot \mathrm{BER}_{\mathrm{target}}}{2(M-1)}\right) \tag{3.3}$$

where $R_b$ is the bit rate and $R$ is the photodiode responsivity. For our analysis and simulation in this paper, we assume $R_b = 107$ Gbit/s, $R = 1$ A/W, and TIA with input-referred noise of $\bar{I}_{n,in} = 30\mathrm{pA}/\sqrt{\mathrm{Hz}}$. Hence, for $\mathrm{BER}_{\mathrm{target}} = 1.8 \times 10^{-4}$, the reference receiver sensitivity is $\bar{P}_{req,ref} \approx -13$ dBm.

In an APD-based receiver signal power scales with $G_{\mathrm{APD}}^2$, while shot noise variance scales with $G_{\mathrm{APD}}^3$, as can be seen from (3.1). This implies that increasing the APD gain after shot noise becomes dominant hurts receiver sensitivity. Therefore, there is an optimal APD gain that minimizes receiver sensitivity, and this gain lies below the range at which shot noise becomes completely dominant.

Moreover, in the avalanche buildup time limited regime, the APD gain and bandwidth are coupled and related by the GBP. Hence, to determine the optimal APD gain we must account for APD bandwidth limitations.

The APD filters both signal and shot noise. Consequently, the total noise (shot plus thermal noise) is not white. From Fig. 3.5, the shot noise component of the decision variable is given by

$$y_{n,sh}[k] = h(t) * n_{sh}(t)|_{t=kT_s}, \tag{3.4}$$

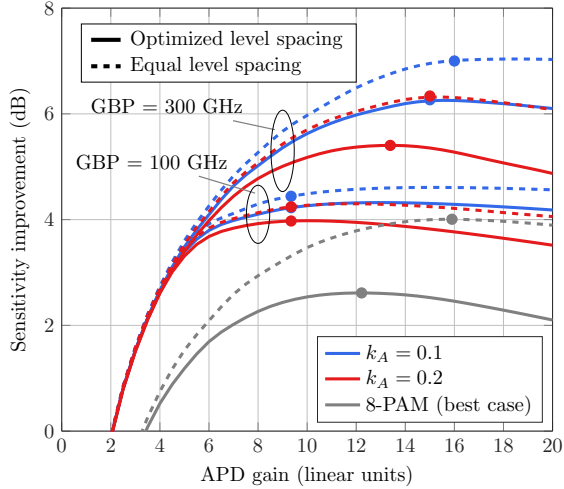

Figure 3.6: Receiver sensitivity improvement versus APD gain for 4-PAM and $k_A = 0.1$ (Si) and $k_A = 0.2$ (InAlAs) and two values of GBP: 100 GHz and 300 GHz. The 8-PAM best-case scenario of GBP = 300 GHz and $k_A = 0.1$ is shown for reference. Results assume parameters from Table 3.2.

where $h(t) = h_{\mathrm{APD}}(t) * h_c(t)$, and the electrical filter $h_c(t)$ comprises of a noise-whitening filter, a matched filter matched to the received pulse shape, and the continuous-time equivalent of the discrete-time LE. $n_{sh}(t)$ is the shot noise, whose PSD is given by (3.1). It thus follows that

$$\mathrm{Var}(y_{n,sh}[k]) = [2qG_{\mathrm{APD}}^2 F_A(G_{\mathrm{APD}})(RP_{rx}(t) + I_d)] * |h(t)|^2 \Big|_{t=kT_s}, \quad (3.5)$$

Note that the received intensity waveform $P_{rx}(t)$ already includes ISI caused by the modulator. Thus, $P_{rx}(t) = x(t) * h_{mod}(t)$, where $x(t)$ is the ISI-free modulator drive signal, and $h_{mod}(t)$ is the modulator impulse response.

If shot noise were signal-independent, the convolution in (3.5) would reduce to simply scaling the noise variance by the energy of $h(t)$, leading to the well-known noise enhancement penalty. Here, however, the convolution in (3.5) makes the noise variance dependent on the sequence of symbols within the memory length of $|h(t)|^2$. Fortunately, the memory length of $|h(t)|^2$ is fewer than five symbols, even when the APD bandwidth is as low as 20 GHz. The impact of $|h(t)|^2$ on the shot noise variance is particularly noticeable on the lowest intensity levels. As an example, the variance of the shot noise component $y_{n,sh}[n]$ is nonzero even when the symbol $P_0 = 0$ is transmitted (modulation with an ideal extinction ratio), due to shot noise from neighboring symbols.

In performing level spacing optimization, we adopt a conservative approach and calculate (3.5) considering the worst-case scenario, where all the symbols in the memory of are the highest level.

The effect of thermal noise can be computed in terms of its equivalent one-sided noise bandwidth, since thermal noise is not signal-dependent. Hence, $\sigma_{th}^2 = N_0 \Delta f$, where $\Delta f_{th} = \int_0^\infty |H_e(f)|^2 df$, and $H_e(f)$ is the electric filter frequency response normalized such that $H_e(0) = 1$.

Fig. 3.6 shows the sensitivity improvements for 4-PAM as a function of the APD gain for two different GBP scenarios: 100 GHz and 300 GHz. Curves for 8-PAM are included for the best scenario only. We assume the APD has the same responsivity as the reference system, i.e., $R = 1$ A/W, but the results in Fig. 3.6 can be easily converted to $R \neq 1$ A/W by appropriately shifting the curves vertically. For instance, for $R = 0.5$ A/W, the sensitivity improvements would be 3 dB lower than those presented in Fig. 3.6.

For GBP = 100 GHz, avalanche buildup time limits the bandwidth when $G_{\text{APD}} \geq 5$. In this regime, increasing the gain further reduces the APD bandwidth, but this does not translate into an increased noise enhancement penalty, since the APD filters both signal and shot noise. As a result, the sensitivity improvement remains almost constant.

For GBP = 300 GHz, the avalanche buildup time-limited regime is reached at higher gains, when $G_{\text{APD}} \geq 15$, and hence higher sensitivity improvements can be achieved. For $k_A = 0.1$ we observe sensitivity improvements up to 6.3 dB for equal level spacing, and up to 7 dB for optimized levels. Note, however, that there is very little penalty by operating the APD at gains substantially smaller than the optimal gain.

Although 8-PAM is more spectrally efficient than 4-PAM, its poorer noise tolerance leads to significantly smaller sensitivity improvements. Hence, as in systems using PIN-based receivers [8], 4-PAM outperforms 8-PAM.

## 3.3 WDM system performance

In this subsection, we evaluate the performance of 100 Gbit/s-per-wavelength WDM links. The performance of individual channels could be different due to wavelength-dependent characteristics of the system such as chromatic dispersion (CD). The WDM analysis is particularly important for receivers with SOAs to evaluate the effects of wavelength-dependent gain and nonlinear crosstalk caused by cross-gain modulation [31].

The amount of dispersion that each channel experiences depends on the channel spacing as well as the number of channels. The typical channel spacing for short-reach applications is 20 nm for coarse wavelength-division multiplexing (CWDM) links and 4.5 nm for LAN-WDM. Although CWDM is often preferred, as it generally does not require temperature-controlled lasers, LAN-WDW can accommodate more channels close to the zero-dispersion wavelength, which is necessary for 1 Tbit/s and possibly 1.6 Tbit/s systems.

The number of channels in the short-reach WDM links is typically constrained by transmit power limit and CD. Here, we consider two systems: 10-channel LAN-WDM with 4.5-nm channel spacing and 4-channel CWDM with 20-nm channel spacing. The average optical power launched in to the fiber is conservatively limited to 9.4 dBm due to eye-safety restrictions of Class 1 lasers, which are considered inherently safe. The industry also uses Class 1M lasers with eye safety limit of about 14 dBm near 1310 nm, but these lasers require more precaution [19]. Therefore, for 4- and 10-channel WDM links, the average transmitted optical power per channel cannot exceed 3.4 dBm and $-0.6$ dBm, respectively. Thus, even if the 4- and 10-channel systems have relatively similar receiver sensitivities, the 10-channel LAN-DWM has $\sim 4$ dB lower link margin than the 4-channel CWDM.

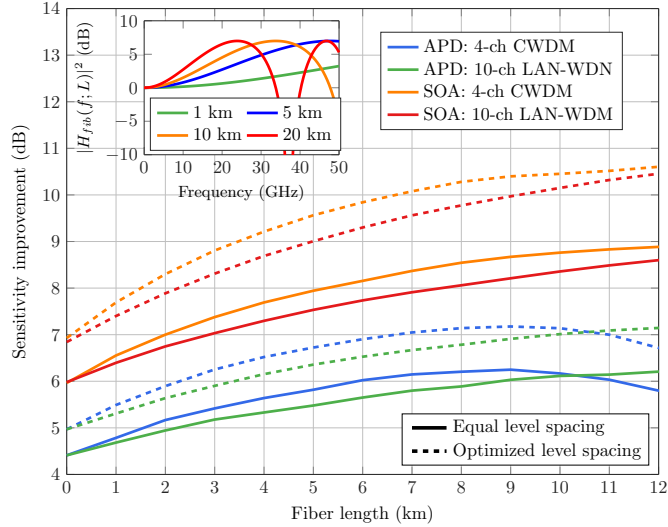

Figure 3.7: Receiver sensitivity improvement versus fiber length for 4-PAM. SOA curves from [31] are shown for comparison. The APD parameters assumed are listed in Table 3.2.

As discussed in Section 2.1, CD in intensity-modualated direct-detected (IM-DD) systems leads to power fading. When the modulator introduces transient chirp, the power fading is mitigated by satisfying $\alpha D(\lambda) < 0$, so that the fiber small-signal frequency response provides some gain, which can potentially compensate for the limited bandwidth of the modulator or APD. Therefore, we assume that all WDM channels are placed at wavelengths shorter than the zero-dispersion wavelength (1310 nm for standard SMF), such that $\alpha D(\lambda) < 0$ is satisfied for modulators with positive chirp.

Fig. 3.7 shows the receiver sensitivity improvement versus the optical fiber length for both WDM links. The Monte Carlo simulation parameters are shown in Table 3.2. For comparison, Fig. 3.7 also includes the curves of SOA-based systems from [31]. For each WDM system, we only show the performance at the wavelength subject to the highest dispersion (1250 nm for CWDM and 1270 nm for LAN-WDM, assuming one of the channels is at the zero-dispersion wavelength). The $x$-axis of Fig. 3.7 can be simply scaled to evaluate the performance of a channel at a different wavelength.

For both WDM links, as shown in Fig. 3.7, SOA-based receiver outperforms their APD-based counterparts, mainly due to the limited responsivity of the APDs. Another observation in Fig. 3.7 is that for short link lengths, the combined effect of transient chirp and CD actually improve the sensitivity of both types of receivers. This is due to the gain in the dispersion-induced frequency response of the channel, which partially compensates the LE noise enhancement penalty. For a 5-km link, the receiver sensitivity is improved by about 8 dB and 6 dB for SOA- and APD-based CWDM receivers respectively with equally spaced levels. The improvement is less significant for LAN-WDM receivers as these systems experience less dispersion. For longer link lengths, however, modulator chirp and CD cause severe nonlinear distortion that cannot be compensated by a LE.

Table 3.2: Simulation parameters for Monte Carlo simulation of APD-based system.

|          |                                         |                                |
|----------|-----------------------------------------|--------------------------------|
| System   | Bit rate ($R_b$)                        | 107 Gbit/s                     |
|          | PAM order                               | 4                              |
|          | Target BER                              | $1.8 \times 10^{-4}$           |
| Modulator| Bandwidth ($f_{3dB}$)                   | 30 GHz                         |
|          | Extinction ratio ($r_{ex}$)             | $-10$ dB                       |
|          | Relative intensity noise (RIN)          | $-150$ dB/Hz                   |
|          | Transient chirp ($\alpha$)              | 2                              |
| SMF      | Dispersion slope ($S_0$)                | 0.092 ps/(nm$^2$km)            |
|          | Zero-dispersion wavelength ($\lambda_0$)| 1310 nm                        |
| APD [45] | Responsivity ($R$)                      | 0.74 A/W                       |
|          | Low-gain bandwidth                      | 20 GHz                         |
|          | Dark current ($I_d$)                    | 100 nA @ $G_{\text{APD}} = 10$ |
| TIA      | Input-referred thermal noise ($I_n$)    | 30 pA/$\sqrt{\text{Hz}}$       |

## 3.4 Practical considerations

Temperature sensitivity and power consumption are important practical considerations in data center applications.

Temperature sensitivity in APDs is more manageable than in SOAs. Breakdown voltage variations over temperature can be compensated through active APD bias control. The breakdown voltage thermal coefficient is 70%/°C for InAlAs-based APDs, but only 0.05%/°C for Si-based APDs [51]. APD-based systems can operate over wide temperature ranges with small sensitivity variations, e.g., a commercial 10 Gbit/s receiver can operate over a 0°– 75°C range with only 1-dB penalty at $BER = 10^{-12}$ [52].

Compared to SOAs, APDs are low-power devices with power consumption of the order of 1 mW for typical values of input optical power $\bar{P}_{rx} \sim -15$ and bias voltage $V_{\text{bias}} \sim -25$ V. Commercial SOAs with TECs have power consumption of $\sim 1$ W [60, 61] while uncooled SOAs have power consumption of few hundreds of mW [60].

## 3.5 Summary

We have evaluated the performance of 4-PAM and 8-PAM 100 Gbit/s per-wavelength links using APDs to improve receiver sensitivity and using LE to compensate for ISI caused by the modulator and APD.

APD-based receivers may provide sensitivity improvements up to 4 dB for GBP = 100 GHz, and up to 6.2 dB for GBP = 300 GHz, assuming a low-gain bandwidth of 20 GHz, $k_A = 0.1$ (Si). APDs fabricated in InAlAs ($k_A = 0.2$) with otherwise the same characteristics may provide sensitivity improvements up to 5.3 dB. These sensitivity improvement values are with respect to an ideal thermal-noise limited PIN receiver with same responsivity. Unfortunately, however, current APDs still have responsivities below 1 A/W, which reduce the achievable sensitivity, e.g., by 3 dB for Ge-Si APDs having $R = 0.5$ A/W and by 1.3 dB for InGaAs-InAlAs having $R = 0.74$ A/W (Table 3.1). Hence, current resonant-cavity or waveguide InGaAs-InAlAs APDs offer better tradeoff between, responsivity, and GBP than Ge-Si APDs. As an example, the resonant-cavity InGaAs-InAlAs from [45] has a sensitivity improvement of 4.5 dB over the reference system.

Optimization of the PAM level spacing and receiver decision thresholds can provide about 1 to 2 dB additional sensitivity improvement for either SOA and APD-based receivers. Moreover, by appropriately selecting wavelengths such that $\alpha D(\lambda) < 0$, the receiver sensitivity is improved after a few km of SMF, due to the combined effect of CD and modulator chirp.

Practical SOA-based receivers offer sensitivity improvement of about 6 dB, which is higher than APD-based receivers due particularly to the poor responsivity of current APDs. Moreover, SOAs can amplify multiple WDM channels, helping amortize their higher cost and power dissipation. Important practical considerations such as cost, temperature sensitivity, and power consumption may nonetheless favor APDs in practical systems.

## Chapter 4

# Low-Power DSP-Free Coherent Receivers

Chapters 2 and 3 showed that four-level pulse amplitude modulation (4-PAM) systems currently adopted by the industry already face tight optical signal-to-noise ratio (OSNR) and power margin constraints in amplified and unamplified systems, respectively. This is concerning because next-generation interconnects will likely need to accommodate increased optical losses due to fiber plant, wavelength demultiplexing of more channels, and possibly optical switches. To alleviate some of these constraints, both mature and emerging technologies can help on a number of fronts. High-bandwidth, low-power modulators [62] will reduce intersymbol interference (ISI) and improve signal integrity. And as discussed in Chapter 3, avalanche photodiodes (APD) and semiconductor optical amplifiers (SOA) may improve receiver sensitivity of 100 Gbit/s 4-PAM systems by 4.5 and 6 dB, respectively. Improved laser frequency stability, either using athermal lasers [63] or frequency combs [64], will enable dense wavelength-division multiplexing (DWDM) within the data center, possibly yielding a multi-fold increase in capacity.

These technologies will extend the lifetime of 4-PAM, but they do not address the fundamental problem of such intensity-modulated direct-detections (IM-DD) systems, which is that they only exploit one degree of freedom of optical signals, namely, their intensity. Stokes vector detection has been proposed to enable up to three independent dimensions [65], while avoiding a local oscillator (LO) laser and coherent detection. Nonetheless, Stokes vector receivers rely on power-hungry analog-to-digital converters (ADCs) and digital signal processing (DSP) and do not address the problem of high required OSNR in amplified links or poor receiver sensitivity in unamplified links.

Coherent detection is more scalable, as it enables four degrees of freedom of the single-mode fiber (SMF), namely two quadratures in two polarizations, and improves sensitivity by up to 20 dB by mixing a weak signal with a strong local oscillator (LO) [66]. Coherent detection based on high-speed

Table 4.1: Impairments and constraints for intra- and inter-data center links.

| Application | Reach (km) | Wavelength (nm) | Wavelength multiplexing | Main impairments | Amp. | Priorities |
|---|---|---|---|---|---|---|
| Intra-data center | $\leq 10$ | 1310 | LAN-WDM, CWDM | CD | No | Power consumption, power margin, bit rate |
| Inter-data center | $\leq 100$ | 1550 | DWDM | CD | Yes | Bit rate, power consumption |
| Long-haul | $\leq 1000$s | 1550 | DWDM | PMD, CD, Nonlinearities | Yes | Bit rate, reach |

DSP is a mature technology in long-haul systems, but it may be currently unsuitable for data center links. Table 4.1 summarizes the different constraints and impairments of intra- and inter-data center, in contrast with long-haul systems. In long-haul systems, the high cost and power consumption of complex transceiver designs are amortized by extending the maximum reach. For instance, a 3-dB improvement in receiver sensitivity may double the reach and nearly halve the number of required repeaters, thus substantially reducing the overall cost of the system. Data center links, however, have other design priorities such as transceiver cost, power consumption, and port density, and they face fewer propagation impairments, as polarization mode dispersion (PMD) and Kerr nonlinearity are typically negligible over short propagation distances.

These fundamental differences may favor low-power architectures based on analog signal processing that avoid high-speed ADCs and DSP altogether. DSP-based coherent receivers optimized for short-reach applications [67, 68] will inevitably require high-speed ADCs and DSP for basic operations such as polarization demultiplexing, carrier recovery (CR), and timing recovery, which, combined, consume roughly 17 W in 40-nm complementary metal-oxide semiconductor (CMOS) for a 100 Gbit/s dual-polarization (DP) quaternary phase-shift keying (QPSK) receiver [44].

In this chapter, we propose and evaluate homodyne DSP-free coherent receiver architectures for dual-polarization quadrature phase-shift keying (DP-QPSK). This study was done in collaboration with Dr. Anujit Shastri, who proposed polarization demultiplexing based on optical phase shifters that are controlled by low-frequency marker tone detection circuitry. CR is based on either an optical or an electrical phase-locked loop (PLL). We propose a novel multiplier-free phase detector based on exclusive-OR (XOR) gates. We also study the relative performance of homodyne DP-differential QPSK (DP-DQPSK), whereby information is encoded in phase transitions, hence avoiding CR circuitry.

The estimated power consumption of the high-speed analog electronics of our most power-hungry architecture is nearly 4 W for 200 Gbit/s DP-QPSK, assuming 90-nm CMOS. Moreover, near zero chromatic dispersion (CD), the proposed DSP-free systems exhibit $\sim 1$ dB power penalty compared to their DSP-based counterparts. The DSP-based receiver used as a benchmark employs a newly proposed $2 \times 2$ multiple-input multiple-output (MIMO) equalizer based on a small-differential group

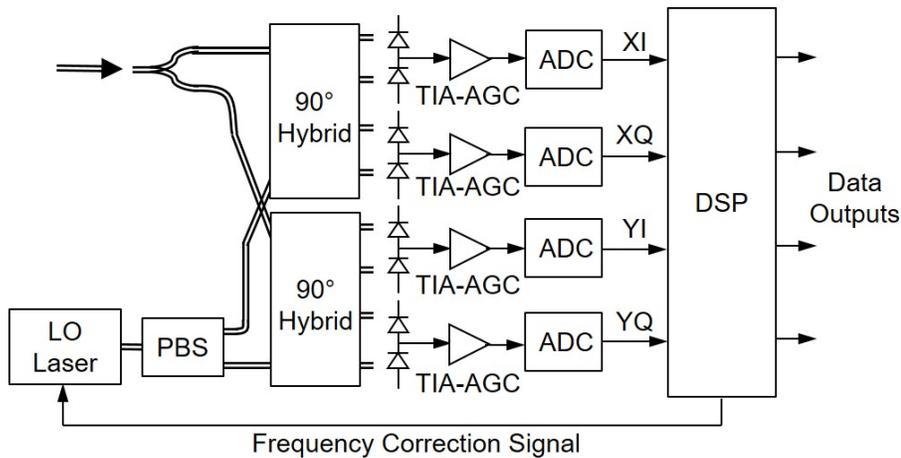

Figure 4.1: Block diagram of a DSP-based coherent receiver. Acronyms: local oscillator (LO), polarization beam splitter (PBS), transimpedance amplifier (TIA), automatic gain control (AGC), analog-to-digital converter (ADC), digital signal processor (DSP).

delay (DGD) approximation, halving the number of required real operations.

The remainder of this chapter is organized as follows. In Section 4.1, we start by reviewing DSP-based coherent receivers used as the benchmark to our proposed DSP-free receivers. In Section 4.2 present the proposed architecture for a DP-QPSK receiver based on analog signal processing and describe polarization demultiplexing, CR, and a startup protocol. In Section 4.3, we present a homodyne DP-DQPSK receiver architecture that does not require CR. Section 4.4 compares the performance of these different analog receivers to a simplified DSP-based receiver. Section 4.5 compares the complexity and power consumption of the different receiver architectures proposed. Section 4.6 summarizes the main conclusions of this chapter.

## 4.1 DSP-based coherent receiver (DP-$M$-QAM)

Coherent detection based on high-speed DSP is a mature technology in long-haul systems, but it may be currently unsuitable for data center links, where cost and power consumption are paramount. DSP-based coherent solutions may eventually become viable for short-reach applications by leveraging more power-efficient CMOS processes and optimized implementations for short-reach applications, where fiber impairments are less severe.

Fig. 4.1 shows a typical implementation of a dual-polarization DSP-based coherent receiver. The incoming optical signal is split and combined with orthogonal polarizations of the LO laser in two independent 90° hybrids. After balanced photodetection, transimpedance amplifiers (TIAs) with automatic gain control (AGC), and low-pass filtering (LPF) to minimize noise and aliasing, the four outputs are sampled by high-speed ADCs. The DSP stage performs functions such as polarization

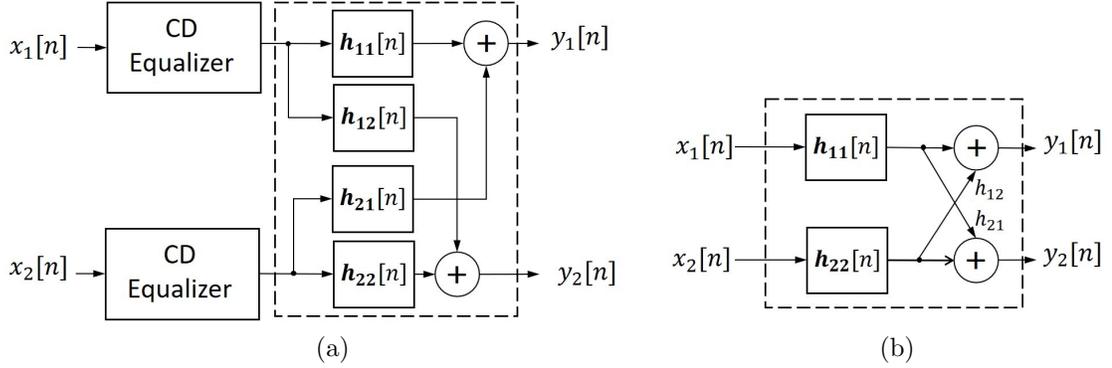

Figure 4.2: Block diagram of (a) CD and $2 \times 2$ MIMO equalizers used in conventional coherent receivers, and (b) simplified equalizer for short-reach applications assuming small-CD and small-DGD approximation.

Table 4.2: Update equations using CMA or LMS algorithm for the simplified polarization demultiplexer.

| Algorithm | Error measure | Update equations |
|---|---|---|
| CMA | $e_1[n] = 2 - \|\|y_1[n]\|\|^2$ | $\boldsymbol{h_{11}} \leftarrow \boldsymbol{h_{11}} + \mu e_1[n] y_1[n] \boldsymbol{x_1^*}$ <br> $h_{12} \leftarrow h_{12} + \mu e_1[n] y_1[n] \boldsymbol{h_{11}^H} \boldsymbol{x_1^*}$ |
| LMS | $e_1[n] = y_1[n] - \left[y_1[n]\right]_D$ | $\boldsymbol{h_{11}} \leftarrow \boldsymbol{h_{11}} - 2\mu e_1[n] \boldsymbol{x_1^*}$ <br> $h_{12} \leftarrow h_{12} - 2\mu e_1[n] y_1[n] \boldsymbol{h_{11}^H} \boldsymbol{x_1^*}$ |

Variables in boldface are vectors, $[\cdot]_D$ denotes the decision operator for a QAM symbol, $\boldsymbol{x^*}$ denotes element-wise complex conjugate and $\boldsymbol{x^H}$ denotes the Hermitian (transpose conjugate) of a vector.

demultiplexing, PMD compensation, CD compensation, carrier recovery and clock recovery. Some implementations place the DSP chip on the line card itself with an analog interface to the pluggable transceivers, referred to as analog coherent optics (ACO). While this can increase transceiver port density, it essentially offloads the power consumption to elsewhere in the system.

The power consumption of the various operations performed by the receiver was extensively studied in [44]. The most power-hungry operations are CD equalization and polarization demultiplexing with PMD compensation, which together amount to roughly 55% of the receiver power consumption [44]. Fig. 4.2a shows the block diagram of CD equalization and polarization demultiplexing with PMD compensation stages typically used in long-haul systems. First, CD equalization is performed using nearly static frequency-domain equalizers with hundreds of taps. This is followed by a $2 \times 2$ MIMO equalizer comprised of filters with typically less than 15 taps that are updated frequently to mitigate PMD and track changes in the received state of polarization [67].

The CD equalizers may be omitted if CD is small enough such that the filters in the $2 \times 2$ MIMO equalizer can compensate for it. Moreover, if the skew between the two polarizations is much smaller than the sampling rate, the coefficients of filter $\boldsymbol{h_{11}}$ are approximately proportional to those of $\boldsymbol{h_{12}}$, and similarly for filters $\boldsymbol{h_{21}}$ and $\boldsymbol{h_{22}}$. Hence, we can simplify the $2 \times 2$ MIMO as shown in Fig. 4.2b, which nearly halves the require number of DSP operations compared to the $2 \times 2$ MIMO equalizer in Fig. 4.2a. The filters $\boldsymbol{h_{11}}$ and $\boldsymbol{h_{22}}$ mitigate ISI caused by CD, PMD, and component bandwidth limitations. The cross terms $h_{12}$ and $h_{21}$ remove the Y component from X and vice-versa. Filter coefficient update equations using either least-mean squares (LMS) or constant-modulus amplitude (CMA) algorithms are given in Table 4.2. This simplification only holds when the mean differential group delay (DGD) between the two polarizations is much smaller than the sampling rate, so that the two polarizations appear synchronized at the receiver. Assuming a sampling rate of 70 GS/s (oversampling ratio of 5/4 at 56 Gbaud), and PMD of 0.1 ps/$\sqrt{\text{km}}$, the small-DGD approximation holds up to $\sim 200$ km.

To simplify the complexity of the CD equalizers, Martins et al. [69] have proposed a distributive finite-impulse response (FIR) equalizer that leverages the high multiplicity of the quantized FIR filter coefficients to sharply reduce the number of required operations. Compared to a conventional frequency-domain CD equalizer, their distributive FIR equalizer requires 99% fewer multiplications and 30% fewer additions [69].

Assuming that ISI is effectively mitigated and that phase error after carrier recovery is negligible, the BER for square $M$-QAM signals is approximately

$$\text{BER} \approx \frac{4}{\log_2 M} \frac{\sqrt{M}-1}{\sqrt{M}} Q\Big(\sqrt{\frac{3\log_2 M}{M-1}\text{SNR}}\Big). \tag{4.1}$$

In unamplified systems, the receiver noise is dominated by shot-noise due to the strong LO laser signal, while in amplified systems the ASE noise is dominant. The SNR for each of these scenarios is given by

$$\text{SNR} = \begin{cases} \frac{R\bar{P}_{rx}}{4q\Delta f}, & \text{shot-noise limited} \\ \frac{R\bar{P}_{rx}}{2N_A n_{sp}h\nu\Delta f}, & \text{ASE-limited} \end{cases}, \tag{4.2}$$

where $\bar{P}_{rx}$ is the average received optical power, $R$ is the photodiodes responsivity, $q$ is the electron charge, $h$ is Planck's constant, $\nu$ is the optical signal frequency, $N_A$ is the number of amplifiers, and $\Delta f$ is the receiver equivalent noise bandwidth, which in a ideal receiver would be $\Delta f = R_s/2$, where $R_s$ is the symbol rate. Note that a 1-dB penalty in SNR corresponds to a 1-dB penalty in the receiver sensitivity. In DSP-based systems, the combination of anti-aliasing filtering followed by fractionally spaced adaptive equalization achieves similar performance to the optimal receiver consisting of analog matched filtering and symbol-rate equalizer. In this case, $\Delta f \approx R_s/2$. The difference between $\Delta f$ and $R_s/2$ corresponds to the noise enhancement penalty. For DSP-free receivers, discussed in the following Section, the noise bandwidth is determined solely by the receiver LPF.

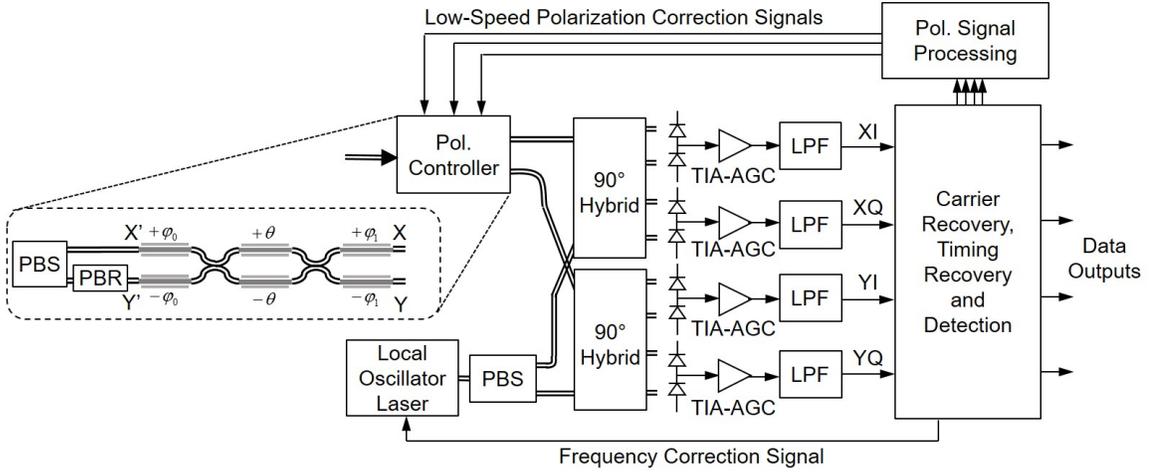

Figure 4.3: Block diagram of DP-QPSK receiver based on analog signal processing. The polarization controller is composed of optical phase shifters detailed in Fig. 4.5. The block diagram corresponding to carrier recovery, timing recovery and detection is detailed in Fig. 4.4 for carrier recovery based on EPLL and OPLL. This diagram is also used for DP-DQPSK, but the polarization controller and carrier recovery blocks are replaced by those shown in Fig. 4.10. Acronyms: polarization beam splitter (PBS), polarization-beam rotator (PBR), trans-impedance amplifier with automatic gain control (TIA-AGC), and low-pass filter (LPF).

## 4.2 DSP-free coherent receiver (DP-QPSK)

Coherent detection using analog signal processing was studied extensively in the 1980s and early 1990s [70], but the advent of the EDFA and later DSP-based coherent detection diminished its popularity.

Fig. 4.3 shows the proposed implementation of a DSP-free coherent receiver for DP-QPSK signals. Polarization demultiplexing is performed by optical phase shifters that are controlled by low-speed circuitry. The polarization controller, shown by the inset in Fig. 4.3, must recover the transmitted state of polarization by inverting the fiber polarization transfer matrix. Three cascaded phase shifter pairs can perform any arbitrary polarization rotation [71].

After balanced photodetection, TIAs with AGC, and low-pass filtering (LPF) to reduce noise, the signals reach the high-speed analog electronics stage, where CR, timing recovery and detection are performed. Timing recovery and detection may be realized using conventional clock and data recovery (CDR) techniques [72]; thus, we do not discuss them further herein. Polarization recovery and CR are performed using only analog waveforms and do not depend on timing information. The high-speed analog electronics stage is detailed in Fig. 4.4 for CR based on optical PLL (OPLL) and electrical PLL (EPLL).

In an OPLL (Fig. 4.4a), the LO laser is frequency-modulated by the frequency correction signal generated by the CR stage. Hence, an OPLL requires a LO laser with wideband frequency modulation (FM) response and short propagation delay in the LO path to minimize the overall loop delay. Minimizing the loop delay is one of the main challenges in OPLL design, since the loop includes the LO laser, 90° hybrid, photodiodes, and all the subsequent electronics in CR, which may not be realized within the same chip. Notably, Park et al have demonstrated loop delays of only 120 ps for a highly integrated 40 Gbit/s binary PSK (BPSK) coherent receiver [73].

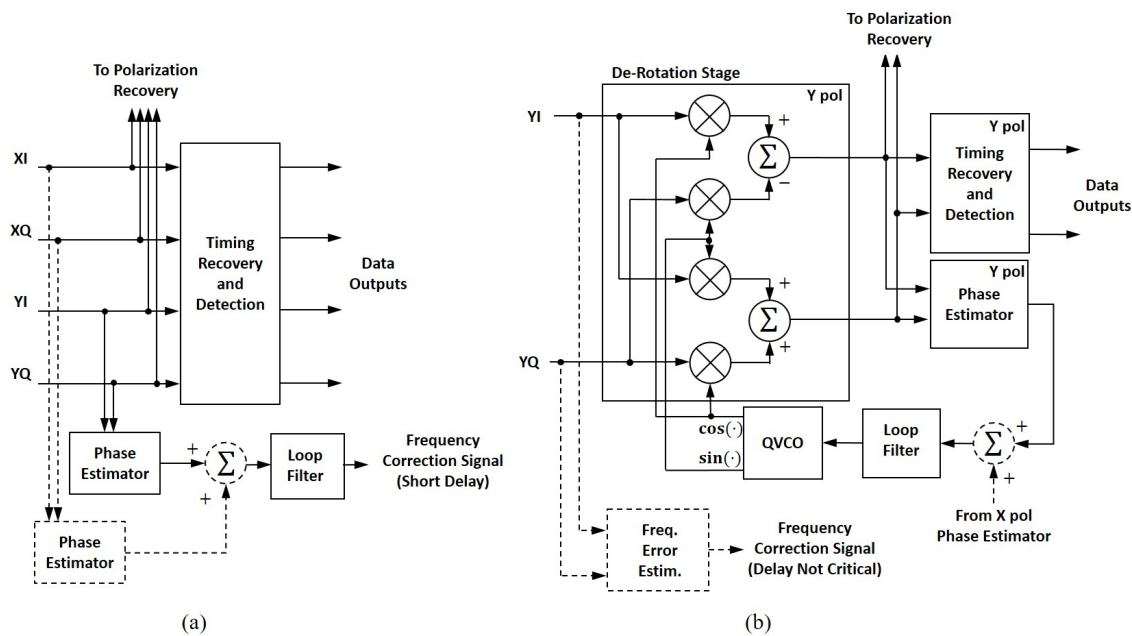

Figure 4.4: Block diagram of carrier recovery based on (a) OPLL and (b) EPLL (shown for one polarization only). Phase estimates in the two polarizations may be optionally combined in the adder depicted in dashed lines in both diagrams. In an OPLL implementation, the delay of the frequency correction signal must be as short as possible, which means that the LO laser must be physically close to that output. Note that an EPLL implementation requires a de-rotation stage (single-sideband mixer) in each polarization, since the transmitter and LO lasers are not phase locked. However, only one quadrature VCO (QVCO) and loop filter are necessary. The EPLL implementation may also require a frequency error estimator if the laser frequency drift exceeds the VCO frequency range. The phase estimator block diagram is detailed in Fig. 4.6.

An EPLL (Fig. 4.4b) implementation eliminates requirements on LO laser FM response and on propagation delay at the cost of more complex analog electronics. Specifically, an EPLL requires a single-side band mixer in each polarization to de-rotate the incoming signals (see Fig. 4.4b), since the transmitter and LO lasers are not phase locked. Additionally, the frequency offset between transmitter and LO lasers must always be within the lock-in and hold-in ranges of the EPLL, which are typically limited by the tuning range of the voltage-controlled oscillator (VCO) [74]. The VCO tuning range can be on the order of several GHz (e.g., 11.8 GHz for a ring oscillator VCO [75]).

The constraint on frequency offset can be satisfied by strict laser temperature control, whose cost and power consumption could be shared among several channels by using frequency combs [64] for both the transmitter and LO. Alternatively, a frequency error estimation stage (Fig. 4.4b), based on relatively simple frequency discriminator circuitry [76], may be used to keep the LO laser frequency sufficiently close to the transmitter laser.

We restrict our analysis to the feedback CR techniques OPLL and EPLL, which are governed by the same underlying theory, as described in Section 4.2.2. Feedforward CR (FFCR) has been widely used in DSP-based coherent receivers [77], and it is also feasible in analog signal processing [78]. However, analog FFCR has several implementation drawbacks. First, phase estimation in analog FFCR is limited to non-data-aided (NDA) methods, e.g., raising the signal to $M$th power (for M-PSK), which have poorer performance than decision-directed methods [79] and restrict modulation to PSK. Second, compared to feedback techniques, FFCR requires more complex analog circuitry to implement an $M$th-power operation and frequency division. Furthermore, analog FFCR would offer virtually no improvement over EPLL, since commercial distributed feedback (DFB) lasers already have narrow linewidths on the order of 300 kHz [80], and the loop delay in an EPLL is very small, as the loop can be realized within a single chip.

### 4.2.1 Polarization demultiplexing

In DSP-based coherent receivers, a $2 \times 2$ MIMO equalizer performs polarization demultiplexing and compensates for PMD and polarization-dependent loss (PDL) [81].

Fortunately, PMD effects are negligible up to 80 km at 56 Gbaud with modern standard SMF [82]. With PDL causing only small power penalties at these distances, polarization rotation becomes the only impairment that needs to be compensated. Polarization rotation through a fiber varies on a time scale of the order of milliseconds [83], becoming slower on shorter link lengths [84], and can be compensated at the receiver by an optical polarization controller driven by low-speed ($< 100$ kHz) circuitry.

For the DSP-free receiver we propose polarization rotation compensation by cascaded phase shifters controlled by low-speed circuitry. Fig. 4.5 shows the block diagram of this system. A low-frequency ($< 50$ kHz) marker tone is added to one of the tributaries at the transmitter, e.g., in-phase component of the X polarization (XI). After propagation through the fiber, the received state of polarization is unknown and consequently the marker tone will be detected in all tributaries XI, XQ, YI, and YQ. The polarization controller at the receiver sequentially adjusts the individual phase shifts of each phase shifter to minimize the presence of the marker tone on the other tributaries XQ, YI, and YQ. Thus, maximizing the marker tone in the XI tributary and inverting the polarization rotations caused by the fiber. As a result, the polarizations are demultiplexed into the signals transmitted on the X and Y polarizations at the two output ports of the polarization controller, at which point they are guided to the 90° hybrid.

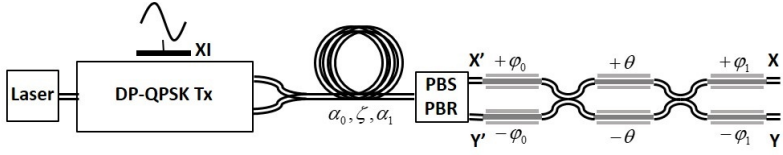

Figure 4.5: Schematic diagram of polarization recovery. A marker tone is added to the in-phase tributary of the X polarization at the transmitter. Propagation through the fiber causes random polarization rotation, thus the received state of polarization is unknown. The individual phase shifts of the three cascaded phase shifters in the polarization controller are adjusted to minimize the marker tone in the other tributaries (XQ, YI, and YQ), thus compensating for the fiber polarization rotation. Image credit: Anujit Shastri [85].

Further details about the phase shifters and phase tuning algorithm is given in [85]. Importantly, we discuss how to achieve endless phase excursion, despite individual phase excursion limits of each phase shifter. This can be realized by cascading a fourth phase shifter in Fig. 4.5, or by periodically resetting the relative phase shifts of each phase shifter. Resetting will cause burst errors during the switching period. For phase shifting speeds on the order of 1 ns for $\pi$ phase shifts, typical of lithium niobate phase shifters used for high-speed data modulation [86], the burst errors can be corrected by 7% FEC with current interleaving standards at 56 Gbaud [87]. With phase shifting speeds on the order of 1 $\mu$s phase shifts, typical of Silicon photonics phase shifters tuned thermally [88], additional buffering of $\sim$200 kbits would be required at 56 Gbaud, increasing latency on the order of the shifting time.

### 4.2.2 Carrier recovery

CR architectures based on an OPLL or an EPLL consist of three basic stages: phase estimator, loop filter, and oscillator. The oscillator is the LO laser in an OPLL, and an electronic VCO in an EPLL. The phase estimator stage wipes off the modulated data in order to estimate the phase error, which is then filtered by the loop filter, producing a control signal for the oscillator frequency. We consider a second-order loop filter [74] whose Laplace transform is given by

$$F(s) = 2\zeta\omega_n + \omega_n^2/s, \qquad (4.3)$$

where $\zeta$ is the damping coefficient, typically chosen to be $1/\sqrt{2}$ as a compromise between fast response and small overshoot. Here, $\omega_n = 2\pi f_n$ is the loop natural frequency, which must be optimized to minimize the phase error variance. A second-order loop filter is typically preferred, as it has ideally infinite d.c. gain, resulting in zero steady-state error for a frequency step input.

Fig. 4.6. Block diagram of carrier phase estimators for QPSK inputs based on (a) Costas loop and (b) a multiplier-free approach based on XORs.

Fig. 4.6 shows two possible implementations of a phase estimator for QPSK inputs. Fig. 4.6a shows the block diagram of a conventional Costas loop [79], which requires two linear and wideband analog multipliers per polarization. We propose a novel multiplier-free phase detector based on XOR gates, as shown in Fig. 4.6b. Multiplier-free Costas loop alternatives based on XOR gates have been proposed for BPSK [89] and for QPSK [90]. The latter relies on precisely delaying and adding the in-phase and quadrature components prior to the XOR operation. Using simple operations, our proposed phase detector estimates the sign of the phase error rather than its actual value. When XI and XQ form a QPSK signal, the output of the second XOR $O_{\text{XOR 2}}$ reduces to the sign of phase error: $O_{\text{XOR 2}}(t) = \text{sgn}(\phi_e(t))$. After loop filtering and negative feedback, this output counteracts the phase error. When the loop has made the phase error small, $O_{\text{XOR 2}}$ oscillates very rapidly, but these fast oscillations are virtually eliminated after low-pass filtering by the loop filter.

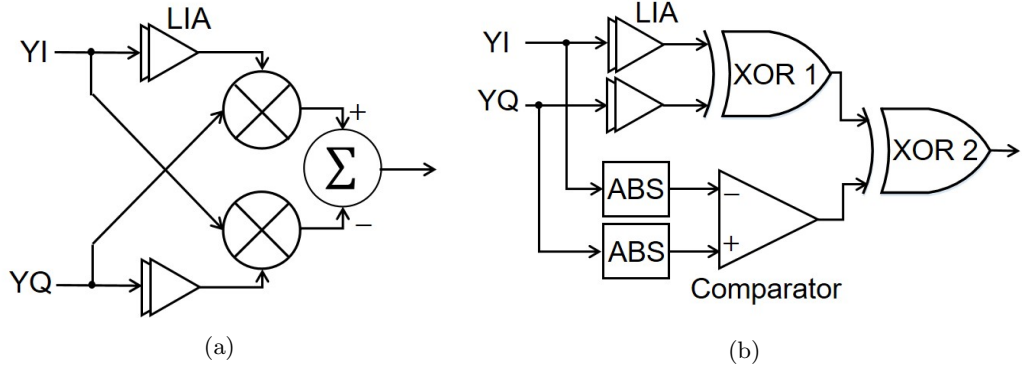

Figure 4.6: Block diagram of carrier phase estimators for QPSK inputs based on (a) Costas loop and (b) a multiplier-free approach based on XORs. LIA denotes limiting amplifier, and ABS denotes full-wave rectifier. Though not explicitly shown, the comparator may be clocked in order to facilitate circuit design.

Fig. 4.7 shows an equivalent block diagram of Costas and XOR-based loops of Fig. 4.6. They differ only in the nonlinear characteristic within the loop. While the Costas loop nonlinear function is simply $\sin(\cdot)$, for the XOR-based loop it is $\text{sgn}\sin(\cdot)$. The delay accounts for lumped and distributed delays of components and signal paths in the EPLL or OPLL.

Similarly to [91], we use the small-signal approximation to linearize the loop transfer function in Fig. 4.7 and obtain the phase error variance:

$$\begin{aligned}\sigma_e^2 =& \Delta\nu_{tot} \int_{-\infty}^{\infty} |j\omega + e^{-j\omega\tau_d}F(j\omega)|^{-2} d\omega \\ &+ 2(2\pi)^2 k_f \int_0^{\infty} |\omega|^{-1} |j\omega + e^{-j\omega\tau_d}F(j\omega)|^{-2} d\omega \\ &+ \frac{T_s}{2N_{PE}\text{SNR}} \frac{1}{2\pi} \int_{-\infty}^{\infty} \left|\frac{F(j\omega)}{j\omega + e^{-j\omega\tau_d}F(j\omega)}\right|^2 d\omega,\end{aligned} \quad (4.4)$$

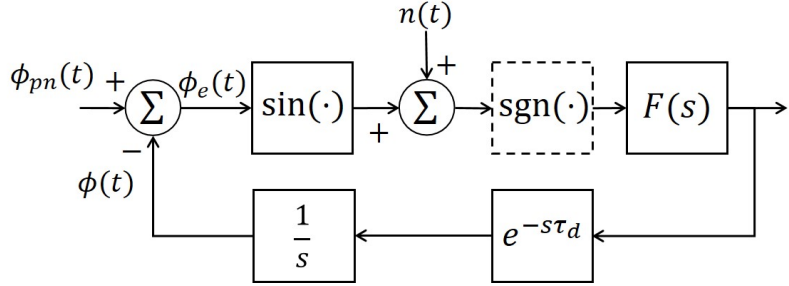

Figure 4.7: Equivalent block diagram for Costas loop, without sign operation sgn(·), and XOR-based loop including sgn(·).

where $\Delta\nu_{tot}$ denotes the sum of the transmitter laser and LO laser linewidths, $k_f$ characterizes the magnitude of flicker noise [92], $T_s$ is the symbol time, and SNR is the signal-to-noise ratio (SNR). $N_{PE} = 1$, if phase estimation is performed using only one polarization, and $N_{PE} = 2$, if phase estimation is performed in both polarizations and summed, as illustrated in Fig. 4.4ab. The terms in (4.4) account for phase error contribution due to the intrinsic laser phase noise caused by spontaneous emission, flicker noise and additive white Gaussian noise (AWGN), respectively. The loop filter, and in particular $f_n$, should be optimized to minimize (4.4).

It is important to highlight that $\Delta\nu_{tot}$ refers to the intrinsic laser linewidth due to spontaneous emission. Low-frequency flicker noise caused by electrical noise in the tuning sections of tunable lasers may lead to an apparent broader linewidth. Indeed, as reported in [93], a typical sampled grating (SG) distributed Bragg reflector (DBR) laser with linewidth below 1 MHz had an apparent linewidth ranging from 10 to 50 MHz. However, as indicated in (4.4), the flicker noise component on the phase error variance is smaller than intrinsic phase noise component, since the flicker noise term integral decays with an additional $|\omega|^{-1}$ factor. Not considering this effect would lead to a suboptimal choice of $f_n$.

The SNR is given in (4.2) and it depends on whether the receiver is shot-noise limited, e.g., in unamplified intra-data center links, or ASE-limited, e.g., in amplified inter-data center links:

As shown in [94], the bit error probability of a PSK signal with phase error distributed according to $\mathcal{N}(0, \sigma_e^2)$ is

$$\text{BER} = Q(\sqrt{2\text{SNR}}) + \sum_{l=0}^{\infty}(-1)^l H_l\left(1 - \cos((2l+1)\frac{\pi}{4}\right)\exp\left(-\frac{(2l+1)^2\sigma_e^2}{2}\right) \quad (4.5)$$

where $\sigma_e^2$ is given by (4.4) and

$$H_l = \frac{\sqrt{\text{SNR}}e^{-\text{SNR}/2}}{\sqrt{\pi}(2l+1)}\left(I_l\left(\frac{\text{SNR}}{2}\right) + I_{l+1}\left(\frac{\text{SNR}}{2}\right)\right) \geq 0, \quad (4.6)$$

where $I_l(x)$ is the modified Bessel function of the first kind and order $l$. Using equations (4.4)–(4.6),

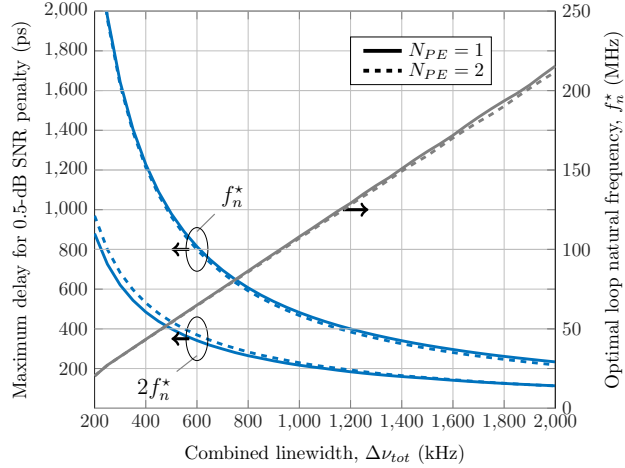

Figure 4.8: Maximum loop delay for 0.5-dB SNR penalty as a function of the combined linewidth. Curves are shown for loop natural frequency optimized at every point, and when loop natural frequency is twice the optimal.

we can compute the receiver sensitivity penalty as a function of $f_n$, $\tau_d$ and $\Delta\nu_{tot}$. Fig. 4.8 shows the maximum delay for a 0.5-dB SNR penalty as a function of the combined linewidth for $N_{PE} = 1, 2$ with respect to a system with no phase noise. The loop natural frequency is optimized at each point. The maximum delay is significantly reduced at wider linewidths or when the natural frequency is suboptimal.

An example of this is shown in Fig. 4.8 by the curve where the natural frequency is twice the optimal. Interestingly, there is virtually no penalty for using only one of the polarizations for phase estimation in CR, as the optimal value of $f_n$ is reached when the phase noise component in (4.4) is dominant. Fig. 4.8 assumes $k_f = 1.7 \times 10^{10}$ Hz$^2$, which is typical of DFB lasers [80], but for $k_f = 3.4 \times 10^{11}$ Hz$^2$, observed in digital supermode DBR (DS-DBR) lasers [80], the flicker noise effects become significant for $\Delta\nu_{tot} < 1$ MHz.

Although (4.4) was derived using the small-signal approximation for the Costas loop, the performance of the XOR-based loop is similar to the Costas loop for the same loop filter parameters optimized using (4.4)–(4.5). Fig. 4.9 compares the performance of Costas and XOR-based loops as a function of the combined linewidth. The analysis curves were obtained using equations (4.4)–(4.6), while the curves for Costas loop and XOR-based loop were obtained through Monte Carlo simulations. Interestingly, although the XOR-based loop does not allow a small-signal approximation to be made in analysis, its performance is very similar to the Costas loop. They differ by less than 0.5 dB for $N_{PE} = 1, 2$.

Both Costas and XOR-based phase estimators exhibit a 90° phase ambiguity. This ambiguity is typically resolved by either transmitting a known training sequence at the beginning of transmission, or by differentially decoding the bits. Although differentially decoding the bits doubles the bit-error ratio (BER) [95], near the FEC threshold this corresponds to less than 0.5 dB SNR penalty. Moreover, using a training sequence would require retraining whenever there is a cycle slip. If the bits are differentially decoded, however, a cycle slip only causes a few more error events that could be corrected by the FEC.

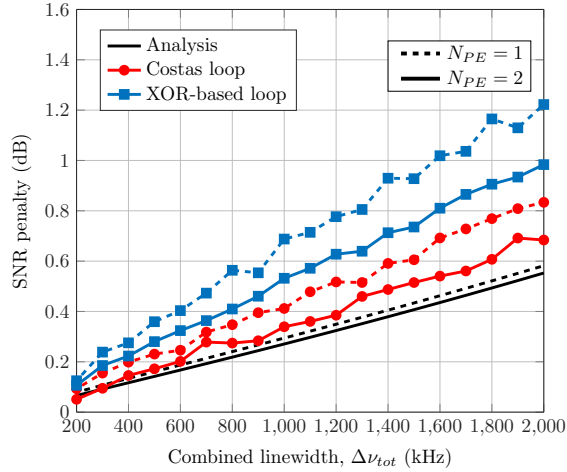

Figure 4.9: Comparison of SNR penalty vs combined linewidth for Costas loop and XOR-based loop. Simulation curves include thermal noise and ISI penalties, while theory curves do not.

### 4.2.3 Proposed startup protocol

At startup, the receiver cannot perform polarization demultiplexing and CR simultaneously. For instance, marker tone detection is only possible after CR, so that the marker tone is at the expected frequency. CR, in turn, requires that the received signals in each polarization branch must be QPSK, which is not the case for any given received state of polarization. To circumvent these problems, we have devised a startup protocol, which can also be used to recover from a continuous loss of the marker tone in the relevant tributary caused by a discontinuous polarization change.

First, the transmitter sends the same data in both polarizations so that the received signal in each polarization branch is QPSK regardless of the received state of polarization. The transmitted sequence needs to be known at the receiver only if the bits are not differentially decoded, in which case a training sequence is required to resolve the 90° phase ambiguity. Once phase lock is acquired, the polarization estimation algorithm can adjust the phase shifters to demultiplex the two polarizations, as described in section II.A, with the marker tone now at the appropriate frequency. Once the polarizations have been demultiplexed, as determined by the polarization recovery processing detecting sufficiently low marker tone amplitudes in the XQ, YI and YQ tributaries, data transmission in both polarizations can start.

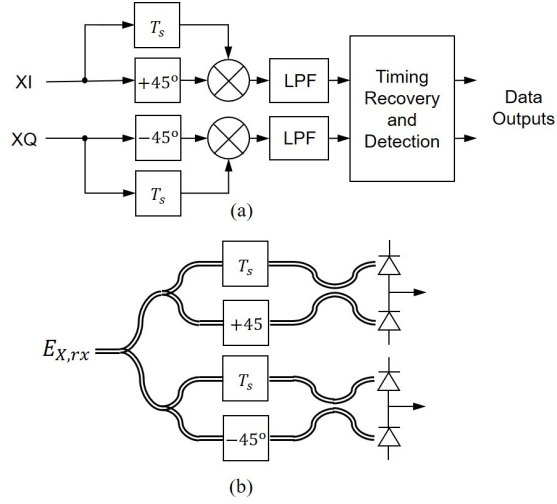

Figure 4.10: Block diagrams of differentially coherent detection methods (a) with a local oscillator and (b) without a local oscillator. The inputs to the differentially coherent detection method in (a) are XI and XQ from Fig. 4.3. Optical delay interferometers are used for (b).

## 4.3 DSP-free differentially coherent (DP-DQPSK)

In DQPSK transmission, the information is encoded in the phase transitions between two consecutive symbols. Hence, DQPSK detection does not require an absolute phase reference and CR is not necessary, which significantly simplifies the receiver. Homodyne DQPSK, however, has some disadvantages compared to homodyne QPSK. First, DQPSK has an inherent $\sim 2.4$ dB SNR penalty due to differential detection compared to coherent detection [30]. Second, differential detection restricts modulation to PSK, which limits its spectral efficiency compared to quadrature-amplitude modulation (QAM).

Differential detection may be performed in the electrical domain or in the optical domain. Fig. 4.10a shows one implementation of differentially coherent detection, whereby the phase difference between two symbols is realized in the electrical domain. The XI and XQ signals in this figure correspond to the XI and XQ in Fig. 4.3, in which a LO laser is used to perform homodyne detection.

The polarization controller shown in Fig. 4.3 would only need two phase shifters, as the residual phase difference between the two polarizations that is compensated for by the third phase shifter is no longer needed, since the two polarizations are detected separately. One method to control the phase shifters is to minimize the radio frequency (RF) PSD of the optical signal after the final phase shifter. Minimization of this value ensures demultiplexing of the polarizations [96].

Since the receiver does not perform carrier recovery, the frequency difference between the LO and transmitter laser may be large. The BER of homodyne $M$-DPSK in the presence of frequency

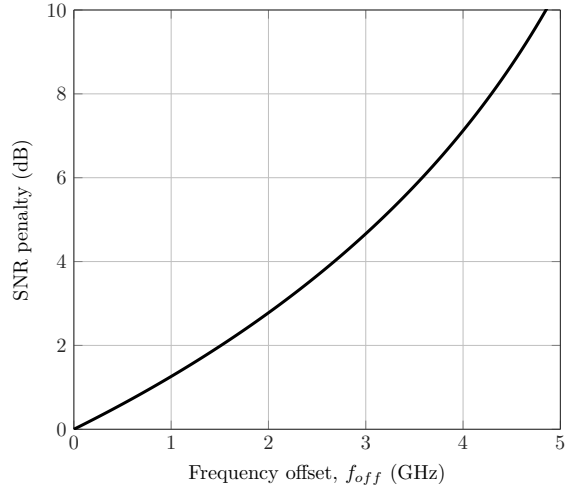

Figure 4.11: SNR penalty as a function of frequency offset between transmitter and LO lasers for a 224 Gbit/s DP-DPQSK system.

error is given by [97]:

$$\text{BER} = \frac{2}{\log_2 M}(F(\pi) - F(\pi/M)) \tag{4.7}$$

$$F(\varphi) = \frac{\text{SNR} \sin(\Delta\Psi - \varphi)}{4\pi} \int_{-\pi/2}^{\pi/2} \frac{\exp\left(-(\text{SNR} - \text{SNR}\cos(\Delta\Psi - \varphi)\cos t)\right)}{\text{SNR} - \text{SNR}\cos(\Delta\Psi - \varphi)\cos t} dt,$$

where $\Delta\Psi = 2\pi f_{off} T_s$ is the phase error due to frequency offset $f_{off}$ during a symbol period. As shown in [85], a 2-GHz frequency offset between transmitter and LO laser incurs nearly 3-dB SNR penalty.

Fig. 4.11 shows the SNR penalty as a function of the frequency offset. The SNR penalty grows roughly quadratically with frequency offset and reaches 3 dB at $f_{off} \approx 2$ GHz. As in the EPLL-based receivers discussed in Section II, frequency combs [64] at both the transmitter and LO can be used to amortize the high cost and power consumption of strict laser temperature control. Alternatively, frequency-locking techniques based on frequency discriminators can be employed [76] to minimize the frequency offset penalty.

The computation of the phase difference between two consecutive symbols may also be realized in the optical domain by using delay interferometers, as illustrated in Fig. 4.10b. The receiver electronics, in this case, must only perform timing recovery. This configuration does not employ a LO laser, which simplifies the receiver significantly. This architecture is particularly interesting for amplified links (e.g., inter-data center), where the LO gain is not critical. The delay caused by the delay interferometer is sensitive to the wavelength. As a result, the transmitter laser's frequency drifts can cause a penalty if not properly compensated by tuning the delay interferometer [98]. For DP-DQPSK, at 224 Gbit/s without delay interferometer tuning, a frequency drift of ±800 MHz would incur a 2-dB penalty. The BER for a DQPSK signal can be calculated from (4.7) by setting $M = 4$ and $\Delta\Psi = 0$.

## 4.4 Performance comparison

In this section, we compare the performance of the proposed receiver architectures based on analog signal processing with their DSP-based counterparts. In the DSP-based receiver, equalization and polarization demultiplexing are simplified, as discussed in Appendix II. CR is performed using the Viterbi-Viterbi method [99], a feedforward method that uses a simple averaging filter rather than the optimal Wiener filter [77].

We target a bit rate of 200 Gbit/s per wavelength, resulting in 224 Gbit/s after including 7% hard-decision FEC overhead [100], and 5% Ethernet overhead. As in previous Chapters, the FEC is assumed to be hard-decision RS(255, 239) or similar, which leads to a FEC threshold of $1.8 \times 10^{-4}$.

Fig. 4.12 shows the performance of various coherent and differentially coherent systems as a function of dispersion (or residual dispersion after optical dispersion compensation). The simulation parameters are shown in Table 4.3. The curves in Fig. 4.12 for DSP-based receivers are flat across dispersion values, as CD is effectively compensated by electronic equalization. DSP-based coherent detection systems can use higher-order modulation, such as 16-QAM, to reduce the bandwidth required of electro-optic components. For intra-data center links or inter-data center links with optical dispersion compensation, DSP-free solutions can significantly reduce power consumption.

At small dispersion, DSP-free exhibit a penalty with respect to their DSP-based counterparts due to imperfect receiver filtering. In our simulations, the LPF is a fifth-order Bessel filter with bandwidth of 39.2 GHz ($0.7R_s$ for 224 Gbit/s DP-QPSK), for which $\Delta f = 40.7$ GHz. Hence, the imperfect receiver filtering results in a 1.6 dB penalty compared to DSP-based receiver. As dispersion increases, the receiver sensitivity decreases or OSNR required increases sharply, since the receiver does not equalize CD. Nonetheless, the sensitivity would allow unamplified eye-safe systems near 1310 nm to reach up to 40 km. In fact, systems with 100 GHz wavelength spacing could support 49 channels with 5 dB of margin, and systems with 200 GHz wavelength spacing could support 25 channels with 8 dB of margin.

As shown by Fig. 4.12, DQPSK without an LO has significantly poorer receiver sensitivity in unamplified systems, such as intra-data center links. However, the OSNR required in amplified systems remains approximately the same as that of a LO-based DQPSK receiver. This makes LO-free DQPSK an attractive option for amplified inter-data center links that have optical CD compensation, as they have the lowest receiver complexity among coherent and differentially coherent receivers. Note that since the outputs of the balanced photodetection for differentially coherent detection without a LO laser are no longer linear in signal electric field values, CD and PMD cannot be equalized using DSP.

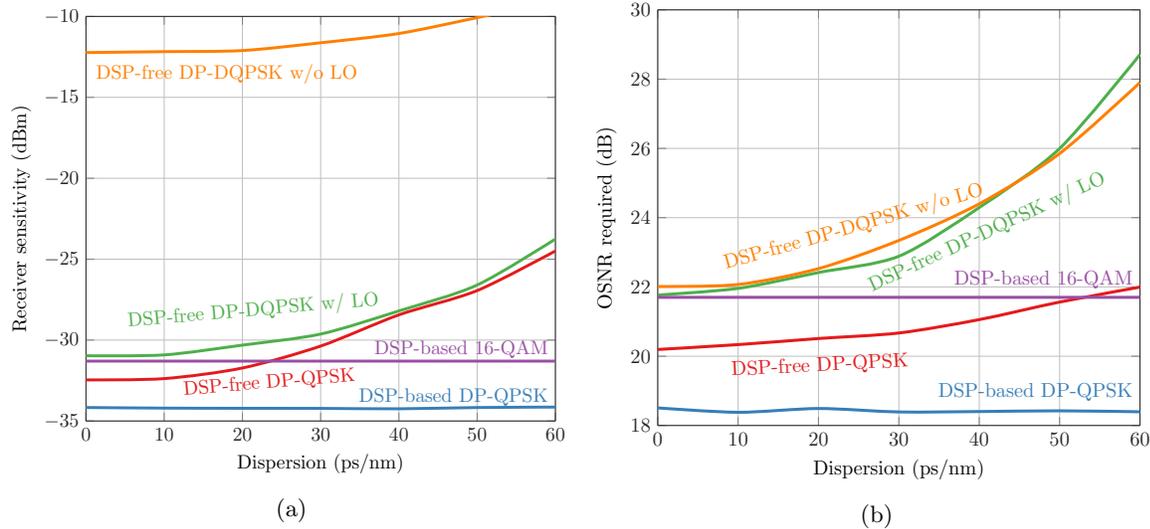

Figure 4.12: Comparison of performance of coherent detection schemes vs. dispersion at 224 Gbit/s. Unamplified systems are characterized in terms of (a) receiver sensitivity, while amplified systems are characterized in terms of (b) OSNR required. The $x$-axis may be interpreted as total dispersion in intra-data center links or residual dispersion after optical CD compensation in inter-data center links similarly to the resultin in Chapter 2.

## 4.5 Complexity comparison

The previous sections compared the performance of the various modulation formats and detection techniques in terms of receiver sensitivity and OSNR required. This section focuses on the overall complexity and power consumption of these schemes.

Table 4.4 summarizes the main complexity differences between the various schemes discussed in this paper. This comparison covers spectral efficiency, modulator type, complexity of the optical receiver, number of ADCs and their sampling rate and ENOB, capability to electronically compensate for CD, and DSP operations required at the receiver.

Fig. 4.13 shows a coarse estimate of power consumption in 28-nm CMOS for various modulation schemes at 200 Gbit/s. The power consumption of DSP-based techniques is estimated using the power consumption models presented in [44]. First, the number of real additions and real multiplications is counted for all DSP operations (summarized in Table 4.4). Then, the power consumption is obtained by computing how much energy a given operation consumes. For instance, a real addition in 28-nm CMOS with 6-bit precision consumes 0.28 pJ, while a real multiplication with 6-bit precision consumes 1.66 pJ [44]. The power consumption estimates for DACs and ADCs assume

Table 4.3: Coherent and differentially coherent systems simulation parameters. Monte Carlo simulations used $2^{17}$ symbols.

| | | |
|---|---|---|
| Tx | Bit rate ($R_b$) | 224 Gbit/s |
| | Target BER | $1.8 \times 10^{-4}$ |
| | Laser linewidth | 200 kHz |
| | Relative intensity noise | $-150$ dB/Hz |
| | Modulator bandwidth | 30 GHz |
| | Chirp parameter ($\alpha$) | 0 |
| | Extinction ratio ($r_{ex}$) | $-15$ dB |
| Rx | Photodiode responsivity ($R$) | 1 A/W |
| | TIA input-referred noise ($\sqrt{N_0}$) | 30 pA/$\sqrt{\text{Hz}}$ |
| Optical Amplifier | Gain ($G_{\text{AMP}}$) | 20 dB* |
| | Noise figure ($F_n$) | 5 dB |
| | Number of amplifiers ($N_A$) | 1 |
| LO Laser | Linewidth | 200 kHz |
| | Output power | 15 dBm |
| | Relative intensity noise | $-150$ dB/Hz |
| DSP | ADC effective resolution | 4 bits |
| | Oversampling rate ($r_{os}$) | 5/4 |
| | Equalizer number of taps ($N_{taps}$) | 7 |
| | Filter adaptation algorithm | CMA |
| Analog Carrier Recovery | Loop filter damping factor ($\xi$) | $\sqrt{2}/2$ |
| | Loop delay ($\tau_d$) | 213 ps |
| | Optimal natural frequency ($f^\star$) | 123 MHz |

* 30 dB for LO-free DP-DQPSK

that the power consumption scales linearly with resolution and sampling rate. The DAC figure of merit is 1.56 pJ/conv-step, while the ADC figure of merit is 2.5 pJ/conv-step [44]. The resolution of the DACs and ADCs, as well as the DSP arithmetic precision, is assumed equal to ENOB + 2, where ENOB is given in Table 4.4. For all cases, the oversampling ratio assumed is $r_{os} = 5/4$, even though Stokes vector receivers have only been reported with $r_{os} = 2$.

The DSP-free receiver power consumption is estimated at 90-nm CMOS as detailed below. Power consumption of the analog receiver is harder to estimate, since there is more variability in the choice of the functional block implementation and transistor technology. For instance, CMOS transistors would offer lower manufacturing costs, while bipolar transistors would offer improved linearity and lower power consumption. The most complex and power hungry parts of the proposed analog circuitry are analog mixers and XORs. Both can be realized using Gilbert cells [90, 101]. A 9-to-50-GHz Gilbert-Cell down-conversion mixer built in 130-nm CMOS had a total power consumption of 97 mW [102], while a 25–75 GHz broadband Gilbert-Cell mixer using 90-nm CMOS had a total power consumption of 93 mW [103]. Passive mixers would exhibit even lower power consumption. An EPLL implementation requires eight analog mixers, two XORs, four adders, two limiting amplifiers, two

Table 4.4: Complexity comparison of modulation schemes allowing more than one degree of freedom of the optical channel.

| Scheme | SE (b/s/Hz) | Mod. type | Optical receiver | ADC (GS/bit) | # ADCs / ENOB | Digital CD comp. | DSP operations |
|---|---|---|---|---|---|---|---|
| DSP-based DP-QPSK | 4 | DP I&Q | $2 \times 90°$ OH, LO, 4 PD | $0.25 r_{os}$ | 4 / 4 | High | EQ, $2 \times 2$ MIMO, CR |
| DSP-based DP-16-QAM | 8 | DP I&Q | $2 \times 90°$ OH, LO, 4 PD | $0.125 r_{os}$ | 4 / 5 | High | EQ, $2 \times 2$ MIMO, CR |
| DSP-free DP-QPSK | 4 | DP I&Q | $2 \times 90°$ OH, LO, 4 PD | NA | 0 | None | None |
| DSP-free DP-DQPSK | 4 | DP I&Q | $2 \times 90°$ OH, LO, 4 PD | NA | 0 | None | None |

Acronyms: spectral efficiency (SE), optical hybrid (H), photodiode (PD), time-domain equalizer (TD-EQ), frequency-domain equalizer (FD-EQ), phase estimation (PE), single-input single output (SISO), carrier recovery (CR), and not applicable (NA).

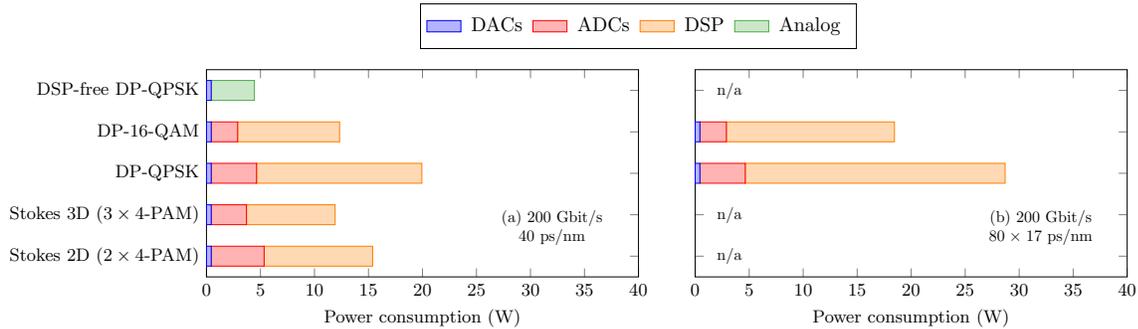

Figure 4.13: Coarse estimate of power consumption of high-speed DACs, ADCs, and DSP for various modulation schemes at 200 Gbit/s. DSP power consumption estimates are made for 28-nm CMOS using the models presented in [44]. DSP-free receiver power consumption is estimated for 90-nm CMOS [85]. The graphs to the left assume that CD is compensated optically and the residual CD is at most 80 ps/nm at 100 Gbit/s (a) and 40 ps/nm at 200 Gbit/s (c). The graphs to the right assume uncompensated transmission up to 80 km near 1550 nm. In this regime, most techniques cannot work due to the high uncompensated CD.

full-wave rectifiers, one comparator, one loop filter, and one QVCO. Under the conservative assumption that the power consumption of each individual component is equal to the power consumption of a Gilbert cell (93 mW in 90-nm CMOS), the aggregate power consumption of all functional blocks would be nearly 2 W. This estimate does not account for layout and interconnects, which typically double the power consumption of high-speed analog integrated circuits. Hence, we estimate that the power consumption of the high-speed analog electronics for an EPLL implementation would be 4 W. More accurate estimates may only be obtained after circuit-level design, which is beyond the scope of this work. An OPLL-based DP-QPSK receiver and a DP-DQPSK receiver have even lower power consumption, as they do not require a de-rotation stage.

Other receiver operations such as polarization demultiplexing and CDR are also power-efficient. For instance, three phase shifting sections can have a total power consumption of approximately 75 mW [104]. Moreover, a 40 Gb/s CDR in 90 nm CMOS consumes 48 mW [72], excluding output buffers.

Fig. 4.13 compare schemes with higher degrees of freedom at 200 Gbit/s for (a) a CD-compensated link where the residual CD is at most 40 ps/nm, and for (b) an 80-km uncompensated link. In this comparison, we have also included Stokes vector receivers that do not require LO laser, but do require high-speed ADCs and DSP [27]. DSP-free coherent is more power efficient as it avoids high-speed ADCs and DSP, which comes at the expense of small tolerance to CD. In the small residual CD regime (Fig. 4.13c), DSP-based coherent receivers have similar power consumption to that of Stokes vector receiver. The LO laser in coherent receivers provides improved receiver sensitivity, and it may account for up to 2.5 W of the total receiver power consumption [44]. In the high-uncompensated-CD regime (Fig. 4.13d), DSP-based coherent is the only viable option. The results of Fig. 4.13(c and d) also illustrate that it is more power efficient to operate with higher constellation sizes and more degrees of freedom in order to minimize the symbol rate.

The power consumption estimates of Fig. 4.13 illustrate that optical CD compensation either by DCFs or FBGs allow different receiver architectures that are more power efficient than DSP-based coherent receivers. As discussed in Section 2.1, to minimize link dispersion, future data centers may also leverage dispersion shifted fibers (DSFs) with zero-dispersion wavelength near 1550 nm or dispersion-flattened fibers with zero-dispersion wavelengths near both 1310 nm and 1550 nm bands.

## 4.6 Summary

We proposed and evaluated DSP-free analog coherent receiver architectures for unamplified intra-data center links and amplified inter-data center links. We showed that using a marker tone-based polarization demultiplexing scheme with an optical polarization controller, the analog coherent receiver can recover and track the transmitted polarization-multiplexed signals for a receiver operating at baseband. This technique can be extended to higher order QAM formats like 16-QAM and above, and can also be extended to higher-order IM formats such as 4-PAM and above. We also showed how CR can be conducted using a multiplier-free phase detector based on XOR gates and that its performance is within 0.5 dB of a Costas loop-based phase detector. Our proposed multiplier-free phase estimator is limited to QPSK inputs, however. Finally, we showed that DSP-free analog coherent receivers would have ∼1 dB penalty at small CD relative to their DSP-based counterparts. The SNR-penalty for DSP-free systems increases quadratically with CD and reaches 5 dB at roughly

±35 ps/nm. The power consumption of polarization demultiplexing and high-speed electronics is estimated to be nearly 4 W in 90 nm CMOS. Moreover, the improved receiver sensitivity due to coherent detection would allow 40-km unamplified and eye-safe transmission of up to 49 DWDM channels near 1310 nm, potentially blending intra- and inter-data center applications. The high spectral efficiency enabled by coherent detection, combined with its improved receiver sensitivity, will potentially blur distinctions between intra- and inter-data center links.

# Part II

# Submarine Optical Systems

# Chapter 5

# Importance of Amplifier Physics in Maximizing the Capacity of Submarine Links

Submarine transport cables interconnect countries and continents, forming the backbone of the global Internet. Over the past three decades, pivotal technologies such as erbium-doped fiber amplifiers (EDFAs), wavelength-division multiplexing (WDM), and coherent detection employing digital compensation of fiber impairments have enabled the throughput per cable to jump from a few gigabits per second to tens of terabits per second, fueling the explosive growth of the information age.

Scaling the throughput of submarine links is a challenging technical problem that has repeatedly demanded innovative and exceptional solutions. This intense technical effort has exploited a recurring strategy: to force ever-larger amounts of information over a small number of single-mode fibers [1]. This strategy is reaching its limits, however, as the amount of information that can be practically transmitted per fiber approaches fundamental limits imposed by amplifier noise and Kerr nonlinearity [105, 106]. In submarine cables longer than about 5,000 km, this strategy faces another fundamental limit imposed by energy constraints, as the electrical power available to the undersea amplifiers ultimately restricts the optical power and throughput per fiber.

Insight from Shannon's capacity offers a different strategy: employ more spatial dimensions (fibers or modes), while transmitting less data in each [107–109]. In fact, numerous recent works have studied how this new strategy improves the capacity and power efficiency of ultra-long submarine links [3, 5–7]. But a fundamental question remained unanswered: what is the optimal way of utilizing each spatial dimension? Formally, what is the channel power allocation that maximizes the information-theoretic capacity per spatial dimension given a constraint in the total electrical power? In this paper, we formulate this problem mathematically and demonstrate how to solve it. Our formulation accounts for amplifier physics, Kerr nonlinearity, and power feed constraints. Modeling amplifier physics is critical for translating energy constraints into parameters that govern the channel capacity such as amplification bandwidth, noise, and optical power.

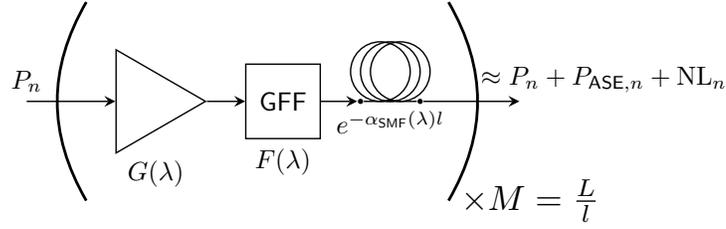

Figure 5.1: Equivalent block diagram of each spatial dimension of submarine optical link including amplifier noise and nonlinear noise.

Although the resulting optimization problem is not convex, the solutions are robust, i.e., they do not seem to depend on initial conditions. This suggests that the optimization reaches the global minimum or is consistently trapped in an inescapable local minimum. In either case, the solutions are very promising. The optimized power allocation increases the theoretical capacity per fiber by 70% compared to recently published results that employ spatial-division multiplexing (SDM) and flat power allocation. This improvement in capacity may be achieved without modifying the submerged plant and may only require altering the terminal equipment.

The optimization also provides insights into power-limited submarine link design and operation. For instance, in agreement to prior work [7], overall cable capacity is maximized by employing tens of spatial dimensions in each direction. Moreover, although EDFAs exhibit higher power conversion efficiency (PCE) in the highly saturated regime, the additional pump power to achieve that regime could be better employed in new spatial dimensions. Furthermore, the channel power optimization balances amplifier and nonlinear noise, and in the case of high pump power ($> 200$ mW) nonlinear noise is about 4 dB below ASE power.

## 5.1 Problem formulation

A submarine transport cable employs $S$ spatial dimensions in each direction, which could be modes in a multimode fiber, cores of a multi-core fiber, or simply multiple SMFs. Throughout this paper, we assume that each spatial dimension is a SMF, since this is the prevailing scenario in today's submarine systems. Each of those fibers can be represented by the equivalent diagram shown in Fig. 5.1.

The link has a total length $L$, and it is divided into $M$ spans, each of length $l = L/M$. An optical amplifier with gain $G(\lambda)$ compensates for the fiber attenuation $A(\lambda) = e^{\alpha_{\text{SMF}}(\lambda)l}$ of each span, and a gain-flattening filter (GFF) with transfer function $0 < F(\lambda) < 1$ ensures that the amplifier gain matches the span attenuation, so that at each span we have $G(\lambda)F(\lambda)A^{-1}(\lambda) \approx 1$. In practice, this condition has to be satisfied almost perfectly, as a mismatch of just a tenth of a dB would accumulate to tens of dBs after a chain of hundreds of amplifiers. As a result, in addition to GFF per span, periodic power rebalancing after every five or six spans corrects for any residual mismatches.

The input signal consists of $N$ potential WDM channels spaced in frequency by $\Delta f$, so that the channel at wavelength $\lambda_n$ has power $P_n$. Our goal is to find the power allocation $P_1, \ldots, P_N$ that maximizes the information-theoretic capacity per spatial dimension. We do not make any prior assumptions about the amplifier bandwidth, hence the optimization may result in some channels not being used i.e., $P_n = 0$ for some $n$.

Due to GFFs and periodic power rebalancing, the output signal power remains approximately the same. But the signal at each WDM channel is corrupted by amplifier noise $P_{\text{ASE},n}$ and nonlinear noise $\text{NL}_n$. Thus, the $\text{SNR}_n$ of the $n$th channel is given by

$$\text{SNR}_n = \begin{cases} \dfrac{P_n}{P_{\text{ASE},n} + \text{NL}_n}, & G(\lambda_n) > A(\lambda_n) \\ 0, & \text{otherwise} \end{cases}. \tag{5.1}$$

Note that only channels for which the amplifier gain is greater than the span attenuation can be used to transmit information, i.e., $P_n \neq 0$ only if $G(\lambda_n) > A(\lambda_n)$.

The optical amplifiers for submarine links generally consist of single-stage EDFAs with redundant forward-propagating pump lasers operating near 980 nm. In ultra-long links, the pump power is limited by feed voltage constraints at the shores. From the maximum power transfer theorem, the total electrical power available to all undersea amplifiers is at most $\text{P} = \text{V}^2/(4L\rho)$, where V is the feed voltage, and $\rho$ is the cable resistance. To translate this constraint on the total electrical power into a constraint on the optical pump power $P_p$ per amplifier, we use an affine model similar to the one used in [3, 5]:

$$P_p = \eta\Big(\frac{\text{P}}{2SM} - \text{P}_\text{o}\Big), \tag{5.2}$$

where $\eta$ is an efficiency constant that translates electrical power into optical pump power, and $\text{P}_\text{o}$ accounts for electrical power spent in operations not directly related to optical amplification such as pump laser lasing threshold, monitoring, and control. The factor of $2S$ appears because there are $S$ spatial dimensions in each direction.

This constraint on the pump power limits the EDFA output optical power and bandwidth, thus imposing a hard constraint on the fiber throughput. As an example, increasing $P_n$ may improve the SNR and spectral efficiency of some WDM channels, but increasing $P_n$ also depletes the EDF and reduces the amplifier overall gain. As a result, the gain of some channels may drop below the span attenuation, thus reducing the amplifier bandwidth and the number of WDM channels that can be transmitted. Further increasing $P_n$ may reduce the SNR, as the nonlinear noise power becomes significant. These considerations illustrate how forcing more power per fiber is an ineffective strategy in improving the capacity per fiber of power-limited submarine cables.

### 5.1.1 Amplifier physics

Returning to (5.1), we need to compute the amplifier gain and noise for a given pump power $P_p$ (5.2) and input power profile $P_1, \ldots, P_N$.

The steady-state pump and signal power evolution along an EDF of length $L_{EDF}$ is well modeled by the standard confined-doping (SCD) model [110], which for a two-level system is described by a set of coupled first-order nonlinear differential equations:

$$\frac{d}{dz}P_k(z) = u_k(\alpha_k + g_k^*)\frac{\bar{n}_2}{\bar{n}_t}P_k(z)$$
$$- u_k(\alpha_k + l_k)P_k(z) + 2u_k g_k^* \frac{\bar{n}_2}{\bar{n}_t}h\nu_k \Delta f \quad (5.3)$$

$$\frac{\bar{n}_2}{\bar{n}_t} = \frac{\sum_k \frac{P_k(z)\alpha_k}{h\nu_k \zeta}}{1 + \sum_k \frac{P_k(z)(\alpha_k + g_k^*)}{h\nu_k \zeta}} \quad (5.4)$$

where the subindex $k$ indexes both signal and pump i.e., $k \in \{p, 1, \ldots, N\}$, $z$ is the position along the EDF, and $\mu_k = 1$ for beams that move in the forward direction i.e., increasing $z$, and $\mu_k = -1$ otherwise. Here, $l_k$ denotes the background loss (or excess loss), $\alpha_k$ is the absorption coefficient, $g_k^*$ is the gain coefficient, and $\bar{n}_2/\bar{n}_t$ denotes the population of the second metastable level normalized by the Er ion density $\bar{n}_t$. $\zeta = \pi r_{Er}^2 \bar{n}_t/\tau$ is the saturation parameter, where $r_{Er}$ is the Er-doping radius, and $\tau \approx 10$ ms is the metastable lifetime. According to this model, the amplifier characteristics are fully described by three macroscopic parameters, namely $\alpha_k$, $g_k^*$, and $\zeta$. Fig 5.2 shows $\alpha_k$ and $g_k^*$ for the EDF used in our simulations for this paper.

The first term of (5.3) corresponds to the medium gain, the second term accounts for absorption, and the third term accounts for amplified spontaneous emission (ASE) noise.

To compute the amplifier gain and noise using (5.3), we must solve the boundary value problem (BVP) of $N + 1 + 2N$ coupled equations, where we have $N$ equations for the signals, one for the pump, and the noise at the signals' wavelengths is broken into $2N$ equations: $N$ for the forward ASE, and $N$ for the backward ASE.

Fig. 5.3 compares the gain and ASE power predicted using the theoretical model in (5.3) with experimental measurements for several values of pump power $P_p$. The amplifier consists of a single 8-m-long EDF pumped by a forward-propagating laser near 980 nm with power $P_p$. The incoming signal to the amplifier consists of 40 unmodulated signals from 1531 to 1562. The power of each signal is $-13$ dBm, resulting in a total of 3 dBm. The theoretical results use (5.3) with experimentally measured values of the absorption and gain coefficients $\alpha$ and $g^*$. The nominal experimentally measured values have been scaled up by 8% to achieve the best fit between theory and experiment. The experimental error in these values was estimated independently to be about 5%.

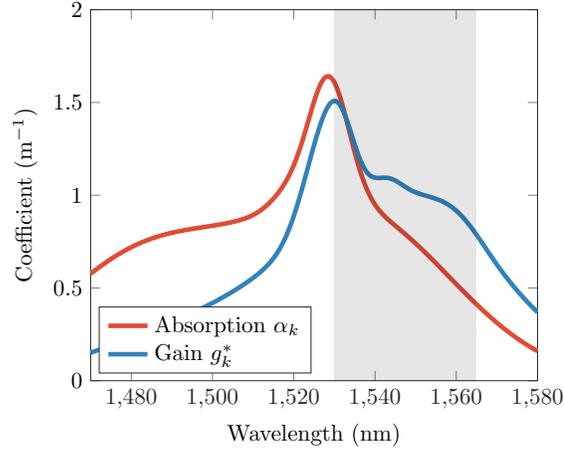

Figure 5.2: Absorption and gain coefficients for the EDF used in simulations. C-band is highlighted. For the pump at 980 nm, $\alpha_p = 0.96$ m$^{-1}$, and $g_p^* = 0$ m$^{-1}$. Other relevant parameters are $r_{Er} = 1.38\mu$m and $\bar{n}_t = 5.51 \times 10^{18}$ cm$^3$.

Although (5.3) is very accurate, the optimizations require evaluation of the objective function hundreds of thousands of times, which would require solving the BVP in (5.3) that many times. Hence, approximations for the gain and noise are necessary.

#### 5.1.1.1　Approximated amplifier noise power

By assuming that the amplifier is inverted uniformly, equation (5.3) can be solved analytically resulting in the well-known expression for ASE power in a bandwidth $\Delta f$ for a single amplifier:

$$P_{ASE,n} = 2n_{sp,n}(G_n - 1)h\nu_n\Delta f \tag{5.5}$$

where $n_{sp}$ is the excess noise factor [110, equation (32)]. The excess noise factor is related to the noise figure $NF_n = 2n_{sp,n}\frac{G_n-1}{G_n}$, where the commonly used high-gain approximation $\frac{G(\lambda_n)-1}{G(\lambda_n)} \approx 1$ may be replaced by the more accurate approximation $\frac{G(\lambda_n)-1}{G(\lambda_n)} \approx 1 - e^{-\alpha_{\text{SMF}}l}$, since in submarine systems the amplifier gain is approximately equal to the span attenuation, which is on the order of 10 dB. This approximation conveniently makes the amplifier noise figure independent of the amplifier gain.

Thus, the amplifier noise $P_{\text{ASE},n}$ in a bandwidth $\Delta f$ after a chain of $M$ amplifiers is given by

$$P_{\text{ASE},n} = M\text{NF}_n h\nu_n\Delta f. \tag{5.6}$$

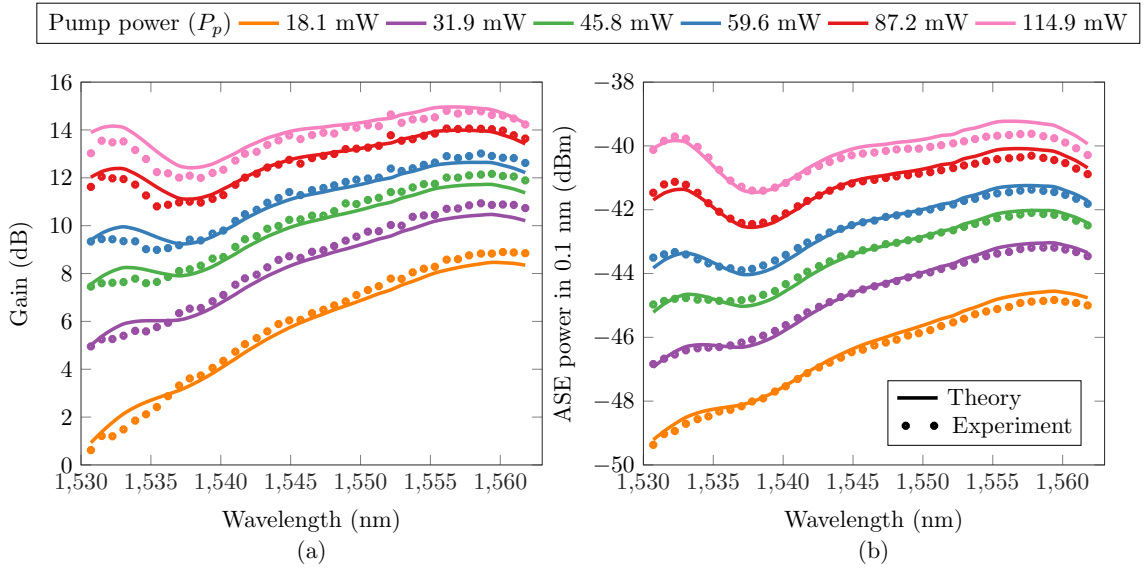

Figure 5.3: Comparison between experiment and theory for (a) gain and (b) ASE power in 0.1 nm for different values of pump power. Theoretical gain and ASE curves are computed according to (5.3).

For amplifiers pumped at 980 nm, the noise figure is approximately gain-and-wavelength independent, and it can be computed from theory or measured experimentally. Although we focus on end-pumped single-mode EDFAs, similar models exist for multicore EDFAs [111].

#### 5.1.1.2 Approximated amplifier gain

By assuming that the amplifier is not saturated by ASE, equation (5.3) reduces to a single-variable implicit equation [112], which can be easily solved numerically. According to this model, the amplifier gain is given by

$$G_k = \exp\left(\frac{\alpha_k + g_k^*}{\zeta}(Q^{in} - Q^{out}) - \alpha_k L_{\text{EDF}}\right) \tag{5.7}$$

where $Q_k^{in} = \frac{P_k}{h\nu_k}$ is the photon flux in the $k$th channel, and $Q^{in} = \sum_k Q_k^{in}$ is the total input photon flux. The output photon flux $Q^{out}$ is given by the implicit equation:

$$Q^{out} = \sum_k Q_k^{in} \exp\left(\frac{\alpha_k + g_k^*}{\zeta}(Q^{in} - Q^{out}) - \alpha_k L_{\text{EDF}}\right) \tag{5.8}$$

Therefore, to compute the amplifier gain using the semi-analytical model, we must first solve (5.8) numerically for $Q^{out}$, and then compute the gain using (5.7). This procedure is much faster than solving (5.3). In this calculation, we assume that the input power to the amplifier is equal to $P_n + (M - 1)\text{NF}_n h\nu_n \Delta f$. That is, the signal power plus the accumulated ASE noise power at the input of the last amplifier in the chain. As a result, all amplifiers are designed to operate under the same conditions as the last amplifier. This pessimistic assumption is not critical in systems that operate with high optical signal-to-noise ratio (OSNR), and accounts for signal droop in low-OSNR systems, where the accumulated ASE power may be larger than the signal power, and thus reduce the amplifier useful bandwidth.

### 5.1.2 Kerr nonlinearity

To account for Kerr nonlinearity, we use the Gaussian noise (GN) model, which establishes that the Kerr nonlinearity in dispersion-uncompensated fiber systems is well modeled as an additive zero-mean Gaussian noise whose power at the $n$th channel is given by [113]

$$\text{NL}_n = A^{-1}(\lambda_n) \sum_{n_1=1}^{N} \sum_{n_2=1}^{N} \sum_{q=-1}^{1} \tilde{P}_{n_1} \tilde{P}_{n_2} \tilde{P}_{n_1+n_2-n+q} D_q^{(M \text{ spans})}(n_1, n_2, n), \tag{5.9}$$

for $1 \leq n_1 + n_2 - n + q \leq N$. Here, $\tilde{P}_n$ denotes the launched power of the $n$th channel, which is related to the input power to the amplifier by $\tilde{P}_n = A(\lambda_n) P_n$. The nonlinear noise power is scaled by the span attenuation $A^{-1}(\lambda_n)$ due to the convention in Fig. 5.1 that $P_n$ refers to the input power to the amplifier, rather than the launched power. $D_q^{(M \text{ spans})}(n_1, n_2, n)$ is the set of fiber-specific nonlinear coefficients that determine the strength of the four-wave mixing component that falls on channel $n$, generated by channels $n_1$, $n_2$, and $n_1 + n_2 - n + q$. Here, $q = 0$ describes the dominant nonlinear terms, while the coefficients $q = \pm 1$ describe corner contributions.

The nonlinear coefficients $D_q^{(1 \text{ span})}(n_1, n_2, n)$ for one span of SMF of length $l$, nonlinear coefficient $\gamma$, power attenuation $\alpha_{\text{SMF}}$, and propagation constant $\beta_2$ are given by the triple integral

$$D_q^{(1 \text{ span})}(n_1, n_2, n) = \frac{16}{27} \gamma^2 \iiint_{-1/2}^{1/2}$$
$$\rho((x + n_1)\Delta f, (y + n_2)\Delta f, (z + n)\Delta f)$$
$$\cdot \text{rect}(x + y - z + q) \partial x \partial y \partial z, \tag{5.10}$$

$$\rho(f_1, f_2, f) = \left| \frac{1 - \exp(-\alpha l + j4\pi^2 \beta_2 l (f_1 - f)(f_2 - f))}{\alpha - j4\pi^2 \beta_2 (f_1 - f)(f_2 - f)} \right|^2, \tag{5.11}$$

where $\text{rect}(\omega) = 1$, for $|\omega| \leq 1/2$, and $\text{rect}(\omega) = 0$ otherwise. Equation (5.10) assumes that all channels have a rectangular spectral pulse shape.

As the coefficients $D_q^{(1 \text{ span})}(n_1, n_2, n)$ only depend on the index differences $n_1 - n$ and $n_2 - n$,

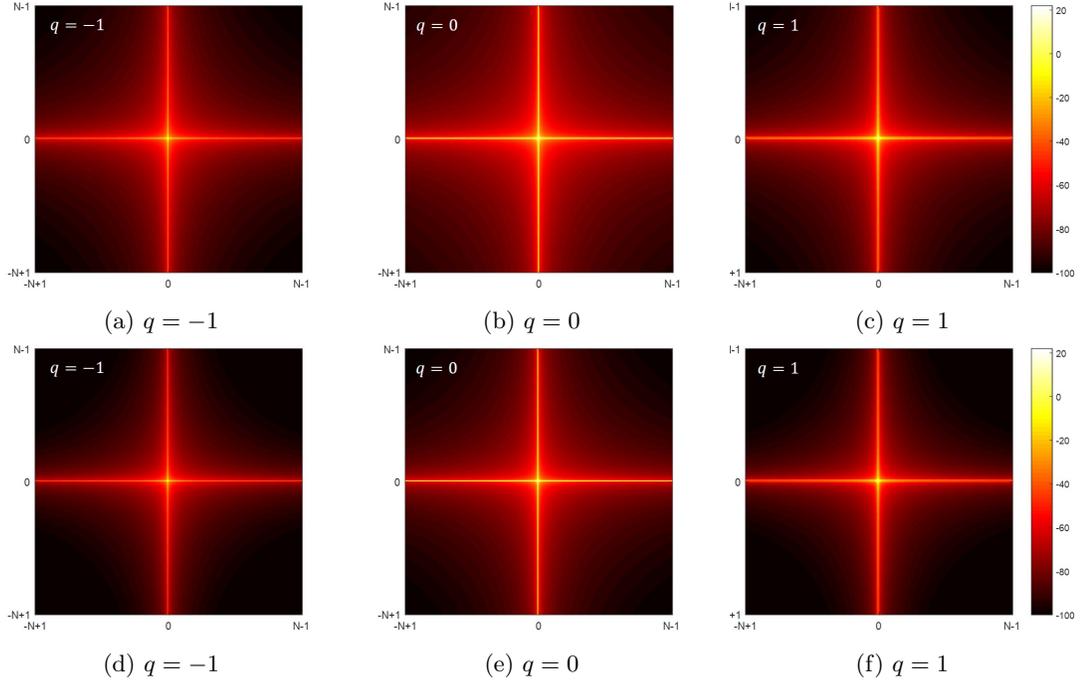

Figure 5.4: Nonlinear coefficients for (top) 50 km of standard SMF, and (bottom) 50 km of large-effective-area fiber used in ultra-long haul optical communications.

we can represent them in a matrix such that $D^{(q)}_{n_1-n, n_2-n} = D_q^{(1\ \text{span})}(n_1, n_2, n)$. Fig. 5.4 shows these coefficients for 50 km of standard SMF and large-effective-area SMF.

The pixel at the center of the images in Fig. 5.4 corresponds to self-phase modulation ($n_1 = n_2 = n$), the horizontal and vertical lines at the center of the images correspond to cross phase modulation ($n_1 = n$ or $n_2 = n$), and the remaining pixels correspond to four wave mixing ($n_1 \neq n_2 \neq n$).

Computing $D_q^{(1\ \text{span})}(n_1, n_2, n)$ is computationally less intensive than $D_q^{(M\ \text{spans})}(n_1, n_2, n)$, since the highly oscillatory term $\chi(f_1, f_2, f)$ in $D_q^{(M\ \text{spans})}(n_1, n_2, n)$ [113] is constant and equal to one in $D_q^{(1\ \text{span})}(n_1, n_2, n)$. The nonlinear coefficients for $M$ spans can be computed by following the nonlinear power scaling given in [114]:

$$D_q^{(M\ \text{spans})}(n_1, n_2, n) = M^{1+\epsilon} D_q^{(1\ \text{span})}(n_1, n_2, n), \tag{5.12}$$

where the parameter $\epsilon$ controls the nonlinear noise scaling over multiple spans, and for bandwidth of $\sim 40$ nm (e.g., 100 channels spaced by 50 GHz), it is approximately equal to 0.06 [114]. The parameter $\epsilon$ may also be computed from the approximation [114, eq. (23)].

We do not include stimulated Raman scattering (SRS) in our modeling for two reasons. First, long-haul submarine cables employ large-effective-area fibers, which reduces SRS intensity. Second, the optimized amplifier bandwidth is not larger than 45 nm, while the Raman efficiency peaks when the wavelength difference is $\sim 100$ nm.

### 5.1.3 Optimization problem

Using equations (5.1), (5.6), (5.7), and (5.9), we can compute Shannon's capacity per fiber by adding the capacities of the individual WDM channels:

$$C = 2\Delta f \sum_{n=1}^{N} \mathbb{1}\{\mathcal{G}(\lambda_n) \geq \mathcal{A}(\lambda_n)\} \log_2(1 + \Gamma \text{SNR}_n), \tag{5.13}$$

where $0 < \Gamma < 1$ is the coding gap to capacity and $\mathcal{G}(\lambda_n), \mathcal{A}(\lambda_n)$ denote, respectively, the amplifier gain and span attenuation in dB units. The indicator function $\mathbb{1}\{\cdot\}$ is one when the condition in its argument is true, and zero otherwise. As we do not know a priori which channels contribute to capacity ($P_n \neq 0$), we sum over all channels and let the indicator function indicate which channels have gain above the span attenuation.

Since the indicator function is non-differentiable, it is convenient to approximate it by a differentiable sigmoid function such as

$$\mathbb{1}\{x \geq 0\} \approx 0.5(\tanh(Dx) + 1), \tag{5.14}$$

where $D > 0$ controls the sharpness of the sigmoid approximation. Although making $D$ large better approximates the indicator function, it results in vanishing gradients, which retards the optimization process.

Hence, the optimization problem of maximizing the capacity per fiber given an energy constraint that limits the amplifier pump power $P_p$ can be stated as

$$\underset{L_{\text{EDF}}, \mathcal{P}_1, \ldots, \mathcal{P}_N}{\text{maximize}} \quad C$$
$$\text{given } P_p \tag{5.15}$$

In addition to the power allocation $\mathcal{P}_1, \ldots, \mathcal{P}_N$ in dBm units, we optimize over the EDF length $L_{\text{EDF}}$, resulting in a $(N+1)$-dimensional non-convex optimization problem. $L_{\text{EDF}}$ may be removed from the optimization if its value is predefined. It is convenient to optimize over the signal power in dBm units, as the logarithmic scale enhances the range of signal power that can be covered by taking small adaptation steps. Even if we assumed binary power allocation, i.e., $\mathcal{P}_n \in \{0, \bar{\mathcal{P}}\}$, it is not easy to determine the value of $\bar{\mathcal{P}}$ that will maximize the amplification bandwidth for which the gain is larger than the span attenuation.

Note that if we did not have the pump power constraint and the amplifier gain did not change with the power allocation $\mathcal{P}_1, \ldots, \mathcal{P}_N$, the optimization problem in (5.15) would reduce to the convex problem solved in [113]. Therefore, we can argue that to within a small $\Delta \mathcal{P}_n$ that does not change the conditions in the argument of the indicator function, the objective (5.13) is locally concave.

Nevertheless the optimization problem in (5.15) is not convex, and therefore we must employ global optimization techniques. In this paper, we use the particle swarm optimization (PSO) algorithm [115]. The PSO algorithm randomly initializes $R$ particles $X = [L_{\text{EDF}}, \mathcal{P}_1, \ldots, \mathcal{P}_N]^T$. As the optimization progresses, the direction and velocity of the $i$th particle is influenced by the its best known position and also by the best known position found by other particles in the swarm:

$$v_i \leftarrow wv_i + \mu_1 a_i(p_{i,best} - X_i) + \mu_2 b_i(s_{best} - X_i) \qquad \text{(velocity)}$$
$$X_i \leftarrow X_i + v_i \qquad \text{(location)}$$

where $w$ is an inertial constant chosen uniformly at random in the interval $[0.1, 1.1]$, $\mu_1 = \mu_2 = 1.49$ are the adaptation constants, $a_i, b_i \sim \mathcal{U}[0, 1]$ are uniformly distributed random variables, $p_{i,best}$ is the best position visited by the $i$th particle, and $s_{best}$ is the best position visited by the swarm. The PSO algorithm was shown to outperform other global optimization algorithms such as the genetic algorithm in a broad class of problems [116].

To speed up convergence and avoid local minima, it is critical to initialize the particles $X = [L_{\text{EDF}}, P_1, \ldots, P_N]$ to within close range of the optimal solution. From the nature of the problem, we can limit the particles to a very narrow range. The EDF length is limited from 0 to 20 m. Since the amplifier gain will be relatively close to the span attenuation $A(\lambda) = e^{\alpha_{SMF}l}$, we can compute the maximum input power to the amplifier that will allow this gain for a given pump power $P_p$. This follows from conservation of energy [117, eq. 5.3]:

$$P_n < \frac{1}{\bar{N}} \frac{\lambda_p P_p}{\lambda_n A(\lambda)}, \qquad (5.16)$$

where $\lambda_p$ is the pump wavelength, $\lambda_n$ is the signal wavelength, and $\bar{N}$ is the expected number of WDM channels that will be transmitted. The minimum power is assumed to be 10 dB below this maximum value.

When nonlinear noise power is small, the solution found by the PSO does not change for different particle initializations. However, the solutions found by PSO when nonlinear noise is not negligible exhibit some small and undesired variability. To overcome this problem, after the PSO converges, we continue the optimization using the saddle-free Newton's method [118]. According to this algorithm, the adaptation step $X \leftarrow X + \Delta X$ is given by

$$\Delta X = -\mu |H|^{-1} \nabla C, \qquad (5.17)$$

where $\mu$ is the adaptation constant, $\nabla C$ is the gradient of the capacity in (5.13) with respect to $X$, and $H$ is the Hessian matrix, i.e., the matrix of second derivatives of $C$ with respect to $X$. The absolute value notation in (5.17) means that $|H|$ is obtained by replacing the eigenvalues of $H$ with their absolute values.

Table 5.1: Parameters of submarine system considered in the optimization.

| Parameter | Value | Units |
|---|---|---|
| Link length ($L$) | 14,350 | km |
| Span length ($l$) | 50 | km |
| Number of amplifiers per fiber ($M$) | 287 | |
| First channel ($\lambda_1$) | 1522 | nm |
| Last channel ($\lambda_N$) | 1582 | nm |
| Channel spacing ($\Delta f$) | 50 | GHz |
| Max. number of WDM channels ($N$) | 150 | |
| Fiber attenuation coefficient ($\alpha_{\text{SMF}}(\lambda)$) | 0.165 | dB km$^{-1}$ |
| Fiber dispersion coefficient ($D(\lambda)$) | 20 | ps nm$^{-1}$ km$^{-1}$ |
| Fiber nonlinear coefficient ($\gamma$) | 0.8 | W$^{-1}$ km$^{-1}$ |
| Fiber additional loss (margin) | 1.5 | dB |
| Overall span attenuation ($A(\lambda)$) | $8.25 + 1.5 = 9.75$ | dB |
| Nonlinear noise power scaling ($\epsilon$) | 0.07 | |
| Coding gap ($\Gamma$) | $-1$ | dB |
| Sigmoid sharpness ($D$) | 2 | |
| Excess noise factor ($n_{sp}$) | 1.4 | |
| Excess loss ($l_k$) | 0 | dB/m |

Both the gradient and the Hessian can be derived analytically by using the semi-analytical model given in equations (5.7) and (5.8). However, we compute the gradient analytically (See Appendix A) and compute the Hessian numerically using finite differences of the gradient.

## 5.2  Results and discussion

We now apply our proposed optimization procedure to the reference system with parameters listed in Table 5.1. These parameters are consistent with recently published experimental demonstration of high-capacity systems employing SDM [6]. We consider $M = 287$ spans of $l = 50$ km of low-loss large-effective area single-mode fiber, resulting in a total link length of $L = 14,350$ km. The span attenuation is $\mathcal{A}(\lambda) = 9.75$ dB, where 8.25 dB is due to fiber loss, and the additional 1.5 dB is added as margin. For the capacity calculations we assume a coding gap of $\Gamma = 0.79$ ($-1$ dB).

### 5.2.1  Channel power optimization

We first study how the optimized power allocation and the resulting spectral efficiency is affected by the amplifier pump power. We also investigate how Kerr nonlinearity affects the optimized power allocation and when it can be neglected. This discussion does not assume any particular power budget or number of spatial dimensions. In Section 5.2.2, we consider how employing multiple

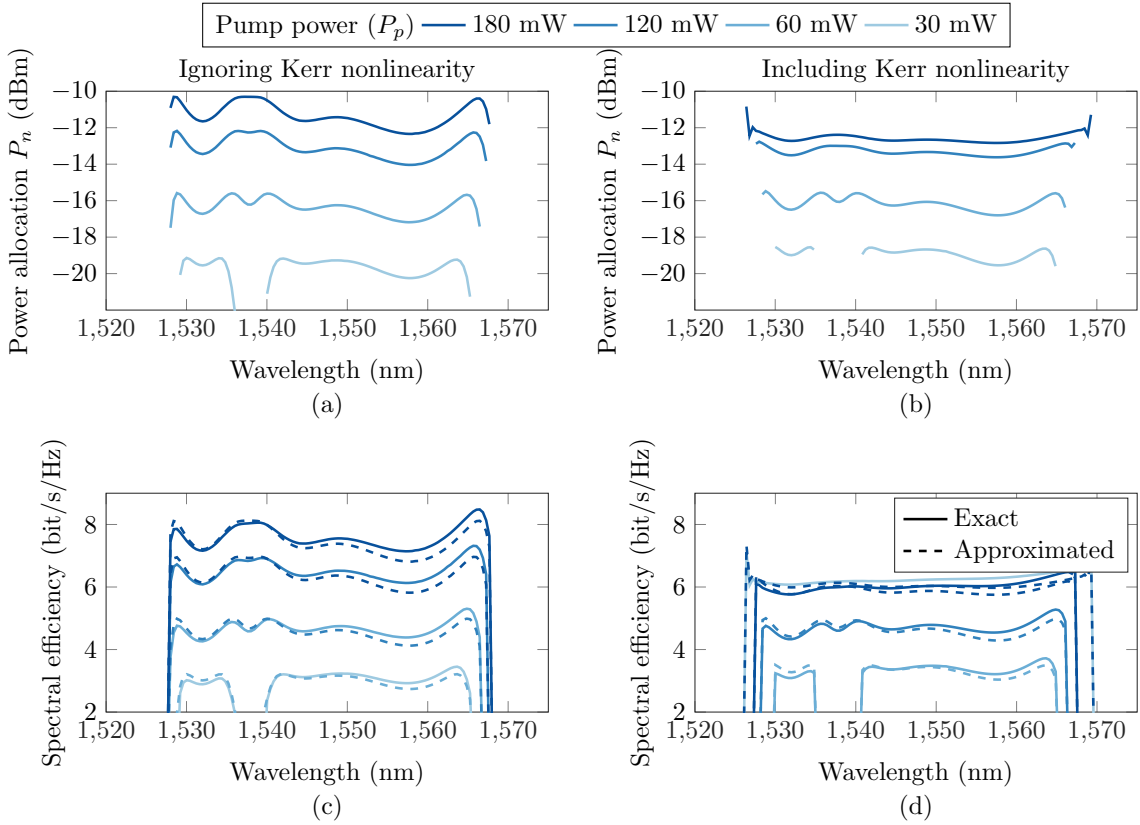

Figure 5.5: Optimized power allocation $P_n$ for several values of pump power $P_p$. Kerr nonlinearity is disregarded in (a) and included in (b). Their corresponding achievable spectral efficiency is shown in (c) and (d). Note that $P_n$ corresponds to the input power to the amplifier. The launched power is $\check{P}_n = G(\lambda_n)F(\lambda_n)P_n = A(\lambda_n)P_n$. Thus, the launch power is 9.75 dB above the values shown in these graphs.

spatial dimensions can lead to higher overall cable capacity.

For a given pump power $P_p$, we solve the optimization problem in (5.15) for the system parameters listed in Table 5.1. The resulting power allocation $P_n$ is plotted in Fig. 5.5 when Kerr nonlinearity is (a) disregarded and (b) included. The corresponding achievable spectral efficiency of each WDM channel is shown in Fig. 5.5cd.

For small pump powers, the optimized power profile is limited by the amplifier, and thus there is a small difference between the two scenarios shown in Fig. 5.5. As the pump power increases and the amplifier delivers more output power, Kerr nonlinearity becomes the limiting factor of the channel power. Interestingly, the optimized power allocation in the nonlinear regime exhibits large oscillations at the extremities because the nonlinear noise is smaller at those channels. Although the optimization is performed for 150 possible channels from 1522 nm to 1582 nm, not all of these WDM channels are utilized, and the useful bandwidth is restricted to 1525 nm to 1570 nm. Note that for $P_p = 30$ mW, part of the amplifier bandwidth cannot be used, as the resulting amplifier gain is below attenuation. The amplifier bandwidth does not change significantly because it is fundamentally limited by the absorption and gain coefficients of the EDF, which depend only on the EDF design and co-dopants. The optimized EDF length does not vary significantly, and it is generally in the range of 6 to 9 m.

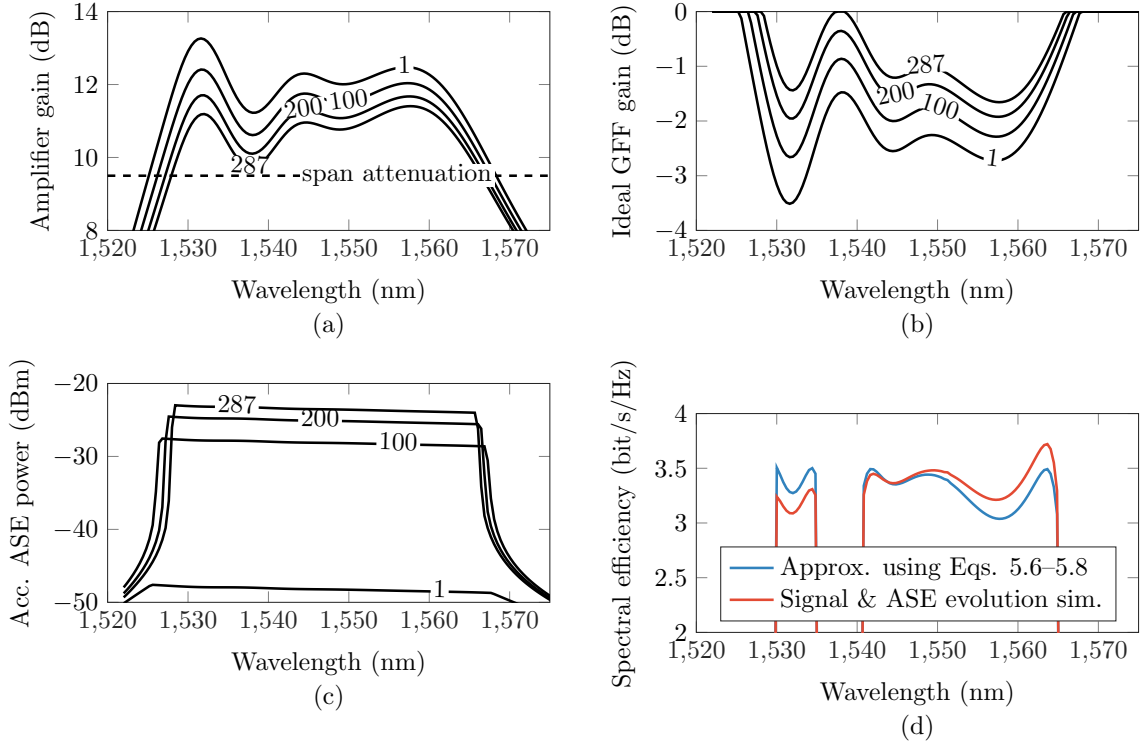

Figure 5.6: Theoretical (a) amplifier gain, (b) ideal GFF gain, and (c) accumulated ASE power in 50 GHz after 1, 100, 200, and 287 spans of 50 km. The pump power of each amplifier is 30 mW, resulting in the optimized power profile shown in Fig. 5.5b for $P_p = 30$ mW and EDF length of 6.6 m. To compute amplifier gain and noise over the entire band, we assume that unused channels have powers of $-126$ dBm. Note that the amplifier gain and ideal GFF gain gradually change as the accumulated ASE power increases. In practice, the ideal GFF gain can be realized by a fixed GFF after each amplifier and periodic power rebalancing after five or more spans. (d) Comparison of the spectral efficiency per channel computed by this signal and ASE evolution simulation to the spectral efficiency predicted by the models and approximations discussed in Section 5.1.

The solid lines in Fig. 5.5c and (d) are obtained from (5.13) by using exact models (5.3) for the amplifier gain and noise, while the dashed lines are computed by making approximations to allow (semi-)analytical calculation of amplifier gain (5.7) and noise (5.6), and speed up the optimization process. Fig. 5.5cd shows that these approximations only cause negligible errors.

For the optimized power profile for $P_p = 30$ mW shown in Fig. 5.5b, we compute the evolution of amplifier gain, accumulated ASE, and the required GFF gain along the 287 spans, as shown in Fig. 5.6. The amplifier gain and ASE power were computed using the exact amplifier model given in (5.3). The accumulated ASE power (Fig. 5.6c) increases after every span, causing the amplifier gain (Fig. 5.6a) and consequently the ideal GFF gain (Fig. 5.6b) to change slightly. In practice, the ideal GFF shape can be achieved by fixed GFF after each amplifier and periodic power rebalancing at intervals of five or so spans.

Recall that in the optimization all amplifiers are designed to operate under the same conditions (noise level) as the last amplifier. As a result, the optimization correctly predicts that channels near 1537 nm should not be used, as otherwise the gain of the last amplifiers in the chain could drop below the span attenuation of 9.75 dB. In the optimization, the span attenuation can be increased to allow higher margin to account for model or device inaccuracies.

At the last span of the signal and ASE evolution simulation, we compute the spectral efficiency per channel and compare it to the approximated results obtained using (5.1)–(5.13). As shown in Fig. 5.5d, the approximations and assumptions made in Section 5.1 only have minor impact in the overall fiber capacity computed by propagating signal and ASE.

Fig. 5.7a shows the total capacity per fiber as a function of the pump power. Once again, for each value of pump power $P_p$, we solve the optimization problem in (5.15) for the system parameters listed in Table 5.1. The capacity per spatial dimension plotted in Fig. 5.7a is computed by summing the capacities of the individual WDM channels. Below about 100 mW of pump power, the system operates in the linear regime. At higher pump powers, the amplifier can deliver higher optical power, but Kerr nonlinearity becomes significant and detains the capacity. Fig. 5.7b details the ratio between ASE power to nonlinear noise power. At high pump powers, ASE is only 4 dB higher than nonlinear noise. This illustrates the diminishing returns of forcing more power over a single spatial dimension.

To gauge the benefits of our proposed optimization procedure, we compare the results of our approach to those of a recently published work [6], which experimentally demonstrated high-capacity SDM systems. In their experimental setup, Sinkin et al used 82 channels spaced by 33 GHz from 1539 nm to 1561 nm. Each of the 12 cores of the multicore fiber was amplified individually by an end-pumped EDFA with forward-propagating pump. Each amplifier was pumped near 980 nm with 60 mW resulting in an output power of 12 dBm [6], thus $-7.1$ dBm per channel. The span attenuation was 9.7 dB, leading to the input power to the first amplifier of $P_n = -16.7$ dBm per channel. We compute the capacity of this system according to (5.13) using the same methods and models for amplifier and Kerr nonlinearity discussed in Section 5.1. Fiber parameters and amplifier noise figure are given in Table 5.1. The EDF length is assumed 7 m, which is the value resulting from our optimization for EDFAs pumped with 60 mW. The resulting achievable spectral efficiency per channel is, on average, 4.8 bit/s/Hz, yielding a maximum rate of about 13 Tb/s per core. This is indicated by the red dot in Fig. 5.7. Naturally, this calculation is oversimplified, but it is consistent with the rate achieved in [6]. Their experimental spectral efficiency is 3.2 bit/s/Hz in 32.6 Gbaud, leading to 106.8 Gb/s per channel, 8.2 Tb s$^{-1}$ per core, and 105 Tb s$^{-1}$ over the 12 cores. The capacity using the optimized power profile is about 22 Tb s$^{-1}$ per core for the same pump power and overall system (ASE + Kerr nonlinearity curve in Fig. 5.7), thus offering 70% higher capacity when compared to the theoretical estimate for a system consistent with [6]. The optimized power profile for $P_p = 60$ mW is plotted in Fig. 5.5b.

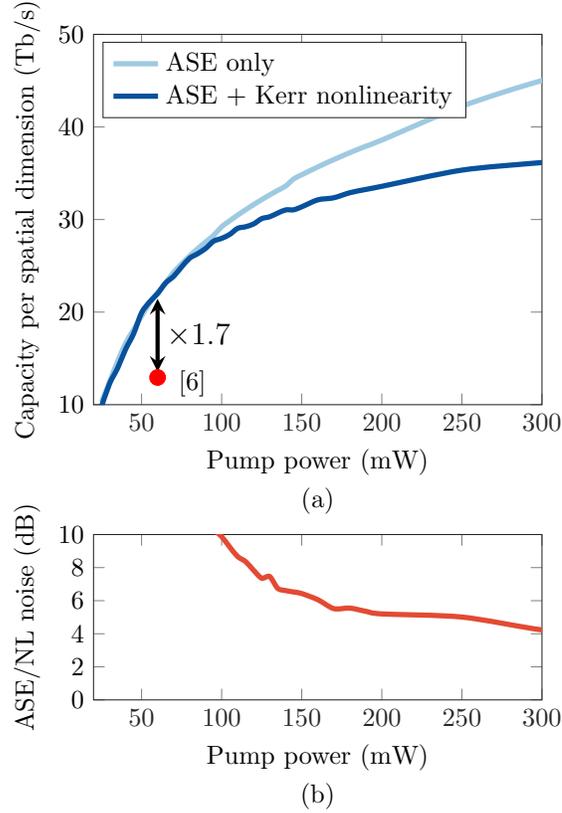

Figure 5.7: (a) Total capacity per single-mode fiber as a function of pump power. The power allocation and EDF length are optimized for each point. The red dot corresponds to the capacity according to (5.13) for a system with parameters consistent with [6]. (b) Ratio between ASE and nonlinear noise power for the optimization in (a).

Fig. 5.8 shows the capacity per fiber as a function of the span length for a fixed power budget. The span attenuation for all cases was calculated as $\mathcal{A} = \alpha_{\text{SMF}} l + 1.5$, where the 1.5 dB of additional attenuation is added as a margin. The total pump power per fiber was assumed $P_{p,total} = 287 \times 50 = 14350$ mW. Hence, making the span length shorter reduces the pump per amplifier. The optimal span length is achieved for 40–50 km, resulting in a span attenuation of 8.1–9.75 dB.

The main benefit of the channel power optimization is to allow the system to operate over a wider amplification bandwidth by appropriately adjusting the channel powers. Given that capacity scales linearly with dimensions (frequency or space) and only logarithmically with power, the optimization will favor power allocations that maximize the useful amplification bandwidth, i.e., bandwidth over which the gain is larger than the span attenuation. The optimization does not necessarily make the amplifiers exhibit higher PCE. In fact, highly saturated optical amplifiers achieve higher PCE, but that does not necessarily mean higher overall amplification bandwidth.

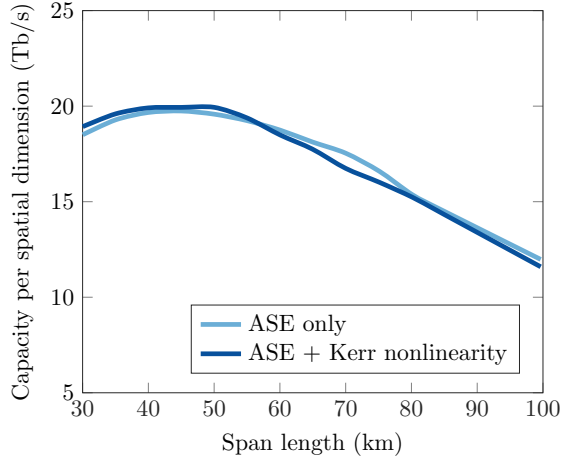

Figure 5.8: Total capacity per spatial dimension as a function of span length for a fixed power budget. ASE only and ASE + Kerr nonlinearity curves overlap, as available power budget restricts operation to the linear regime.

### 5.2.2 Optimal number of spatial dimensions

The optimal strategy is therefore to employ more spatial dimensions while transmitting less power in each one. The optimal number of spatial dimensions depends on the available electrical power budget. As an example, Fig. 5.9 shows the capacity of a cable employing $S$ spatial dimensions in each direction. We consider the feed voltage V = 12 kV, cable resistivity $\rho = 1$ $\Omega$ km$^{-1}$, and the reference link of Table 5.1. Thus, the total electrical power available for all amplifiers is 2.5 kW. From this and assuming efficiency $\eta = 0.4$ and overhead power $P_o$, we can compute the pump power per amplifier $P_p$ according to (5.2), and obtain the capacity per fiber from Fig. 5.7a.

The optimal number of spatial dimensions in each direction $S$ decreases as the overhead power increases, reaching 20, 12, and 8 for the power overhead $P_o = 0.1, 0.2$, and 0.3 W, respectively. This corresponds to amplifiers with pump powers of 43.7, 47.4, and 65.7 mW, respectively. Hence, at the optimal number of spatial dimensions the system operates in the linear regime, as can bee seen by inspecting Fig. 5.7. For small values of $P_o \to 0$, the optimal number of spatial dimensions is very large, illustrating the benefits of massive SDM, as reported in [7].

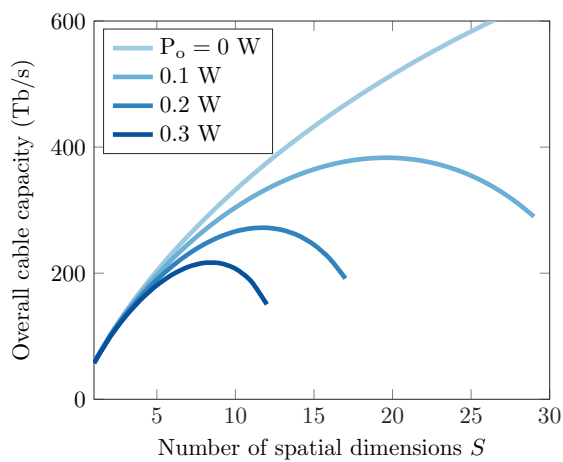

Figure 5.9: Capacity as a function of the number of spatial dimensions for the system of Table 5.1 assuming a power budget of P = 2.5 kW for all amplifiers.

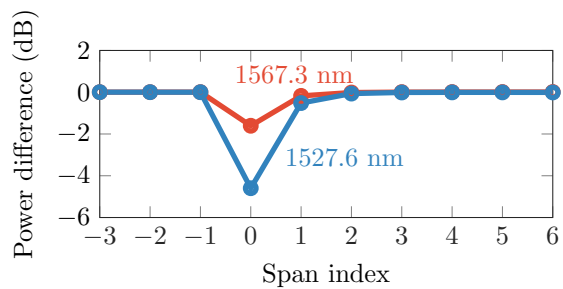

Figure 5.10: Difference in signal power with respect to correct power allocation in the event of a single pump failure at the span indexed by zero. After about two spans the power levels are restored to their correct values.

Fig. 5.9 also illustrates the diminishing returns of operating at a very large number of spatial dimensions. Consider, for instance, the curve for power overhead $P_o = 0.1$ W. The optimal number of spatial dimensions is $S = 20$, resulting in a total capacity per cable of about 383 Tb s$^{-1}$. However, with half of this number of spatial dimensions $S = 10$ (and $P_p = 135$ mW), we can achieve about 80% of that capacity. Thus, systems subject to practical constraints such as cost and size may operate with a number of spatial dimensions that is not very large.

### 5.2.3 Recovery from pump failure

An important practical consideration for submarine systems is their ability to recover when the input power drops significantly due to faulty components or pump laser failure. Thus, submarine amplifiers are designed to operate in high gain compression, so that the power level can recover from these events after a few spans. We show that the optimized input power profile and amplifier operation can still recover from such events. Fig 5.10 illustrates the power variation with respect to the optimized power profile when one of the two pump lasers in an amplification module fails. The failure occurs at the span indexed by zero. The amplifier operates with redundant pumps resulting in $P_p = 50$ mW, and in the event of a single-pump failure the power drops to $P_p = 25$ mW. The signal power in the channels at the extremities of the spectrum are restored with just two spans. Capacity is not significantly affected by a single-pump failure, since the amplifier noise increases by less than 0.5 dB in all channels. Although the power levels could still be restored in the event that the two pump lasers fail, the total amplifier noise power would be about 10 dB higher in some channels.

## 5.3 Summary

We have demonstrated how to maximize the information-theoretic capacity of ultra-long submarine systems by optimizing the channel power allocation in each spatial dimension. Our models account for EDFA physics, Kerr nonlinearity, and power feed limitations. Modeling EDFA physics is paramount to understanding the effects of energy limitations on amplification bandwidth, noise, and optical power, which intimately govern the system capacity. We show that this optimization results in 70% higher capacity when compared to the theoretical capacity of a recently proposed high-capacity system. Our optimization also provides insights on the optimal number of spatial dimensions, amplifier operation, and nonlinear regime operation. Our proposed technique could be used in existing systems, and also to design future systems leveraging SDM.

# Chapter 6

# Conclusions

The first part of this dissertation focused on short-reach optical communication links for data centers. To continue supporting the accelerated Internet traffic growth, next-generation data center transceivers will need to support bit rates beyond 100 Gbit/s per wavelength, while offering high power margin to accommodate additional losses due to longer fiber plant, multiplexing of more wavelengths, and possibly optical switches. These challenges motivated research on spectrally efficient and low-power modulation formats compatible with direct detection, which culminated in the adoption of four-level pulse amplitude modulation (4-PAM) by the IEEE 802.bs task to enable $8 \times 50$ and $4 \times 100$ Gbit/s links. Compared to competing modulation formats such as orthogonal frequency-division multiplexing (OFDM), 4-PAM offers lower transmitter and receiver complexity and higher tolerance to noise. However, as demonstrated in Chapter 2, 4-PAM systems already face tight optical power margin and optical signal-to-noise ratio (OSNR) constraints in unamplified and amplified links, respectively.

To alleviate some of these constraints, emerging technologies can help on a number of fronts. High-bandwidth, low-power modulators based on thin-film lithium niobate, for instance, will reduce intersymbol interference and improve signal integrity. Segmented optical modulators will simplify transmitter-side electronics of multi-level modulation formats. And as discussed in Chapter 3, avalanche photodiodes (APD) and semiconductor optical amplifiers (SOA) may improve receiver sensitivity of 100 Gbit/s 4-PAM systems by 4.5 and 6 dB, respectively. Further work is still necessary in designing APDs that decouple bandwidth from responsivity. Current APD structures are typically made thin to reduce carrier transit time and improve the device bandwidth, but this also reduces the APD responsivity. Another possible fruitful line of research lies on designing Si-based APDs with high-responsivity absorption regions. Si-based APDs offer nearly ideal gain and noise characteristics, but the absorption section of these devices is typically made in Ge, which has poor responsivity.

Aided by those technologies, 4-PAM and possibly other modulation formats compatible with direct detection will meet data center demands in the short-term, but more degrees of freedom are needed to support higher per-wavelength bit rates. Coherent and differentially coherent detection methods enable up to four degrees of freedom while significantly improving receiver sensitivity due to the gain provided by mixing the weak incoming signal with a strong local oscillator laser (LO). However, current coherent receivers rely on power-hungry digital signal processors (DSPs). These DSP-based coherent receivers designed for long-haul transmission, which prioritizes performance, are suboptimal for data center applications, which prioritize cost and power consumption. By reducing receiver complexity and making system performance tradeoffs, the power consumption of coherent links can be made low enough for intra- and inter-data center applications. Following this philosophy, we propose DSP-free coherent and differentially coherent architectures that allow performance comparable to their DSP-based counterparts, while consuming much less power.

Our proposed DSP-free coherent receiver performs three synchronization operations: polarization recovery, carrier recovery, and timing recovery, using analog operations to reduce power consumption. Polarization recovery performs marker tone detection to adjust the individual phase shifts of three (or more) cascaded phase shifters. Carrier recovery is realized by a phase-locked loop (PLL), either optical or electrical. An optical PLL requires frequency modulation of the LO laser as well as highly integrated receivers in order to minimize the loop delay. An electrical PLL eliminates these requirements at the expense of more complex high-speed analog electronics, but the estimated power consumption of the analog electronics remains under 4 W. For either optical or electrical PLL, we proposed a multiplier-free phase detector based on exclusive-OR (XOR) gates. This phase detector is simpler than conventional Costas-type phase detector, but it restricts the modulation to dual-polarization (DP) quadrature phase-shift keying (QPSK). Lastly, timing recovery and detection are performed using conventional clock and data recovery techniques.

Due to their high receiver sensitivity and low power consumption, DSP-free DP-QPSK coherent receivers seem particularly promising for intra-data and inter-data center links. The improved receiver sensitivity would allow 40-km unamplified and eye-safe transmission of up to 49 dense wavelength-division multiplexed (DWDM) channels near 1310 nm, potentially blurring distinctions between intra- and inter-data center links.

In amplified inter-data center links, where receiver sensitivity is not as critical, LO-free differentially coherent receivers for dual-polarization differential QPSK (DP-DQPSK) based on delay interferometers are particularly promising, as the high-speed analog electronics essentially reduces to simple clock and data recovery only.

DSP-free coherent receivers, however, cannot electronically compensate for chromatic dispersion (CD). Hence, they require optical CD compensation by employing dispersion compensating fibers (DCFs). To allow more flexibility and reduce losses introduced by DCFs, future data center links may favor dispersion-shifted fibers (DSFs) with zero dispersion wavelength near 1550 nm, thus allowing low-dispersion amplified links. Nonlinear fiber effects that can be exacerbated by DSFs are negligible in short-reach links. Moreover, dispersion-flattened optical fibers with zero-dispersion wavelengths near both 1310 nm and 1550 nm bands would allow operability of intra-data center links in both bands. If power consumption remains one of the primary concerns in designing optical systems for data centers, these types of fiber should be preferred, since any electronic CD compensation technique will inevitably be more power hungry than passive optics. These same considerations apply to links based on direct detection, as in those systems, CD leads to power fading and limits the system reach.

Future work should solve implementation challenges that will arise in bringing DSP-free coherent receivers to market. On the electronics side, this includes circuit-level design of the proposed receiver functions. Moreover, analog-based phase detectors for higher-order quadrature amplitude modulation (QAM) would allow DSP-free receivers to scale to higher-order formats and provide higher spectral efficiency. On the optics side, commercial coherent receivers today are exclusively designed for 1550 nm operation, but data centers will also require designing polarization and phase hybrids, and similar optical subsystems, for operation at 1310 nm. Moreover, transceivers for data centers applications will likely need to be produced in high volume, which will certainly bring new manufacturing challenges.

Data center links in general will benefit by new advances in photonic integration to reduce cost, power consumption, and to increase port density. Additionally, improved laser frequency stability, either using athermal lasers or frequency combs, will enable DWDM within the data center, possibly yielding a multi-fold increase in capacity per fiber.

As bit rate demands continue to grow, DSP-based receivers may eventually become attractive as they allow higher-order modulation formats and easy compensation of transmission impairments. DSP-based systems may ultimately become economically viable for short-reach data center links by leveraging designs that minimize power consumption and by utilizing more power-efficient complementary metal-oxide semiconductor (CMOS) processes, though that remains unclear as we approach the limits of Moore's law.

The second part of this dissertation focused on long-haul submarine links with an energy constraint. Long-haul submarine links employ hundreds of repeaters (i.e., optical amplifiers) to compensate for fiber loss along the link. These submerged repeaters are designed to operate continuously for over 20 years, and they are powered from the shores, where feed voltage limits the amount of electrical power that can be delivered to the amplifiers. The limited available power per amplifier ultimately limits the amount of optical power and data that can be transmitted per cable.

To mitigate this problem, recent works have turned to an insight from information theory that establishes that in energy-constrained systems, we can maximize capacity by employing more dimensions while transmitting less power, and less data, in each. Dimensions in this context refers to spatial dimensions i.e., modes in multimode fibers (MMFs), cores in multicore fibers (MCFs), or simply multiple single-mode fibers (SMFs). In fact, numerous recent works have studied how employing more spatial dimensions improves the capacity and power efficiency of ultra-long submarine links.

To complement that work, we demonstrated in Chapter 5 how best use each spatial dimension. Specifically, we demonstrated how to maximize the information-theoretic capacity of ultra-long submarine systems by optimizing the channel power allocation in each spatial dimension. Our models account for amplifier physics, Kerr nonlinearity, and power feed limitations. Modeling amplifier physics is paramount to understanding the effects of energy limitations on amplification bandwidth, noise, and optical power, which intimately govern the system capacity.

The main benefit of the channel power optimization is to allow the system to operate over a wider amplification bandwidth by appropriately adjusting the channel powers. Given that capacity scales linearly with dimensions (frequency or space) and only logarithmically with power, the optimization will favor power allocations that maximize the useful amplification bandwidth, i.e., bandwidth over which the amplifier gain is larger than the span attenuation. Interestingly, the optimization does not necessarily make the amplifiers exhibit higher power conversion efficiency.

We show that this optimization improves the capacity by 70% when compared to the theoretical capacity of a recently proposed systems. Our optimization also provides new insights that challenge long-standing assumptions made in designing and analyzing long-haul submarine systems. We show that the optimal number of spatial dimensions can be as large as tens of spatial dimensions in each direction, in contrast with today's systems that typically operate with only eight pairs of fibers. Nonetheless, as the number of spatial dimensions grows, the improvement in capacity diminishes. For instance, we showed that 80% of the cable maximum capacity can be achieved with half the number of dimensions. When the number of spatial dimensions is large, Kerr nonlinearity is negligible. However, systems subject to practical limitations such as cost and size may need to operate with reduced number of dimensions, in which case Kerr nonlinearity may be non-negligible.

In addition to demonstrating the benefits of our proposed optimization experimentally, future work should consider the possibility of using different fibers for long-haul transmission such as MMFs or uncoupled or coupled core MCFs. To become serious contenders MCFs and MMFs need to offer several technical and economical advantages over simply using multiple SMFs. First and foremost, fibers for long-haul transmission must exhibit low loss. This will likely restrict the design of MCFs and MMFs to silica-core fibers, where the attenuation coefficient is as low as 0.16 dB/km. Second, Kerr nonlinearity is smaller in coupled core MCFs and MMFs owing to their large effective area, even when compared to large-effective area SMFs. However, as discussed in Chapter 5, energy constraints favor transmission employing tens of spatial dimensions (modes, cores or fibers) with less power in each dimension. As a result, if practical systems can operate with the optimal number of spatial dimensions, Kerr nonlinearity will become a less important issue. In this scenario, the motivation for coupled core MCFs or MMFs over multiple large-effective area SMFs becomes less apparent. Their higher effective area (lower Kerr nonlinearity) will be less critical, but mode coupling will inevitably require costly, and perhaps impractical, multiple-input multiple-output signal processing at the receiver in order to untangle the modes.

Another important consideration is the electrical power efficiency of optical amplification. In SMF amplifiers, the conversion of optical pump power into signal output power is close to the fundamental limit of 63% for amplifiers pumped near 980 nm. However, the plug power to optical output power efficiency in SMF amplifiers is only 1.5–5%. This poor performance is, among other factors, limited by the pump laser efficiency, which for single-mode lasers is only about 20%. Moreover, each amplifier is pumped by two pump lasers for redundancy.

MCFs and MMFs may make optical amplification of many spatial dimensions more power and cost efficient by leveraging new pumping schemes such as cladding pump, which require fewer pump lasers operating at larger output power. The pump-to-output optical power conversion in these amplifiers will not be superior to SMF amplifiers, as the overlap between the pump mode and the doped cores is smaller than end-pumped SMF amplifiers. However, they potentially can offer higher plug-to-optical power efficiency as they require fewer multi-mode pump lasers, which have efficiency of 46% [111]. Nonetheless, in order to become practical MMF- or MCF-based optical amplifiers need to solve several practical challenges such as low noise figure and flat gain over a wide bandwidth, low cross-talk between the cores or modes, gain-flatness among spatial channels, and low cross-gain modulation due to depletion of a common pump.

# Appendix A

# Derivation of the Gradient of the Channel Capacity with Respect to Channel Power and EDF Length

As discussed in Section 5.1.3, when the particle swarm optimization algorithm ends, a local optimization based on saddle-free Newton's method starts. This method requires knowledge of the Hessian matrix, i.e., matrix of second derivatives.

Although the Hessian can be computed using finite differences method, it is more computationally convenient to compute the gradient analytically and obtain the second derivatives from finite differences of the gradient.

In this appendix we derive analytical equations for the gradient of the amplifier gain with respect to the channel power allocation and erbium-doped fiber (EDF) length. Combining these results with previously published equations for the gradient of the nonlinear noise power, we can derive analytical equations for the gradient of the objective function (spectral efficiency or capacity) with respect to channel power allocation and EDF length. These equations can be used to speed up gradient-descent based simulations Or, for the case of interest in Chapter 5, to compute the Hessian matrix using finite differences of the gradient.

Including nonlinearity, the objective function is given by

$$\text{SE} = 2\sum_{k=1}^{N} s(G_k - A)\log_2(1 + \text{SNR}_k), \qquad (A.1)$$

where $s(x) = 0.5(\tanh(\text{Dx}) - 1)$ is a function that approximates a step function and $D$ determines the sharpness of the step function approximation, $G_k$ is the gain of the $k$-th channel in dB, $A_k$ is the span attenuation in the $k$-th channel in dB, and $\text{SNR}_k$ is the SNR at the $k$-th channel, which is

given by

$$\text{SNR}_k = \frac{P_k}{P_{ASE,k} + \text{NL}_k(\boldsymbol{P})} \tag{A.2}$$

The ASE power $P_{ASE,k} = 2aNn_{sp}(\lambda_k)h\nu_k\Delta f$ does not depend on the signal power $P_k$, but the nonlinear noise power is a function of the power vector $\boldsymbol{P}$ i.e., it depends on the launch power of all channels.

Deriving SE with respect to $P_m$ yields

$$\frac{\partial \text{SE}}{\partial P_m} = \frac{2}{\log(2)} \sum_{k=1}^{N} \left[ \frac{\partial \mathcal{G}_k}{\partial P_m} \log(1 + \text{SNR}_k)s'(\mathcal{G}_k - A) + \frac{\partial \text{SNR}_k}{\partial P_m} \frac{s(\mathcal{G}_k - A)}{1 + \text{SNR}_k} \right] \tag{A.3}$$

The equation above depends on the gradient of the SNR and on the gradient of the gain. Note that the gain $\mathcal{G}$ is in dB in this expression, and therefore the gradient must be computed with respect to the gain in dB. These gradients will be calculated in the next subsections.

## A.1    Gain Gradient

From the semi-analytical derivation the gain is given by the transcendental equation:

$$G_k = \exp\left(a_k(Q^{in} - Q^{out}) - b_k\right) \tag{A.4}$$

$$\implies \frac{\partial G_k}{\partial P_m} = a_k \left( \frac{\partial Q^{in}}{\partial P_m} - \frac{\partial Q^{out}}{\partial P_m} \right) G_k \tag{A.5}$$

where

$$\frac{\partial Q^{in}}{\partial P_m} = \frac{1}{h\nu_m} \tag{A.6}$$

and

$$Q^{out} = \frac{P_p}{h\nu_p} G_p + \sum_i \frac{P_i}{h\nu_i} G_i \tag{A.7}$$

$$\implies \frac{\partial Q^{out}}{\partial P_m} = \frac{1}{h\nu_m} G_m + \frac{P_p}{h\nu_p} \frac{\partial G_p}{\partial P_m} + \sum_i \frac{P_i}{h\nu_i} \frac{\partial G_i}{\partial P_m} \tag{A.8}$$

Note that this gradient includes the pump terms given by the subindex $p$.

Substituting (A.5) in (A.8) and solving for $\frac{\partial Q^{out}}{\partial P_m}$ yields

$$\frac{\partial Q^{out}}{\partial P_m} = \frac{1}{h\nu_m}G_m + \frac{P_p}{h\nu_p}a_p\left(\frac{1}{h\nu_m} - \frac{\partial Q^{out}}{\partial P_m}\right)G_p + \sum_i \frac{P_i}{h\nu_i}a_i\left(\frac{1}{h\nu_m} - \frac{\partial Q^{out}}{\partial P_m}\right)G_i$$

$$\frac{\partial Q^{out}}{\partial P_m}\left(1 + \frac{P_p}{h\nu_p}a_pG_p + \sum_i \frac{P_i}{h\nu_i}a_iG_i\right) = \frac{1}{h\nu_m}\left(G_m + \frac{P_p}{h\nu_p}a_pG_p + \sum_i \frac{P_i}{h\nu_i}a_iG_i\right)$$

$$\frac{\partial Q^{out}}{\partial P_m} = \frac{1}{h\nu_m}\frac{G_m + \frac{P_p}{h\nu_p}a_pG_p + \sum_i \frac{P_i}{h\nu_i}a_iG_i}{1 + \frac{P_p}{h\nu_p}a_pG_p + \sum_i \frac{P_i}{h\nu_i}a_iG_i} \tag{A.9}$$

Now substituting back in (A.5):

$$\frac{\partial G_k}{\partial P_m} = \frac{a_k}{h\nu_m}\left(1 - \frac{G_m + \frac{P_p}{h\nu_p}a_pG_p + \sum_i \frac{P_i}{h\nu_i}a_iG_i}{1 + \frac{P_p}{h\nu_p}a_pG_p + \sum_i \frac{P_i}{h\nu_i}a_iG_i}\right)G_k$$

$$= \frac{a_i}{h\nu_m}\left(\frac{1 - G_m}{1 + \frac{P_p}{h\nu_p}a_pG_p + \sum_i \frac{P_i}{h\nu_i}a_iG_i}\right)G_k$$

$$= \frac{1 - G_m}{h\nu_m}\left(\frac{a_kG_k}{1 + \frac{P_p}{h\nu_p}a_pG_p + \sum_i \frac{P_i}{h\nu_i}a_iG_i}\right) \tag{A.10}$$

To obtain the gradient of the gain in dB we have

$$\frac{\partial \mathcal{G}_k}{\partial P_m} = \frac{10}{\log(10)}\frac{1}{G_k}\frac{\partial G_k}{\partial P_m} \tag{A.11}$$

### A.1.1 Gain derivative with respect to EDF length

It is also necessary to compute the gain gradient with respect to the EDF length.

$$\frac{\partial G_k}{\partial L} = -\left(\alpha_k + a_k\frac{\partial Q^{out}}{\partial L}\right)G_k, \tag{A.12}$$

where

$$\begin{aligned}
\frac{\partial Q^{out}}{\partial L} &= \frac{P_p}{h\nu_p}\frac{\partial G_p}{\partial L} + \sum_i \frac{P_i}{h\nu_i}\frac{\partial G_i}{\partial L} \\
&= -\frac{P_p}{h\nu_p}\left(\alpha_p + a_p\frac{\partial Q^{out}}{\partial L}\right)G_p - \sum_i \frac{P_i}{h\nu_i}\left(\alpha_i + a_i\frac{\partial Q^{out}}{\partial L}\right)G_i \\
\frac{\partial Q^{out}}{\partial L}\left(1 + \frac{P_p}{h\nu_p}a_p G_p + \sum_i \frac{P_i}{h\nu_i}a_i G_i\right) &= -\frac{P_p}{h\nu_p}\alpha_p G_p - \sum_k \frac{P_i}{h\nu_i}\alpha_i G_i \\
\frac{\partial Q^{out}}{\partial L} &= -\frac{\frac{P_p}{h\nu_p}\alpha_p G_p + \sum_i \frac{P_i}{h\nu_i}\alpha_i G_i}{1 + \frac{P_p}{h\nu_p}a_p G_p + \sum_k \frac{P_i}{h\nu_i}a_i G_i}
\end{aligned} \quad (A.13)$$

Substituting in (A.12) yields

$$\frac{\partial G_k}{\partial L} = -\left(\frac{\alpha_k G_k}{1 + \frac{P_p}{h\nu_p}a_p G_p + \sum_i \frac{P_k}{h\nu_i}a_i G_i}\right), \quad (A.14)$$

## A.2  SNR gradient

The SNR depends on the $m$th channel power through the power itself and through the nonlinear noise power. The derivative of the SNR in the $k$th channel with respect to the power in the $m$th channel is given by

$$\frac{\partial \text{SNR}_k}{\partial P_m} = \begin{cases} -\frac{\partial \text{NL}_k}{\partial P_m}\frac{P_k}{(P_{\text{ASE},k}+\text{NL}_k)^2}, & k \neq m \\ -\frac{\partial \text{NL}_k}{\partial P_m}\frac{P_k}{(P_{\text{ASE},k}+\text{NL}_k)^2} + \frac{1}{P_{\text{ASE},k}+\text{NL}_k}, & k = m \end{cases} \quad (A.15)$$

$$= \begin{cases} -\frac{\partial \text{NL}_k}{\partial P_m}\frac{\text{SNR}_k^2}{P_k}, & k \neq m \\ -\frac{\partial \text{NL}_k}{\partial P_m}\frac{\text{SNR}_k^2}{P_k} + \frac{\text{SNR}_k}{P_k}, & k = m \end{cases} \quad (A.16)$$

$$= \mathbb{1}\{k = m\}\frac{\text{SNR}_m}{P_m} - \frac{\partial \text{NL}_k}{\partial P_m}\frac{\text{SNR}_k^2}{P_k} \quad (A.17)$$

Note that this equation is written in terms of $\text{SNR}_k$ for convenience. In practice, it is easier to write out this equation as a function of the total noise power in order to avoid divisions by the power $P_k$, which could go to zero during optimization.

For the nonlinear noise power, we have

$$\text{NL}_n(\boldsymbol{P}) = \sum_{n_1=1}^{N}\sum_{n_1=1}^{N}\sum_{l=-1}^{1}\tilde{P}_{n_1}\tilde{P}_{n_2}\tilde{P}_{i+j-n+l}D_l(n_1,n_2,n), \quad n = 1,\ldots,N, \quad (A.18)$$

where $\tilde{P}$ is the launch power into the fiber at each channel, so that

$$\tilde{P}_k = P_k e^{\alpha_{\text{SMF},k} L}, \tag{A.19}$$

since the gain flattening filter ideally makes the channel gain equal to the fiber attenuation. Therefore, once we know $\frac{\partial NL_k}{\partial \tilde{P}_m}$, we can obtain $\frac{\partial NL_k}{\partial P_m}$ from

$$\frac{\partial \text{NL}_k}{\partial P_m} = e^{\alpha_{\text{SMF},m} L} \frac{\partial \text{NL}_k}{\partial \tilde{P}_m} \tag{A.20}$$

$\frac{\partial NL_k}{\partial \tilde{P}_m}$ can be computed by inspection of (A.18), where each term is differentiated independently. An equation for the gradient of $\text{NL}_n(\tilde{P})$ with respect to logarithmic power is given in [113, Appendix].

## A.3  Spectral efficiency gradient

Returning to (A.3)

$$\frac{\partial \text{SE}}{\partial P_m} = \frac{2}{\log(2)} \sum_{k=1}^{N} \left[ \frac{\partial \mathcal{G}_k}{\partial P_m} \log(1+\text{SNR}_k) s'(\mathcal{G}_k - A) + \frac{\partial \text{SNR}_k}{\partial P_m} \frac{s(\mathcal{G}_k - A)}{1+\text{SNR}_k} \right] \tag{A.21}$$

we can now substitute

$$\frac{\partial \mathcal{G}_k}{\partial P_m} = \frac{10}{\log(10)} \frac{1}{G_k} \frac{1-G_m}{h\nu_m} \left( \frac{a_k G_k}{1 + \frac{P_p}{h\nu_p} a_p G_p + \sum_i \frac{P_i}{h\nu_i} a_i G_i} \right) \tag{A.22}$$

$$\frac{\partial \text{SNR}_k}{\partial P_m} = \mathbb{1}\{k=m\} \frac{\text{SNR}_m}{P_m} - \frac{\partial \text{NL}_k}{\partial P_m} \frac{\text{SNR}_k^2}{P_k} \tag{A.23}$$

Since the optimization is realized with power values in dBm ($\mathcal{P}$), it's necessary to change the differentiation variable:

$$\frac{\partial SE}{\partial P_m} = \frac{\partial SE}{\partial \mathcal{P}_m} \frac{\partial \mathcal{P}_m}{\partial P_m}$$

$$\implies \frac{\partial SE}{\partial \mathcal{P}_m} = \left( \frac{\partial \mathcal{P}_m}{\partial P_m} \right)^{-1} \frac{\partial SE}{\partial P_m} = \frac{P_m \log(10)}{10} \frac{\partial SE}{\partial P_m} \tag{A.24}$$

### A.3.1  Spectral efficiency gradient derivative with respect to EDF length

$$\frac{\partial \text{SE}}{\partial L} = 2 \sum_k \frac{\partial \mathcal{G}_k}{\partial L} s'(\mathcal{G}_k - A) \log_2(1+\text{SNR}_k), \tag{A.25}$$

where the gradient of the gain in dB with respect to the EDF length is given by

$$\frac{\partial \mathcal{G}_k}{\partial L} = -\frac{10}{\log(10)} \left( \frac{\alpha_k G_k}{1 + \frac{P_p}{h\nu_p} a_p G_p + \sum_i \frac{P_k}{h\nu_i} a_i G_i} \right) \tag{A.26}$$

...